\documentclass[usegraphicx,usenatbib,useAMS]{mn2e}

\usepackage{times} %use Adobe fonts - nicer for viewing PDF
\usepackage{aas_macros}     % for journal abbreviations in Bibtex references from ADS
\usepackage{amssymb}        % for math symbols

\newcommand{\gsim}{\gtrsim} % alias for symbol available through amssymb 
\newcommand{\lsim}{\lesssim} % alias for symbol available through amssymb

\newcommand{\Msol}{M_\odot}
\newcommand{\hMsol}{h^{-1}M_\odot}

\newcommand{\Zsol}{Z_\odot}

\newcommand{\ergcms}{{\rm erg\,cm^{-2}s^{-1}}}

\newcommand{\mum}{\mu{\rm m}}
\newcommand{\pc}{{\rm pc}}
\newcommand{\kpc}{{\rm kpc}}
\newcommand{\Mpc}{{\rm Mpc}}

\newcommand{\kms}{{\rm km\,s^{-1}}}
\newcommand{\yr}{{\rm yr}}
\newcommand{\Myr}{{\rm Myr}}
\newcommand{\Gyr}{{\rm Gyr}}

\newcommand{\GALFORM}{{\tt GALFORM}} 
\newcommand{\GRASIL}{{\tt GRASIL}}
\newcommand{\SPITZER}{{\em Spitzer}}

\newcommand{\GALEX}{{\em GALEX}}
\newcommand{\JWST}{{\em JWST}}

\newcommand{\Bau}{Baugh05}
\newcommand{\Bow}{Bower06}

\newcommand{\Vc}{V_{\rm c}}
\newcommand{\Vcrit}{V_{\rm crit}}
\newcommand{\zreion}{z_{\rm reion}}
\newcommand{\Vhot}{V_{\rm hot}}
\newcommand{\alphahot}{\alpha_{\rm hot}}
\newcommand{\ntau}{n_{\tau}}
\newcommand{\tauburstmin}{\tau_{\ast {\rm burst,min}}}
\newcommand{\taudyn}{\tau_{\rm dyn}}
\newcommand{\fdyn}{f_{\rm dyn}}
\newcommand{\tesc}{t_{\rm esc}}
\newcommand{\fcloud}{f_{\rm cloud}}
\newcommand{\Mcloud}{M_{\rm cloud}}
\newcommand{\rcloud}{r_{\rm cloud}}

\newcommand{\MUV}{M_{\rm AB}(1500\AA)-5\log h}

\title[Evolution of Lyman-break galaxies]
{The evolution of Lyman-break galaxies in CDM}

\author[Lacey  et al.]{C. G. Lacey
\thanks{E-mail: Cedric.Lacey@durham.ac.uk (CGL)},$^1$
C. M. Baugh,$^1$ C.S. Frenk,$^1$ A.J. Benson$^2$\\
$^{1}$Institute for Computational Cosmology, Department of Physics,
University of Durham, South Road, Durham, DH1 3LE, UK\\
$^{2}$Mail Code 130-33, California Institute of Technology,
Pasadena, CA 91125, USA}

\begin{document}

%\date{}

\maketitle

\begin{abstract}
We make a detailed investigation of the properties of Lyman-break
galaxies (LBGs) in the $\Lambda$CDM model. We present predictions for
two published variants of the \GALFORM\ semi-analytical model: the
\citet{Baugh05} model, which has star formation at high redshifts
dominated by merger-driven starbursts with a top-heavy IMF, and the
\citet{Bower06} model, which has AGN feedback and a standard Solar
neighbourhood IMF throughout. We show predictions for the evolution of
the rest-frame far-UV luminosity function in the redshift range
$z=3-20$, and compare with the observed luminosity functions of LBGs
at $z=3-10$. We find that the \citeauthor{Baugh05} model is in
excellent agreement with these observations, while the
\citeauthor{Bower06} model predicts too many high-luminosity
LBGs. Dust extinction, which is predicted self-consistently based on
galaxy gas contents, metallicities and sizes, is found to have a large
effect on LBG luminosities. We compare predictions for the size
evolution of LBGs at different luminosities with observational data
for $2\lsim z \lsim 7$, and find the \citeauthor{Baugh05} model to be
in good agreement. We present predictions 
for stellar, halo and gas masses, star formation rates, circular
velocities, bulge-to-disk ratios, gas and stellar metallicities and
clustering bias, as functions of far-UV luminosity and redshift. We
find broad consistency with current observational constraints.
We then present predictions for the abundance and angular sizes of
LBGs out to very high redshift ($z \leq 20$), finding that planned
deep surveys with \JWST\ should detect objects out to $z \lsim
15$. 
%% Finally, we consider the evolution of the star formation and
%% far-UV luminosity densities in the model. 
We predict that the effects of dust extinction on the far-UV
luminosity density should be large ($\sim 2$~mag), even out to high
redshifts. The typical UV luminosities of galaxies are predicted to be
very low at high redshifts, which has implications for detecting the
galaxies responsible for reionizing the IGM; for example, at $z=10$,
50\% of the ionizing photons are expected to be produced by galaxies
fainter than $\MUV \sim -15$.

\end{abstract}


\section{Introduction}

The discovery of Lyman-break galaxies at $z \sim 3$ by
\citet{Steidel96} was a breakthrough in observational studies of
galaxy formation.  It provided, for the first time, a significant
sample of normal galaxies at high redshift whose properties and
population statistics could then be investigated observationally and
compared to the predictions of theoretical models \citep{Baugh98}.

Lyman-break galaxies (LBGs) are star-forming galaxies which are
identified through the Lyman-break feature in their spectra. This
feature is produced by absorption by neutral hydrogen in the
atmospheres of massive stars, in the interstellar medium (ISM) of the
galaxy and in the intergalactic medium (IGM) \citep{Steidel92}. For
ground-based observations, detection of the Lyman break is restricted
to redshifts $z \gsim 3$.  Since the first successful demonstration by
\citet{Steidel96} at $z \sim 3$, the technique has been extended to
identify galaxies at both higher redshifts and lower luminosities,
using ground-based telescopes and the Hubble Space Telescope (HST)
\citep[e.g.][]{Madau96,Steidel99,Bouwens03,Shimasaku05,Yoshida06}.
Finding LBGs at $z \gsim 6$ requires observing in the near-IR, which
was first done using NICMOS on HST, and led to the identification of a
small number of $z \sim 7-8$ objects \citep{Bouwens04b}. More
recently, the WFC3/IR camera on HST has allowed the discovery of much
larger samples of LBGs at $z\sim 7-8$
\citep{Bouwens10a,McLure10,Oesch10a,Bunker10,Yan09} and even a few
candidates at $z\sim 10$ \citep{Bouwens10b}. By observing in the UV
from space, the \GALEX\ satellite has also been used to find LBGs at
$z \sim 1$ \citep{Burgarella06}.

Follow-up observational investigations on LBGs have included estimates
of their star formation rates (SFRs), sizes, morphologies, stellar and
dynamical masses, galactic outflows, metallicities, dust extinctions,
gas masses, IR/sub-mm emission from dust and clustering
\citep{Steidel96,Giavalisco96,Adelberger98,Chapman00,Pettini01,Shapley01,Ferguson04}.
Since LBGs are selected on the basis of their rest-frame far-UV
emission, which is dominated by massive young stars, LBG samples at
different redshifts also provide a means to trace the cosmic star
formation history, a key component in our picture of galaxy formation
\citep{Madau96}, although important uncertainties remain due to the
effects of dust extinction. Observations of LBGs at $z=3-10$ probe
galaxy evolution over the first 3--15\% of the age of the universe.

Other observational techniques have also been used to find normal
galaxies at high redshift. Searches for star-forming galaxies via
their Ly$\alpha$ emission line \citep[e.g.][]{Hu98} cover a similar
redshift range to LBGs. The main drawback with this technique is that
some star forming-galaxies show Ly$\alpha$ absorption rather than
emission \citep{Shapley03}, and consequently are missed in narrowband
surveys. A further complication is that inferring the SFR from the
Ly$\alpha$ emission line is much more uncertain than inferring it from
the far-UV continuum, since the effects of dust extinction are
amplified by resonant scattering of Ly$\alpha$ photons by hydrogen
atoms. Another technique is to search for sub-mm or IR emission from
dust in high-$z$ star-forming galaxies \citep{Smail97,Hughes98}. This
method is currently limited by source confusion at faint fluxes due to
the relatively poor angular resolution of current IR/sub-mm
telescopes, which restricts searches to redshifts $z \lsim 3$ and
mostly to the galaxies with the highest SFRs. Other techniques for
selecting high-$z$ galaxies, which are sensitive also to
non-star-forming galaxies (such as ERO and DRG colour selection) are
limited to even lower redshifts $z \lsim 2$. Overall, the Lyman-break
technique still seems the most effective  for finding large samples
of star-forming galaxies at  $z \gsim 3$ that cover a wide
range of luminosities and SFRs.

The theoretical significance of the discovery of LBGs was highlighted
early on using semi-analytical models of galaxy
formation. \citet{Baugh98} showed that the abundance and observed
properties of the $z \sim 3-4$ LBGs found by \citet{Steidel96} and
\citet{Madau96} could be explained in the framework of CDM, and that
they fitted into a picture in which the cosmic SFR density peaked
around a redshift $z \sim 2$. This evolution of the cosmic SFR density
was driven by the combined effects of the build-up of dark matter
halos, gas cooling and supernova feedback. Their model had star
formation occuring mostly in quiescent disks, and neglected the
effects of dust extinction. Subsequent observational studies found
evidence from UV continuum slopes for significant dust extinction in
LBGs \citep{Meurer99,Steidel99}. \citet{Somerville01} proposed a
different semi-analytical model in which star formation bursts
triggered by galaxy mergers played an important role, and combined
this with an empirical prescription for dust extinction, tuned to
match observational estimates of the extinction for $z\sim 3$
LBGs. 

More recent studies of LBGs in semi-analytical models include
\citet{Guo09} and \citet{LoFaro09}, {which investigated the
effects on inferred luminosity functions and other properties of
applying observational LBG colour selections to model galaxies. The
former used a phenomonelogical model for dust extinction, while the
latter used a physical model similar to that in the present paper.}
LBGs were also studied in gas-dynamical simulations of galaxy
formation \citep{Nagamine02,Weinberg02}, but these simulations had the
drawback that they did not predict properties for the present-day
galaxy population consistent with observations, unlike the
semi-analytical models. Furthermore, none of these models were able to
explain the number counts and redshifts of faint sub-mm galaxies
discovered in surveys at 850$\mum$ \citep{Smail97}, which were
subsequently shown to be dusty starbursts at $z\sim 2-3$
\citep{Chapman03}.

In order to explain within a single framework the sub-mm and
Lyman-break galaxies at high redshift, together with a wide range of
galaxy properties at $z=0$ (including optical and near- and far-IR
luminosity functions, gas fractions, metallicities and galaxy sizes),
\citet{Baugh05} introduced a new semi-analytical model in which the
gas consumption timescale in disks at high redshifts was increased,
with the result that starbursts triggered by galaxy mergers played a
more significant role at high redshift. They assumed, further, that
stars formed in these bursts with a top-heavy initial mass function
(IMF). Unlike previous models of high-redshift galaxies, this model
included a fully self-consistent treatment of both absorption and
emission of radiation by dust, with dust extinction calculated from
radiative transfer based on the predicted gas masses, metallicities
and sizes of galaxies, and the spectrum of the dust emission
calculated by solving for the temperature distribution of the dust
grains in each galaxy. In subsequent papers, we have explored other
predictions from the same model, including stellar and gas
metallicities \citep{Nagashima05a,Nagashima05b}, galaxy colours, sizes
and morphologies in the local universe \citep{Almeida07,Gonzalez09},
the evolution of Ly$\alpha$-emitters
\citep{LeDelliou05,LeDelliou06,Orsi08}, and the evolution of galaxies
at mid- and far-IR wavelengths \citep{Lacey08,Lacey10}, finding
generally good agreement with observational data.

In \citet{Baugh05}, we made only a limited comparison with
observational data on LBGs, focussing on their rest-frame far-UV
luminosity function at $z=3$. Since then, there has been a huge
increase in the amount and quality of observational data on LBGs, in
particular enabling measurements of the luminosity function of LBGs
out to $z\sim 10$. Therefore in this paper we return to studying LBGs,
making detailed predictions for the evolution of their luminosity
functions over a wide redshift range ($z = 3-20$) and for many other
properties. We consider two variants of the \GALFORM\ semi-analytical
model \citep{Cole00}, those of \citet{Baugh05} and
\citet{Bower06}. The two models differ in a number of significant
ways, the most important being that the \citeauthor{Bower06} model
includes AGN feedback, while the \citeauthor{Baugh05} model has a
variable IMF, as already mentioned. We focus here on far-UV-selected
galaxies in the redshift range $z\gsim 3$, where they are
observationally detected using their Lyman-break features. We
investigate the present-day descendants of LBGs in a companion paper
\citep{Gonzalez10b}, and make predictions for the reionization of the
IGM from the same models in \citet{Raicevic10a,Raicevic10b}. The
properties and evolution of far-UV-selected galaxies at lower
redshifts will be the topic of a separate paper.

The plan of this paper is as follows: In \S\ref{sec:GALFORM}, we
briefly review the main features of the two models. In
\S\ref{sec:lf-evoln}, we compare predictions for the evolution of the
far-UV luminosity function with observational data from LBGs, and
investigate the sensitivity of these predictions to various model
parameters. In \S\ref{sec:props}, we investigate the sizes and other
physical properties of UV-selected galaxies, and carry out a detailed
comparison with the observed sizes of LBGs. In \S\ref{sec:high-z}, we
present predictions for LBGs at very high redshifts, which may be
accessible with future telescopes such as \JWST\ and ELTs. In
\S\ref{sec:UVdens}, we show how our predictions for LBGs fit into the
wider picture of the evolution of the cosmic star formation and UV
luminosity densities. We briefly consider the contribution of LBGs to
the reionization of the IGM. Finally, we present our conclusions in
\S\ref{sec:conc}.

\section{The \GALFORM\ galaxy formation model}
\label{sec:GALFORM}

We compute the formation and evolution of galaxies within the
framework of the $\Lambda$CDM model of structure formation using the
semi-analytical galaxy formation model \GALFORM. The general
methodology and approximations behind  \GALFORM\  are set out
in detail in \citet{Cole00} (see also the review by
\citealt{Baugh06}). In summary,  \GALFORM\  follows the main
processes which shape the formation and evolution of galaxies. These
include: (i) the collapse and merging of dark matter halos; (ii) the
shock-heating and radiative cooling of gas inside dark halos, leading
to the formation of galaxy disks; (iii) quiescent star formation in
galaxy disks; (iv) feedback from supernova explosions, from AGN
heating, and from photoionization of the IGM; (v) chemical enrichment
of the stars and gas; (vi) galaxy mergers driven by dynamical friction
within common dark matter halos, leading to the formation of stellar
spheroids, and also triggering bursts of star formation.  The end
product of the calculations is a prediction of the numbers and
properties of galaxies that reside within dark matter haloes of
different masses. The model predicts the stellar and cold gas masses
of the galaxies, along with their star formation and merger histories,
their disk and bulge sizes, and their metallicities. The stellar
luminosities of the galaxies are then computed from their star
formation and chemical enrichment histories using a stellar population
model. Finally, the dust extinctions at different wavelengths are
calculated self-consistently from their gas and metal contents and
sizes using a radiative transfer model.

\subsection{\Bau\ and \Bow\ models}

The two variants of \GALFORM\ considered in this paper,
\citet{Baugh05} and \citet{Bower06} (hereafter \Bau\ and \Bow), have
been used as fiducial models in a number of investigations. We now
briefly compare the main features of the two models. Similar
discussions can be found in \citet{Almeida07,Almeida08},
\citet{Gonzalez09} and \citet{Gonzalez-Perez09}. For full details of
the two models, see the papers by \citet{Cole00}, \citet{Baugh05}
(also \citet{Lacey08}), and \citet{Bower06}. {We give a summary of
previously published results obtained with the two models, and of
their relative performances in explaining different observed
properties of galaxies, in the Appendix.}

\begin{enumerate}

\item {\em Cosmology:}  
The \Bau\ model adopts a $\Lambda$CDM cosmology
with a present-day matter density parameter, $\Omega_{m}=0.3$, a
cosmological constant, $\Omega_{\Lambda}=0.7$, a baryon density,
$\Omega_{b}=0.04$, a Hubble constant $h=0.7$ in units of $100 \kms
\Mpc^{-1}$, and a power spectrum normalization given by
$\sigma_{8}=0.93$. The \Bow\ model instead uses the cosmological model
assumed in the Millennium simulation \citep{Springel05}, where
$\Omega_{m}=0.25$, $\Omega_{\Lambda}=0.75$, $\Omega_{b}=0.045$,
$h=0.73$ and $\sigma_{8}=0.9$.

\item{\em Star formation in disks:} 
In both models, stars can form either
quiescently in disks or in starbursts at a rate $\psi = M_{\rm
gas}/\tau_{\ast}$, where $M_{\rm gas}$ is the mass of cold gas and the
timescale, $\tau_{\ast}$, is different in disks and bursts. The two
models adopt different parameterizations for the dependence of the
quiescent timescale on disk properties. In the \Bau\ model, this
timescale varies simply as a power of the disk circular velocity,
while in the \Bow\ model it is proportional also to the disk dynamical
time. Since the typical dynamical time gets shorter with increasing
redshift, the star formation timescale in the \Bow\ model is shorter
at high redshift than it would be in the equivalent disk in the \Bau\
model. As a consequence, galactic disks at high redshift tend to be
gas poor in the \Bow\ model, but gas rich in the \Bau\ model. This
then results in much more gas being available to fuel starbursts and
in a higher fraction of star formation occuring in bursts at high
redshift in the \Bau\ compared to the \Bow\ model.

\item{\em Starbursts:}
In the \Bau\ model, starbursts are triggered only by galaxy mergers
(both major and minor), while in the \Bow\ model, starbursts are
triggered also by disk instabilities. (We define major mergers as
those in which the mass ratio of the smaller to larger galaxy exceeds
0.3.) Both major mergers and disk instabilities are assumed to
transform the stellar disk(s) into a spheroid, while in minor mergers
only the stars from the smaller galaxy are added to the spheroid,
leaving the stellar disk of the larger galaxy intact. Stars formed in
bursts are always added to the spheroid. The star formation timescale
in bursts is assumed to be
\begin{equation}
\tau_{\ast{\rm burst}} = \max[\fdyn\taudyn,\tauburstmin], 
\label{eq:tauburst}
\end{equation}
so that it scales with the spheroid dynamical time $\taudyn$, but with
a ``floor'' value when this is very short. The parameters $\fdyn$ and
$\tauburstmin$ are different in the two models: $\fdyn=50$ and
$\tauburstmin=0.2\Gyr$ in the \Bau\ model and $\fdyn=2$ and
$\tauburstmin=0.005\Gyr$ in the \Bow\ model, so that the burst SFR
timescale is 25-40 times larger in the \Bau\ model compared to the
equivalent galaxy in the \Bow\ model. {The SFR in a burst then
decays exponentially with time, with a timescale that depends on the
supernova feedback efficiency and gas recycling fraction as well as
the SFR timescale. The burst is assumed to be truncated after $\ntau$
e-folding times, at which point the remaining gas in the burst is
ejected into the halo (see \citet{Granato00} for more details). The
\Bau\ and \Bow\ models both assume $\ntau=3$.}

\item {\em Stellar initial mass function (IMF):}  The \Bow\ model uses
the \citet{Kennicutt83} IMF, consistent with deductions from the Solar
neighbourhood, in all modes of star formation. This has the form ${\rm
d}N/{\rm d}\ln m \propto m^{-x}$ (where $N$ is the number of stars and
$m$ is the stellar mass) with slope $x=0.4$ for $m< \Msol$ and $x=1.5$
for $m> \Msol$ (compared to $x=1.35$ for the Salpeter IMF).  The \Bau\
model also adopts this IMF in quiescent star formation in galactic
disks.  However, in starbursts triggered by galaxy mergers, a
top-heavy IMF is assumed, with slope $x=0$.  The IMF covers the
stellar mass range $0.15<m<120 \Msol$ in all cases. The yield of
metals ($p$) and the fraction of gas recycled per unit mass of stars
formed ($R$) are chosen to be consistent with the form of the
IMF. {We assume $p=0.023$ and $R=0.41$ for the Kennicutt IMF,
and $p=0.15$ and $R=0.91$ for the top-heavy $x=0$ IMF.}

\item {\em Supernova (SN) feedback:} In both models, supernova
explosions are assumed to reheat cold gas (in both disks and
starbursts) and eject it into the halo, at a rate
\begin{equation}
\dot{M}_{\rm eject}=(V_{\rm hot}/V_{\rm gal})^{\alpha_{\rm hot}} \psi, 
\label{eq:snfeedback}
\end{equation}
where $V_{\rm gal}$ is the circular velocity of the disk (for
quiescent star formation) or spheroid (for bursts), and $V_{\rm hot}$
and $\alpha_{\rm hot}$ are parameters.  The SN feedback is much
stronger in the \Bow\ model ($V_{\rm hot}=485 {\rm km s}^{-1}$ and
$\alpha_{\rm hot} = 3.2$, compared with $V_{\rm hot}=300 {\rm km
s}^{-1}$ and $\alpha_{\rm hot} =2$ in the \Bau\ model). A further
difference is that in the \Bow\ model ejected gas is reincorporated
into the halo on a shorter timescale than in the \Bau\ model. This SN
feedback supresses star formation much more effectively in low mass
than high mass galaxies.

\item {\em Superwind vs AGN feedback:} A major difference between the
two models is in the modelling of feedback in massive galaxies. If only
the ``standard'' SN feedback described above is included, then too
much gas cools in massive halos, resulting in the break at the bright
end of the present-day optical and near-IR galaxy luminosity functions
being at too high luminosity. In the \Bau\ model, this problem is
solved by introducing supernova-driven {\em superwinds} (see
\citet{Benson03} and \citet{Lacey08} for details), which eject gas
completely from the halo at a rate proportional to the SFR, and
operate alongside the standard SN feedback. Such winds have been
observed in massive galaxies, with the inferred mass ejection rates
found to be comparable to the star formation rate
\citep[e.g.][]{Heckman90,Pettini01}. The effect of expelling gas from
halos is to reduce the density of hot gas in them and increase the
radiative cooling time, resulting, in particular, in less gas cooling
in massive halos. In contrast, the \Bow\ model solves the same problem
by introducing feedback from AGN. Halos in which the cooling time of
the gas exceeds the free-fall time and which also contain a
sufficiently large central supermassive black hole (SMBH) are assumed
to set up a steady state in which halo gas accreting onto the SMBH
releases energy in the form of relativistic jets which heat the halo
gas, exactly balancing the halo cooling, and preventing any gas from
cooling onto the galaxy disk. The SMBH are assumed to grow by gas
accretion in starbursts triggered by galaxy mergers and disk
instabilities and, less commonly, by black hole/black hole mergers (as
detailed in \citealt{Bower06}, \citealt{Malbon07} and
\citealt{Fanidakis10}). In the \Bow\ model, the disk instabilities
appear to be critical for producing large enough SMBH at early enough
times for the AGN feedback to be effective.

\end{enumerate}

The assumption of a top-heavy IMF in starbursts in the \Bau\ model is
a controversial one. As discussed in detail in \citet{Baugh05}, this
top-heavy IMF was found to be required in order to reproduce the
observed number counts and redshift distributions of the faint sub-mm
galaxies. The top-heavy IMF results both in higher bolometric and
far-UV luminosities for young stellar populations, and greater
production of heavy elements and hence also dust, both effects being
important for reproducing the properties of SMGs in the
model. Furthermore, as shown by \citet{Nagashima05a,Nagashima05b}, the
predicted chemical abundances of the X-ray emitting gas in galaxy
clusters and of the stars in elliptical galaxies also agree better
with observational data in a model with the top-heavy IMF in bursts,
rather than one with a universal Solar neighbourhood IMF. Subsequent
work using the same model has also shown that it predicts galaxy
evolution in the mid-IR in good agreement with observations by \SPITZER\
\citep{Lacey08}. A more detailed comparison of the model with the
properties of observed SMGs has been carried out by
\citet{Swinbank08}, and an investigation of the present-day
descendants of SMGs in \citet{Gonzalez10a}. As shown by
\citet{LeDelliou06} and \citet{Orsi08}, the same model
also reproduces the observed evolution of the luminosity function and
clustering of Ly$\alpha$-emitting galaxies at high redshift.

A variety of other observational evidence has accumulated which
suggests that the IMF in some environments may be top-heavy compared
to the Solar neighbourhood IMF (see \citealt{Elmegreen09} for a recent
review). \citet{Rieke93} argued for a top-heavy IMF in the nearby
starburst M82, based on modelling its integrated properties, while
\citet{Parra07} found possible evidence for a top-heavy IMF in the
ultra-luminous starburst Arp220 from the relative numbers of
supernovae of different types observed at radio wavelengths.  Evidence
has been found for a top-heavy IMF in some star clusters in intensely
star-forming regions, both in M82 \citep[e.g.][]{McCrady03}, and in
our own Galaxy
\citep[e.g.][]{Figer99,Stolte05,Harayama08}. Observations of both the
old and young stellar populations in the central 1~pc of our Galaxy
also favour a top-heavy IMF \citep{Paumard06,Maness07}. In the local
Universe, \citet{Meurer09} find evidence for variations in the IMF
between galaxies from variations in the H$\alpha$/UV luminosity
ratio. \citet{Fardal07} found that reconciling measurements of the
optical and IR extragalactic background with measurements of the
cosmic star formation history also seemed to require an average IMF
that was somewhat top-heavy. \citet{Perez08} compared observational
constraints on the evolution of the star formation rate density and
stellar mass density over cosmic time, and found that reconciling
these two types of data also favours a more top-heavy IMF at higher
redshifts, as had been hinted at by earlier studies. Finally,
\citet{Dokkum08} found that reconciling the colour and luminosity
evolution of early-type galaxies in clusters also favoured a top-heavy
IMF. \citet{Larson98} summarized other evidence for a top-heavy IMF
during the earlier phases of galaxy evolution, and argued that this
could be a natural consequence of the temperature dependence of the
Jeans mass for gravitational instability in gas
clouds. \citet{Larson05} extended this to argue that a top-heavy IMF
might also be expected in starburst regions, where there is strong
heating of the dust by the young stars.

\subsection{Dust extinction.} 

{The attenuation of starlight by dust is a crucial part of our
model, since LBGs are detected by their rest-frame far-UV emission,
which appears to be heavily extincted. In our earlier investigation of
LBGs in \citet{Baugh05}, we calculated  both the absorption and
emission of starlight by dust in model galaxies by coupling \GALFORM\
to the \GRASIL\ spectrophotometric code \citep{Silva98}, which
incorporates a radiative transfer calculation of starlight through a
realistic dust distribution, and a detailed model of heating and
cooling of dust grains (including PAH molecules). The \GRASIL\ model
assumes a composite disk plus spheroid geometry for quiescent
galaxies, with stars in both components, but dust only in the disk. In
the case of ongoing bursts, this is replaced by a stellar spheroid
plus a flattened burst component containing the dust, and having the
same half-mass radius as the spheroid. The dust is assumed to be in two
phases, a diffuse medium and also giant molecular clouds in which the
stars form, and from which they escape after a few \Myr. The coupled
\GALFORM+\GRASIL\ model is described in \citet{Granato00}.  }

{The \GRASIL\ model is computationally expensive, since it
involves a detailed radiative transfer calculation for each model
galaxy, and furthermore in the present paper we only need to calculate
the extinction of starlight from galaxies, and not the IR/sub-mm dust
emission.  In this paper, we therefore calculate dust extinction using
an alternative approach, which assumes the same geometry for the stars
and dust as in \GRASIL, but gains in computational speed by making
some simplifying approximations. This alternative method is described
in detail in \citet{Lacey10b}, where we also show that it gives very
similar results to \GRASIL\ for the dust extinction and for the far-IR
(but not mid-IR) emission from dust. Here we give a brief summary of
how dust extinction is calculated in this alternative approach. }

{The starting point of the approach used here is the model for
extinction by diffuse, smoothly distributed dust described in
\citet{Cole00}, to which we add a treament of extinction of young
stars by molecular clouds. We calculate the extinction by the diffuse
component of the dust using the tabulated radiative transfer models of
\citet{Ferrara99}, which assume that the stars are in both a disk and
a spheroid, and that the dust is uniformly mixed with the stars in the
exponential disk, with a Milky Way extinction law. This already gives
very different extinctions from a foreground screen model with the
same column density of dust. The \citeauthor{Ferrara99} models
give dust attenuation factors separately for the disk and spheroid
light, as functions of wavelength, galaxy inclination, central face-on
V-band dust optical depth, and ratio of spheroid to disk
scalelengths.} 

{The radii of the disk and spheroid of each model galaxy
are predicted directly by \GALFORM, as described in \citet{Cole00},
based on conservation of angular momentum (for the disk) and energy
(for the spheroid), and taking account of their own self-gravity and
that of the dark halo. The \citeauthor{Ferrara99} models assume a
fixed ratio of vertical to radial exponential scalelengths for the
disk stars of 0.1. We assume that the dust has the same radial and
vertical scalelengths as the stars.  The total mass of dust in each
galaxy is calculated from the cold gas mass and metallicity predicted
by \GALFORM, assuming that the dust-to-gas ratio is proportional to
the gas metallicity, and normalized to match the solar neighbourhood
dust-to-gas ratio for solar metallicity ($\Zsol=0.02$). The only
change relative to \citet{Cole00} is that we assume that a fraction
$\fcloud$ of the cold gas and dust are in molecular clouds, so the
mass of dust in the diffuse phase is a fraction $1-\fcloud$ of the
total dust mass. We then calculate the central face-on optical depth
of the diffuse dust from its mass and radial scalelength. Finally, we
choose a random inclination angle for each model galaxy. \GALFORM\
calculates separate disk and spheroid luminosities at each wavelength,
so we interpolate in the tables and then apply the separate dust
attenuation factors for the disk and spheroid light.  }

{We model extinction by molecular clouds based on the same
physical picture as in \GRASIL, but by a more approximate
technique. We assume that a fraction $\fcloud$ of the cold gas and
dust is in molecular clouds of mass $\Mcloud$ and radius
$\rcloud$. All stars are assumed to form inside clouds, and then to
escape on a timescale $\tesc$, such that the fraction inside clouds
drops continuously from 1 for ages $\tau < \tesc$, to 0 for $\tau >
2\tesc$. The fraction of light produced by stars inside clouds depends
on the wavelength, the past star formation and chemical enrichment
history, and the IMF. This fraction is much larger at shorter
wavelengths (for example, the far-UV), for which the light is
dominated by massive stars with short lifetimes. Since it is only for
these stars that the extinction by clouds is significant, we can
approximate the recent SFR history when estimating this fraction. We
assume the recent SFR to be constant in time for the case of a
quiescent disk, while for an ongoing burst we use the actual
exponential SFR history of that burst. The fraction of stars which are
inside clouds are extincted both by the clouds and by the diffuse
medium, while the remaining stars are extincted only by the diffuse
medium. The extinction of stars inside clouds depends on the surface
density of the clouds and on the gas metallicity, with the stars
treated as being at the centre of each cloud. We then calculate the
total dust attenuation by combining the attenuation factors for the
diffuse dust and molecular clouds. For more details see
\citet{Lacey10b}. Important features of our method are that the dust
extinction varies self-consistently with other galaxy properties such
as size, gas mass, and metallicity, and that young stars suffer more
extinction than old stars, even at the same wavelength.}

{Our new extinction model thus depends on the parameters
$\fcloud$, $\Mcloud$, $\rcloud$ and $\tesc$. In fact, the dependence
on $\Mcloud$ and $\rcloud$ is only through the combination
$\Mcloud/\rcloud^2$ which determines the surface density of a single
cloud. Our default values for these parameters are the same as we used
in our \GRASIL\ modelling in \citet{Baugh05}, namely $\fcloud=0.25$,
$\Mcloud=10^6\Msol$, $\rcloud=16\pc$ and $\tesc=1\Myr$. The value
$\tesc$ was adjusted in that paper to match the far-UV LF of LBGs at
$z\sim 3$, while the values of the other three dust parameters were
chosen in \citet{Granato00} by comparison with observations of local
galaxies. With these parameters, the extinction optical depth at
1500\AA\ for stars at the centre of a cloud is around 90 for solar
metallicity, so the far-UV light from such stars is almost completely
extincted. The net attenuation by clouds will therefore be insensitive
to the exact values of $\Mcloud$ and $\rcloud$. We investigate the
effects of variations in $\fcloud$ and $\tesc$ in
\S\ref{sec:vary_dust}. }

\subsection{Photoionization feedback.} 

The reionization of the intergalactic medium (IGM) by a photoionizing
background suppresses the amount of gas cooling in small halos in two
ways: the resulting IGM pressure inhibits collapse of gas into halos,
and the radiative cooling time for gas inside halos is increased by
the photoionizing background. These effects were modelled in detail by
\citet{Benson02}, and found to be reasonably well approximated by a
simple model in which gas cooling is completely suppressed in halos
with circular velocities $\Vc < \Vcrit$ at redshifts $z < \zreion$.
The original \Bau\ and \Bow\ models assumed $\zreion=6$ and
$\Vcrit=60\kms$ or $50\kms$ respectively. This value for the
reionization redshift was motivated by measurements of Gunn-Peterson
absorption in quasar spectra, which imply that reionization is
essentially complete by $z\approx 6$ \citep{Fan00,Becker01}. The
values of the threshold halo circular velocity, $\Vcrit$, were
motivated by the numerical simulations of \citet{Gnedin00}. Since
then, measurements of the Thomson scattering optical depth from
fluctuations in the cosmic microwave background have converged to
imply a higher reionization redshift, $z \sim 10-12$
\citep{Dunkley09}. We therefore adopt a standard value $\zreion=10$ in
the current sudy. More recent numerical simulations of the effect of
reionization on collapse and cooling of gas in low mass halos imply
weaker photoionization feedback effects than found by
\citeauthor{Gnedin00} \citep{Hoeft06,Okamoto08}. In line with these
simulations, we adopt $\Vcrit=30\kms$ as our standard choice in the
present work.  For brevity, we will refer to the slightly modified
\Bau\ and \Bow\ models with $\zreion=10$ and $\Vcrit=30\kms$ simply as
\Bau\ and \Bow\ in what follows.  {We have not changed any
other model parameters from their original values.}

\subsection{Halo merger trees}

\GALFORM\ calculates the evolution of galaxies in halo merger trees
which describe the assembly and merger histories of dark matter
halos. The halo merger trees are either computed using a Monte Carlo
method based on the extended Press-Schechter model \citep{Cole00}, or
extracted from N-body simulations of the dark matter
\citep{Helly03}. The \citet{Baugh05} results were based on Monte Carlo
trees, while those of \citet{Bower06} were based on N-body trees
extracted from the Millennium simulation \citep{Springel05}.
\citet{Cole08} found some differences between the Monte Carlo and
N-body trees, as a result of which \citet{Parkinson08} developed a
modified version of the \citet{Cole00} Monte Carlo algorithm, which
brought the Monte Carlo trees into much better agreement with the
N-body trees from the Millennium simulation. In this paper, we run
both the \Bau\ and \Bow\ models on Monte Carlo trees generated using
the \citet{Parkinson08} method. This has the advantage that we can
resolve much smaller halos at high redshifts than is possible using
the Millennium simulation, which only resolves halos larger than
$2\times 10^{10}\hMsol$. In contrast, for the Monte Carlo trees used
here, we resolve progenitor halos larger than $8\times 10^7\hMsol$ for
model galaxies output at $z=3$, $1.5\times 10^7\hMsol$ at $z=6$, and
$4\times 10^6\hMsol$ at $z=10$. This higher mass resolution is
important for predictions of LBGs at higher redshifts and lower
luminosities. The change to \citeauthor{Parkinson08} Monte Carlo trees
leaves the predictions published in \citet{Baugh05} and
\citet{Bower06} unchanged.

%%%%%%%%%%%%%%%%%%%%%%%%%%%%%%%%%%%%%%%%%%%%%%%%%%%%%%%%%%%%%%%%%%%%%%%%%%%%%%%%%%
% Fig.1
% LF evoln in model
\begin{figure*}

\begin{minipage}{7cm}
\includegraphics[width=7cm]{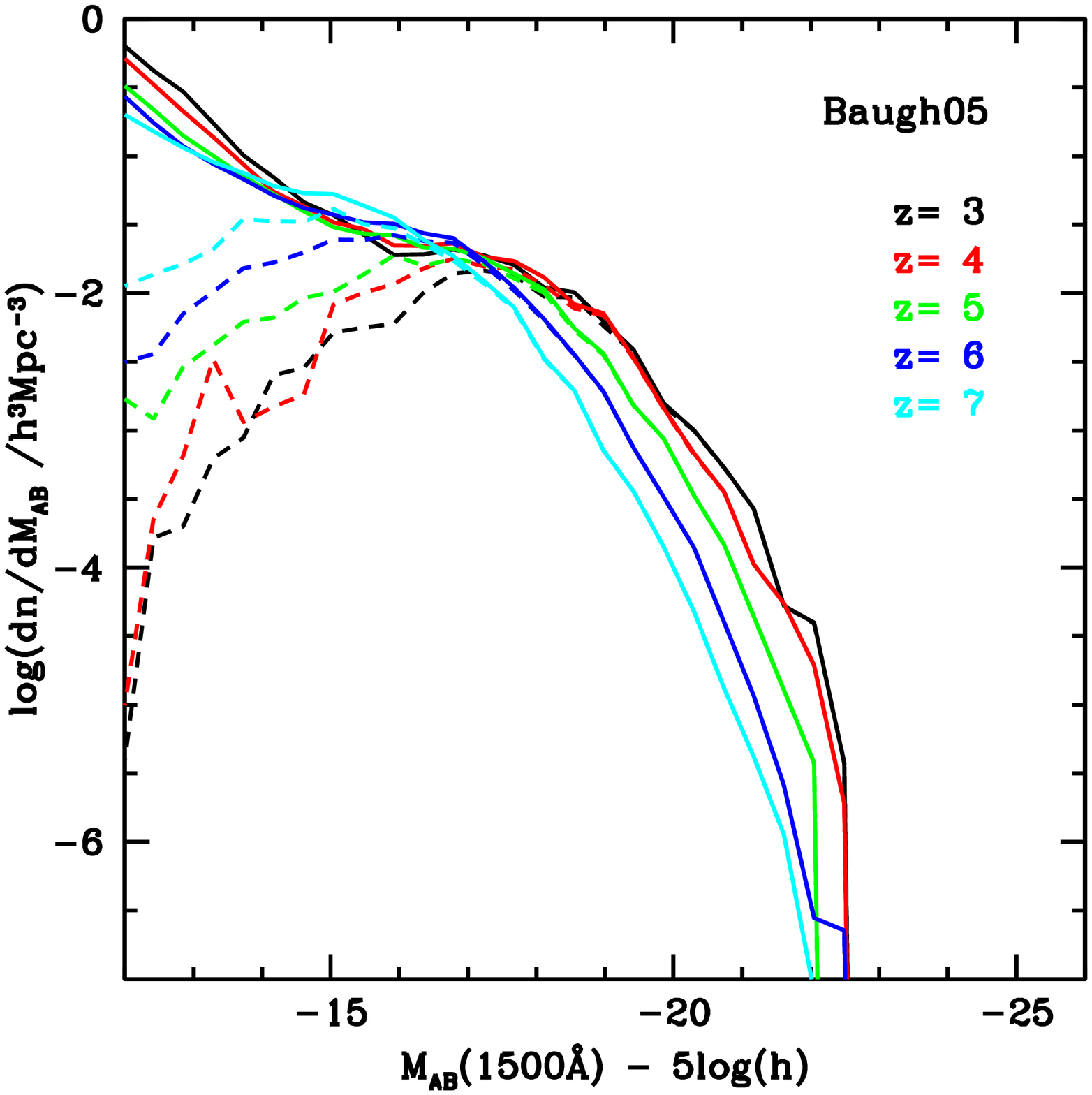}
\end{minipage}
%\hspace{1cm}
\begin{minipage}{7cm}
\includegraphics[width=7cm]{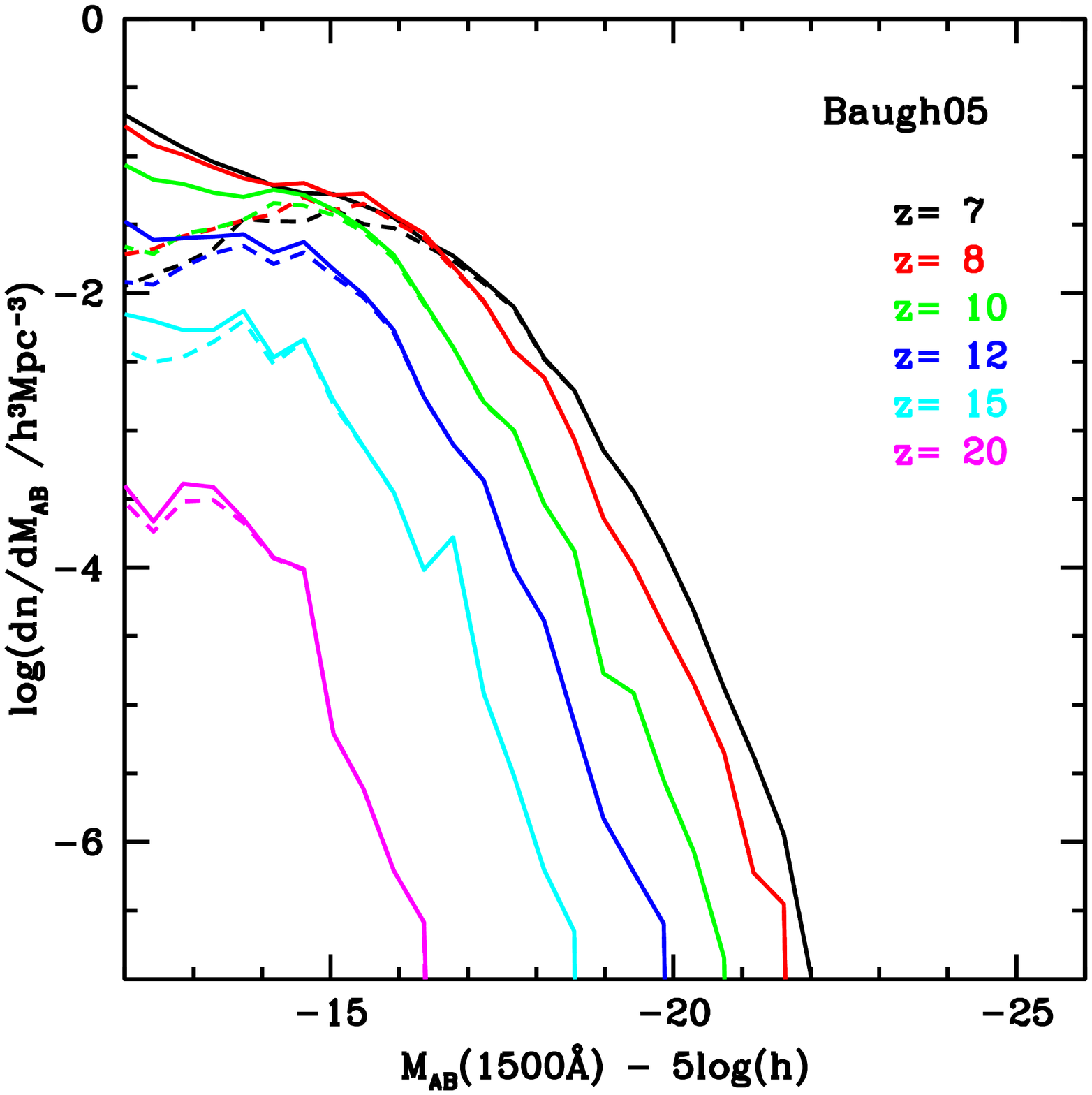}
\end{minipage}

\begin{minipage}{7cm}
\includegraphics[width=7cm]{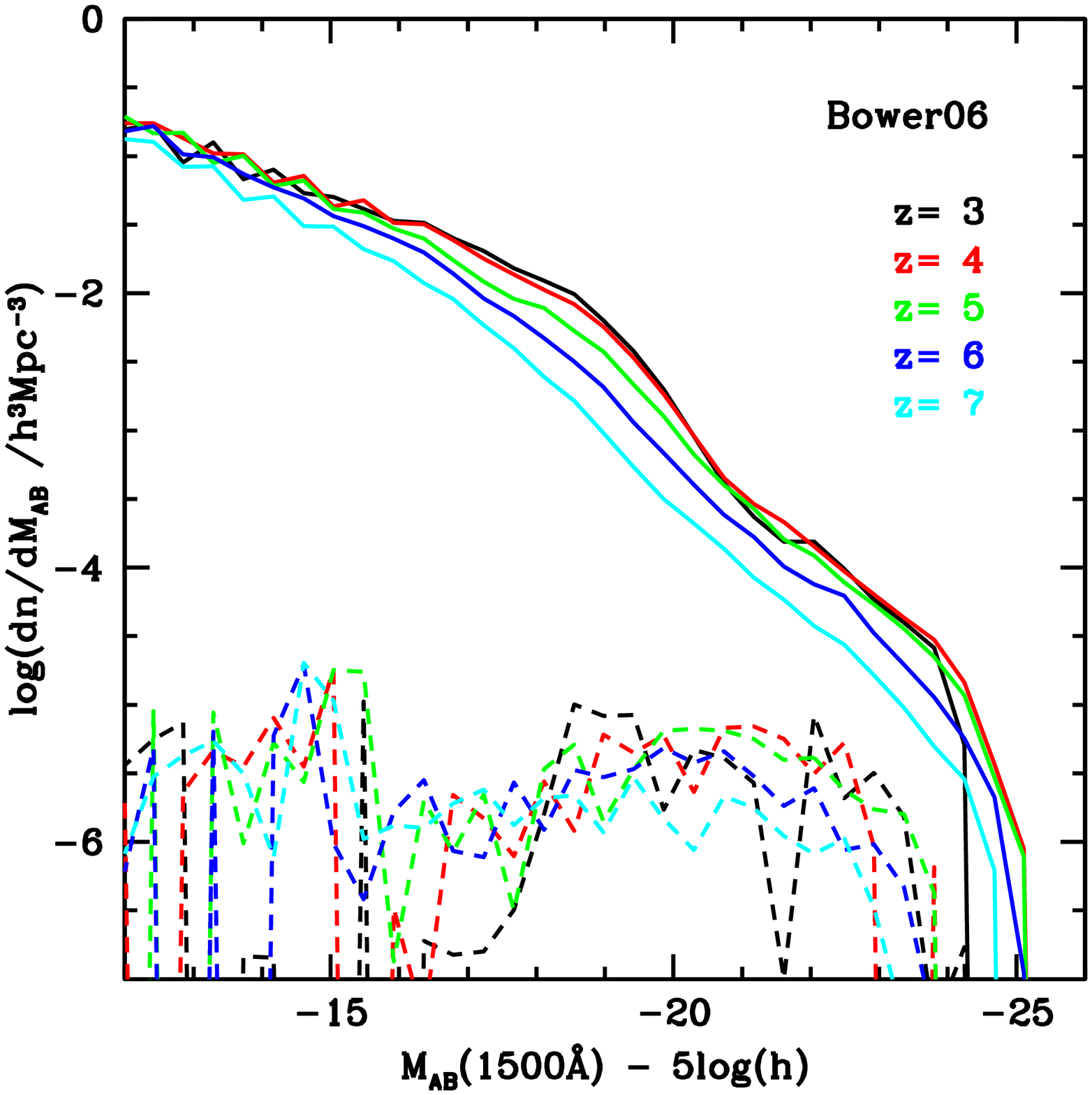}
\end{minipage}
%\hspace{1cm}
\begin{minipage}{7cm}
\includegraphics[width=7cm]{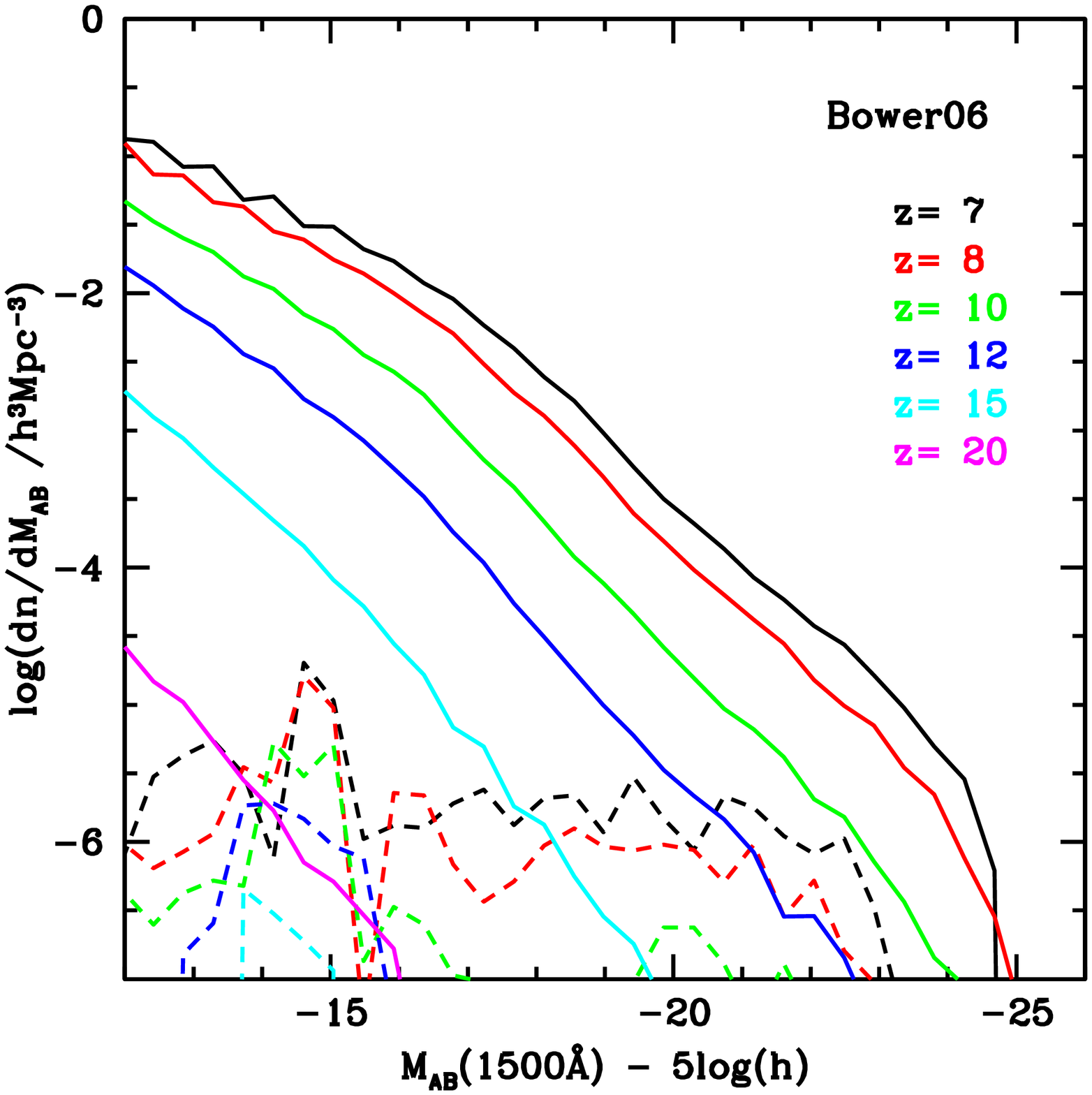}
\end{minipage}

\caption{Predicted evolution of the rest-frame far-UV (1500\AA)
  luminosity function of LBGs in the \Bau\ model (top panels) and the
  \Bow\ model (bottom panels). In both cases, the left panels show the
  evolution for $3<z<7$ and the right panels for $7<z<20$, with
  different redshifts shown in different colours, as indicated by the
  key. The dashed lines show the contribution from ongoing
  starbursts. All of the LFs plotted here include dust extinction.}

\label{fig:lf-evoln}
\end{figure*}

%%%%%%%%%%%%%%%%%%%%%%%%%%%%%%%%%%%%%%%%%%%%%%%%%%%%%%%%%%%%%%%%%%%%%%%%%%%

\section{Evolution of LBG luminosity function}
\label{sec:lf-evoln}

In this section, we show the predictions of the models for the
evolution of the rest-frame far-UV luminosity function (LF), and
compare these with observational data {for LBGs}. The far-UV LF
is one of the most basic observable properties of LBGs, since they are
detected via their rest-frame far-UV emission, and is also of
fundamental importance, since the rest-frame far-UV emission is
closely related to the SFR. {For the purpose of this paper, we
will define model LBGs to include all galaxies with far-UV (1500\AA)
luminosities in the relevant range, regardless of whether they satisfy
the observational colour-selection criteria used in specific surveys, as
discussed futher below.}

\subsection{Comparison of \Bau\ and \Bow\ models}

We start by showing in Fig.~\ref{fig:lf-evoln} the evolution of the
dust-extincted rest-frame 1500\AA\ LFs over a large range in redshift
and luminosity, for both the \Bau\ and \Bow\ models (top and bottom
panels respectively). For clarity, we show the evolution separately
for $3<z<7$ (left panels) and $7<z<20$ (right panels). The solid lines
show the total LF, while the dashed lines show the contribution to the
LF from ongoing bursts. There are a number of interesting features in
these plots. Both models show relatively mild evolution in the far-UV
LF in the redshift range $3 \lsim z \lsim 8$, with a gradual decrease
in comoving number density with increasing redshift at most
luminosities, followed by a much steeper decline in number density at
$z \gsim 8$. This rapid evolution at high redshifts is driven by the
build-up of the dark matter halos hosting the LBGs, and was also found
ealier in \GALFORM\ predictions for the LFs of Ly$\alpha$-emitting
galaxies (LAEs), which are a population closely related to the LBGs
\citep{LeDelliou06}. However, the LFs in the two models have
distinctly different shapes. 

In the \Bau\ model, the LF has a steep, quasi-exponential, cutoff at
high luminosities, with a flattening or bump at lower luminosities,
followed by an upturn and steeper power-law slope at the faintest
luminosities. This behaviour is due to the major role played by
merger-driven starbursts in the \Bau\ model. As can be seen from the
dashed lines in the figure, starbursts dominate the bright end of the
LBG LF, while quiescent galaxies {(with star formation occuring in the
disk)} dominate the faint end. The LF of the starbursts actually
declines at low luminosities, which is what causes the bump seen in
the LF. The steep faint end is due to the quiescent population. In the
\Bow\ model, on the other hand, the LFs show a more power-law
behaviour over the ranges plotted here, with a gradual steepening from
low to high luminosity. There is a sharper break at the highest
luminosities, but this is at much higher luminosities and lower number
densities than in the \Bau\ model. These differences are due to
various effects. {In the \Bow\ model, ongoing starbursts do not
dominate the LBG LF at any luminosity once dust extinction is
included. However, due to the much shorter burst timescales in the
\Bow\ model, there is an important contribution to the far-UV LF from
bursts which have recently terminated. In such cases, star formation
has ceased, and the gas and dust associated with the burst have
already been consumed or dispersed, but many of the massive stars
produced during the burst are still shining in the far-UV. Such
recently terminated bursts in fact dominate the bright end of the
far-UV LF in the \Bow\ model over the whole range of redshifts plotted
in Fig.~\ref{fig:lf-evoln}, and are responsible for the larger number
of objects seen at the highest luminosities compared to the \Bau\
model. The dust extinction in these objects is low, as discussed
further below.}
%% In the \Bow\ model, starbursts play a much
%% smaller role, and the LBG LF is dominated by quiescent galaxies at all
%% luminosities and redshifts once dust extinction is included. The more
%% gradual decline at high luminosities compared to \Bau\ is due to the
%% dust extinctions typically being lower at high luminosities in \Bow,
%% as discussed below. 
The shallower slope seen at the faintest luminosities in the \Bow\
model is due to the stronger SN feedback. Even the LF for ongoing
starbursts has a flatter shape in the \Bow\ model, because bursts are
triggered mainly by disk instabilities rather than galaxy mergers.

%%%%%%%%%%%%%%%%%%%%%%%%%%%%%%%%%%%%%%%%%%%%%%%%%%%%%%%%%%%%%%%%%%%%%%%%%%%%%%%%%%
% Fig.2
% LFs compared to obs at z=3-6
\begin{figure*}

\begin{center}

\begin{minipage}{7cm}
\includegraphics[width=7cm]{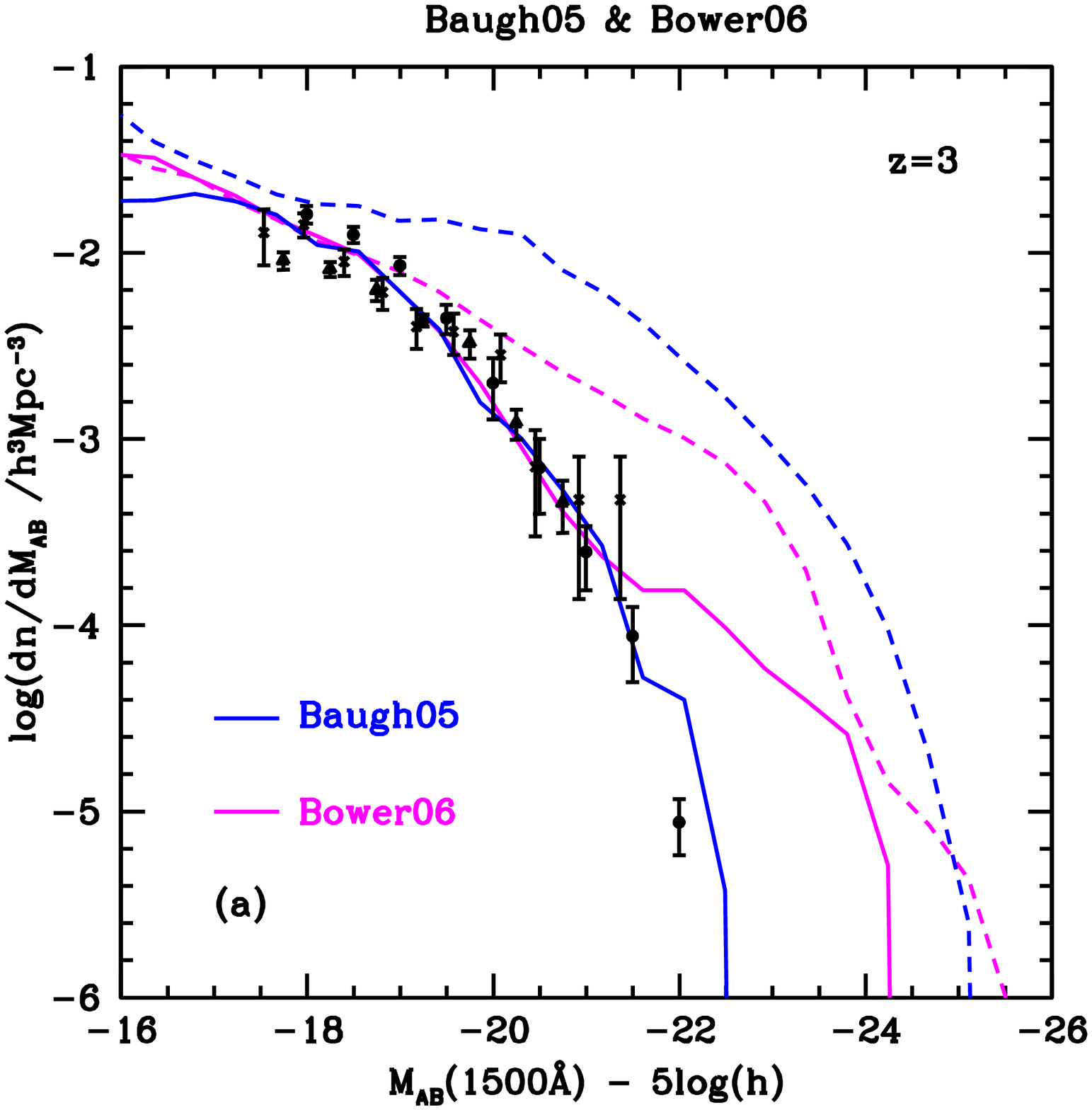}
\end{minipage}
%\hspace{1cm}
\begin{minipage}{7cm}
\includegraphics[width=7cm]{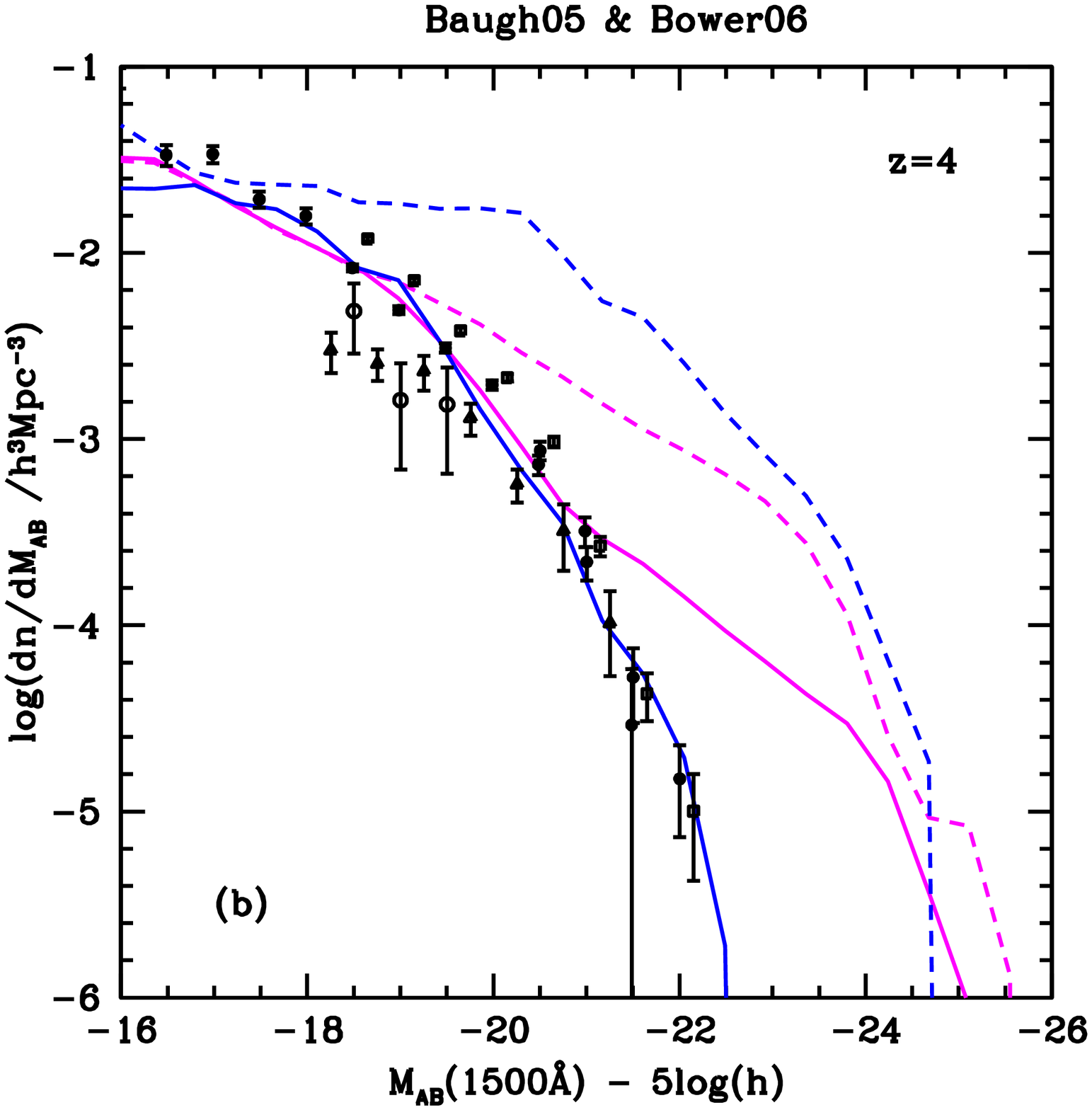}
\end{minipage}

\begin{minipage}{7cm}
\includegraphics[width=7cm]{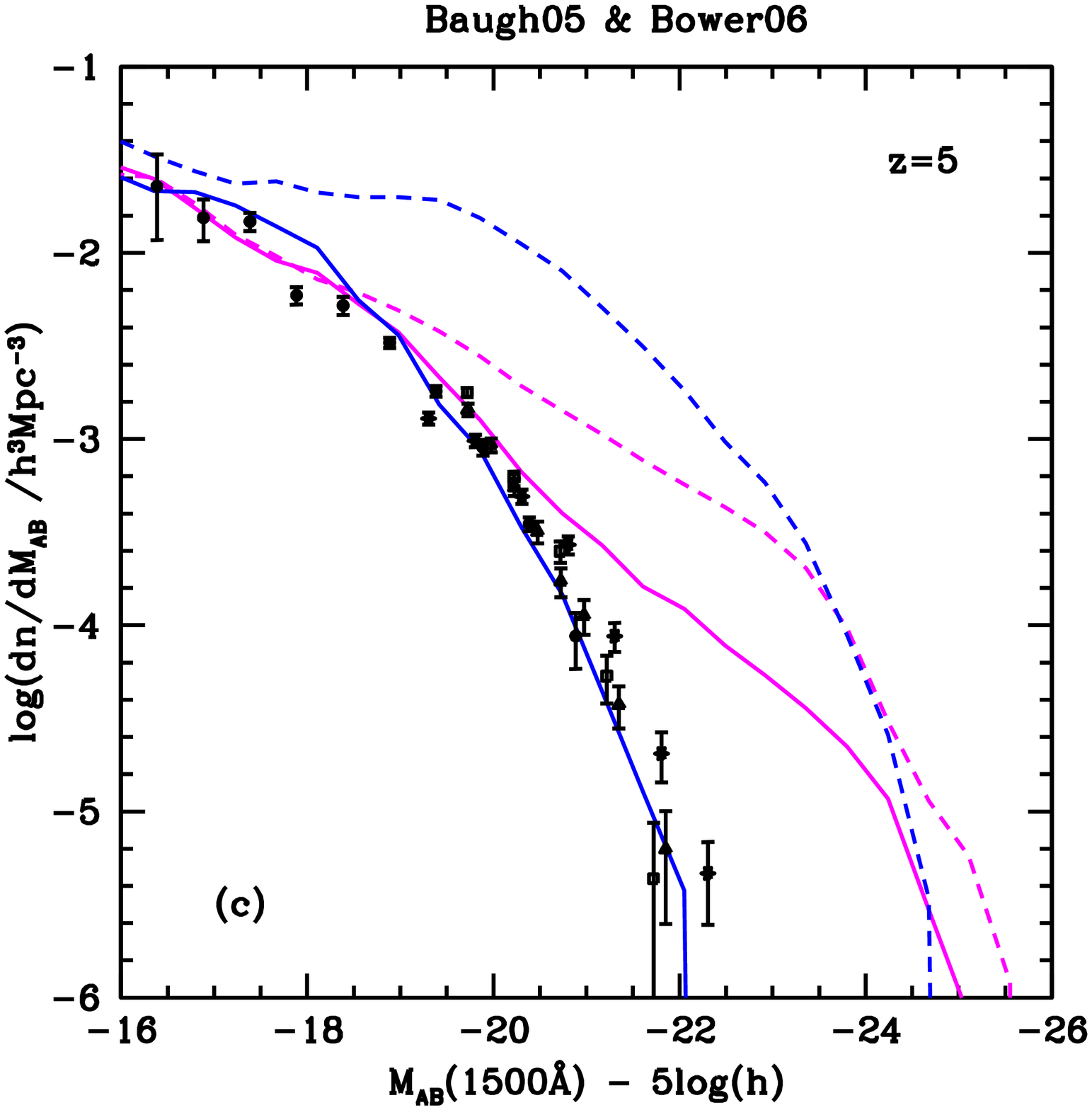}
\end{minipage}
%\hspace{1cm}
\begin{minipage}{7cm}
\includegraphics[width=7cm]{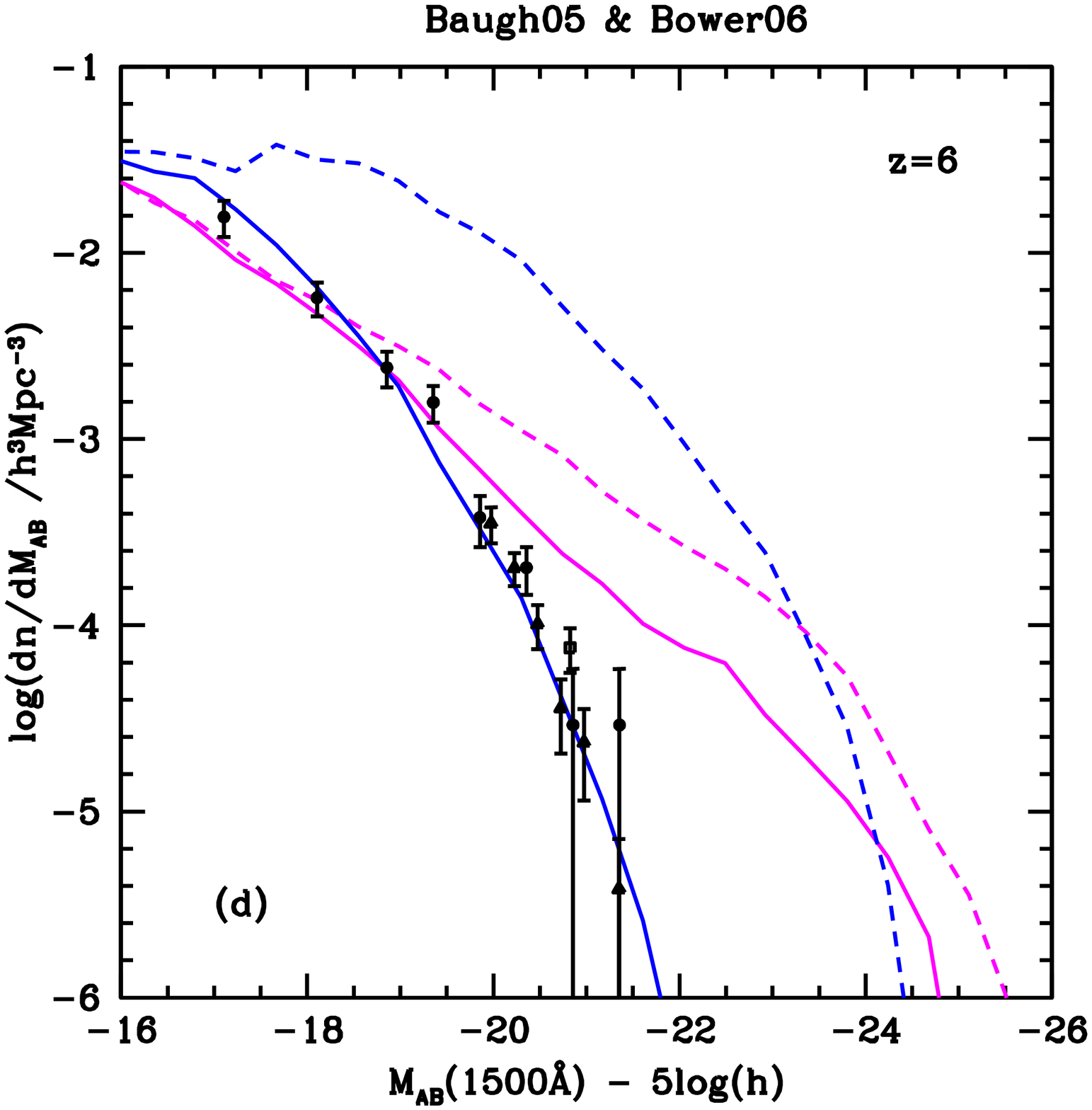}
\end{minipage}

\end{center}

\caption{Predicted rest-frame 1500\AA\ LFs for the Baugh05 model
  (blue) and Bower06 model (magenta) compared to observational data
  from LBG surveys. Solid and dashed lines show LFs with and without
  dust extinction. (a) $z=3$. (b) $z=4$. (c) $z=5$. (d) $z=6$. The
  observational data are as follows (with rest-frame wavelength):
  $z=3$ -- \citet{Arnouts05} (crosses, 1500\AA), \citet{Sawicki06}
  (empty triangles, 1700\AA), \citet{Reddy09} (filled circles,
  1700\AA); $z=4$ -- \citet{Steidel99} (empty circles, 1700\AA),
  \citet{Sawicki06} (empty triangles, 1700\AA), \citet{Yoshida06}
  (empty squares, 1500\AA), \citet{Bouwens07} (filled circles,
  1600\AA); $z=5$ -- \citet{Yoshida06} (empty squares, 1500\AA),
  \citet{Iwata07} (stars, 1600\AA), \citet{Bouwens07} (filled circles,
  1600\AA), \citet{McLure09} (empty triangles, 1500\AA); $z=6$ --
  \citet{Shimasaku05} (open squares, 1400\AA), \citet{Bouwens07}
  (filled circles, 1350\AA), \citet{McLure09} (empty triangles,
  1500\AA).  }

\label{fig:lf-comp-lowz}
\end{figure*}

%%%%%%%%%%%%%%%%%%%%%%%%%%%%%%%%%%%%%%%%%%%%%%%%%%%%%%%%%%%%%%%%%%%%%%%%%%%

%%%%%%%%%%%%%%%%%%%%%%%%%%%%%%%%%%%%%%%%%%%%%%%%%%%%%%%%%%%%%%%%%%%%%%%%%%%%%%%%%%
% Fig.3
% LFs compared to obs at z=7-10
\begin{figure}
\begin{center}
\includegraphics[width=7cm]{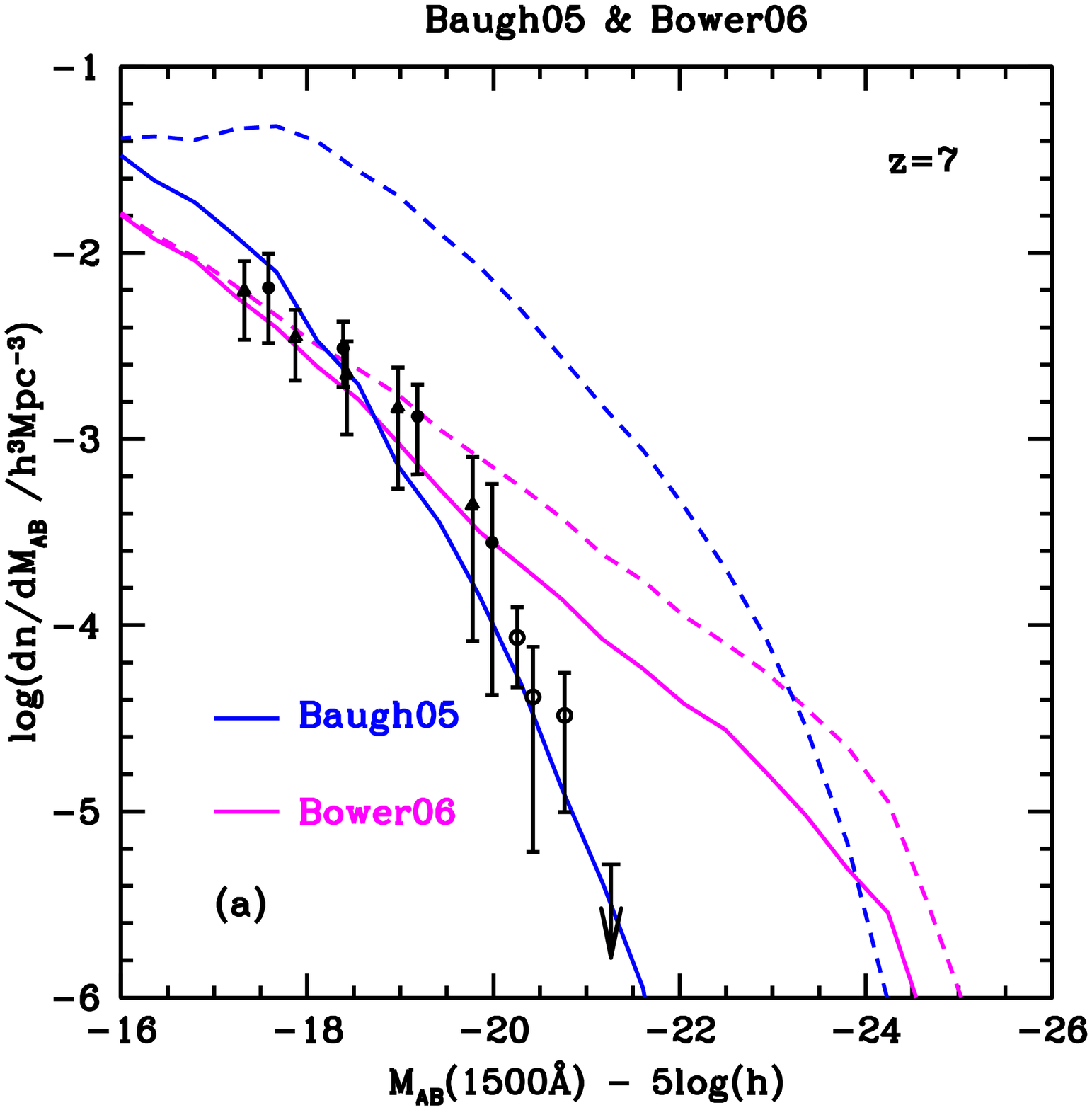}

\includegraphics[width=7cm]{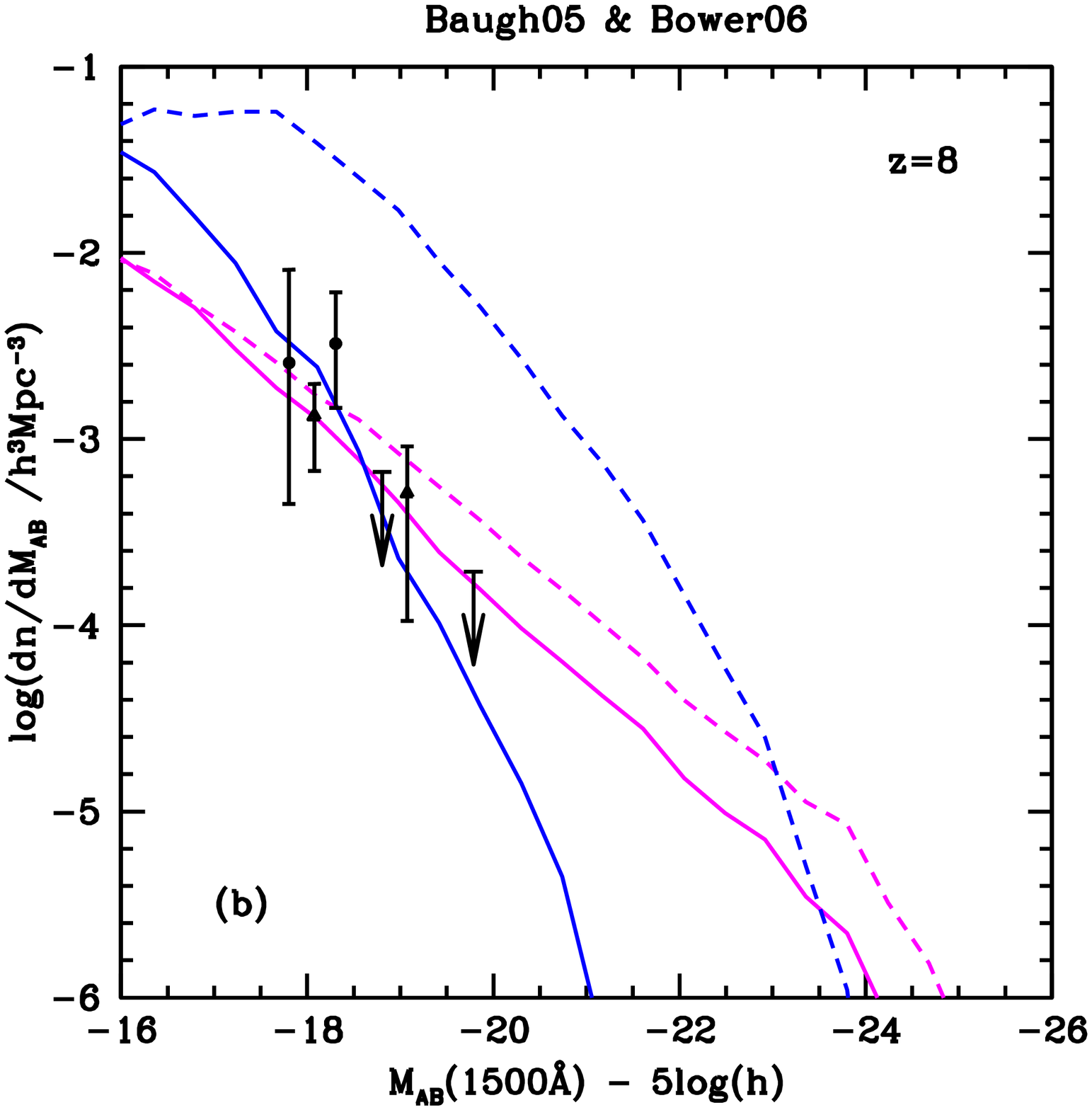}

\includegraphics[width=7cm]{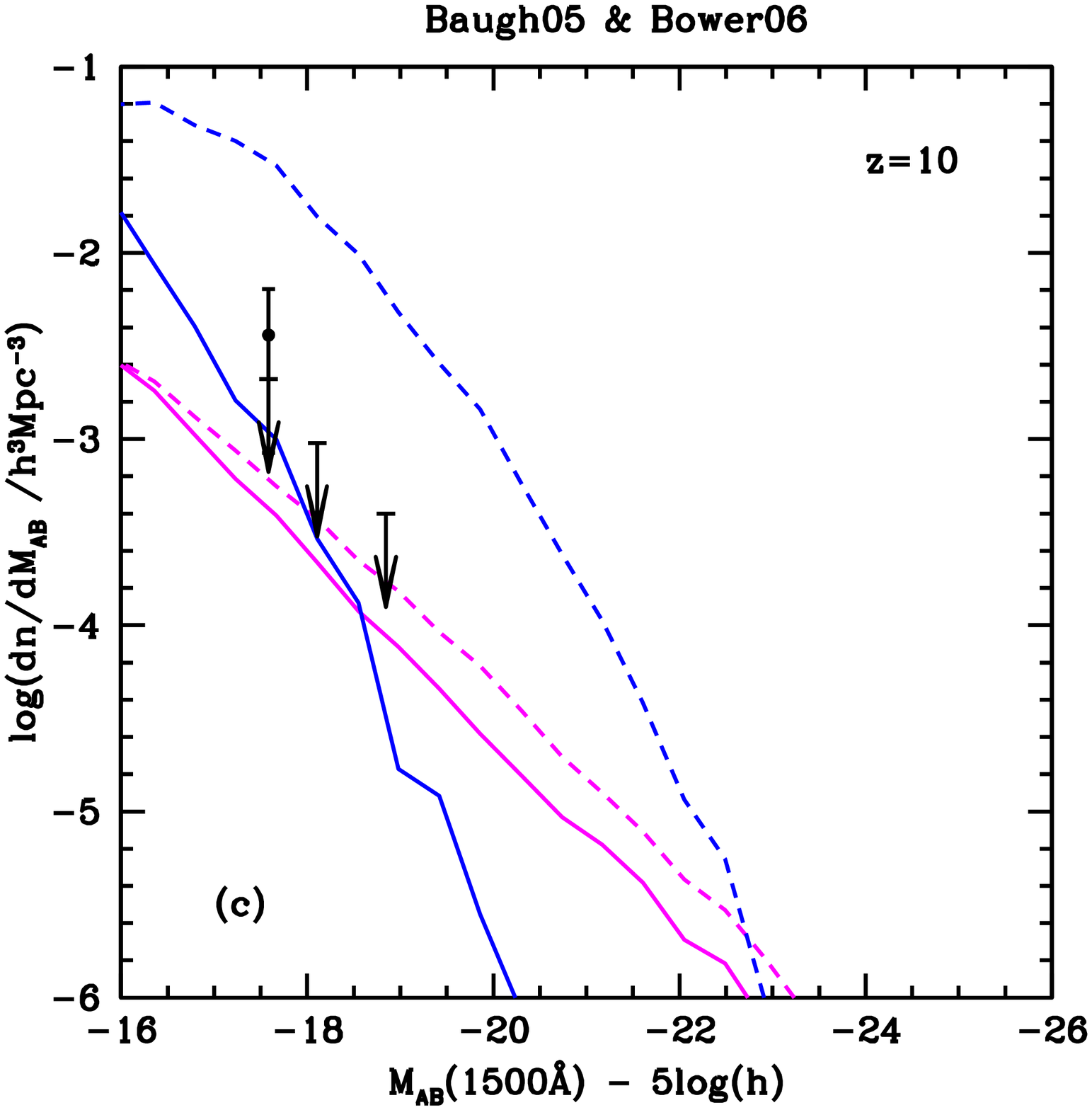}

\end{center}

\caption{Predicted rest-frame 1500\AA\ LFs for the Baugh05 model
  (blue) and Bower06 model (magenta) compared to observational data
  for LBGs. Solid and dashed lines show LFs with and without dust
  extinction. (a) $z=7$. (b) $z=8$. (c) $z=10$. The observational data
  are as follows (with rest-frame wavelength): $z=7$ --
  \citet{Ouchi09} (empty circles and upper limits, 1500\AA),
  \citet{Oesch10a} (filled circles, 1600\AA), \citet{McLure10} (empty
  triangles, 1500\AA); $z=8$ -- \citet{Bouwens10a} (filled circles and
  upper limits, 1700\AA), \citet{McLure10} (empty triangles, 1500\AA);
  $z=10$ -- \citet{Bouwens10b} (filled circles and upper limits,
  1600\AA).  }

\label{fig:lf-comp-hiz}
\end{figure}
%%%%%%%%%%%%%%%%%%%%%%%%%%%%%%%%%%%%%%%%%%%%%%%%%%%%%%%%%%%%%%%%%%%%%%%%%%%%%%%%%%

We next compare the predicted far-UV LFs from the models with
observational estimates of the LFs derived from samples of LBGs at
redshifts $z=3-10$. The observational selection is typically based on
two colours, one of which straddles the Lyman break at the target
redshift, and the other which measures the spectral slope longwards of
the break, and is used to exclude contaminants, principally lower
redshift galaxies and galactic stars, which have redder colours than
expected for a star forming galaxy at the target redshift.  The
observational LFs we plot already include corrections by the original
authors for the completeness as a function of redshift and luminosity,
as well as for absorption by the IGM (where appropriate), but do not
include corrections for dust extinction. We therefore compare our
model LFs, including dust extinction, directly with the LFs inferred
observationally. 
%% after correcting the latter to the same cosmology as
%% assumed in our models, where needed.  
Each LBG survey uses its own set of filters, so in practice they
measure the far-UV LFs at slighly different effective rest-frame
wavelengths, in the range 1350--1700\AA. We plot our model LFs at a
fixed rest-frame wavelength of 1500\AA, so, in principle, we should
correct the luminosities for this small wavelength difference when
comparing with the observational data. However, the spectra of LBGs
are observed to be fairly flat in this wavelength range, so this
correction is small, and we neglect it here. It is also possible that
the observed luminosity functions are effectively missing some
galaxies due to the colour selection, even after the completeness
corrections applied by the original authors (e.g. due to the galaxies
being too red in the far-UV due to dust extinction). Different
completeness corrections have been applied in different surveys, and
this presumably accounts for some of the differences seen in the
inferred LFs. In contrast, our model LFs include all galaxies at a
given 1500\AA\ luminosity, regardless of their far-UV colour. We defer
to a future paper a detailed study of the effect of applying different
LBG colour selections directly to the models, and the effect this has
on the completeness of the galaxy samples at far-UV wavelengths. The
observational LFs are all plotted for the same cosmology
($\Omega_m=0.3$, $\Omega_\Lambda=0.7$) as used in the \Bau\ model
(correcting to this cosmology if needed). The cosmology used in the
\Bow\ model is slightly different, but converting the observational
data to it would have little effect on the comparison, so we ignore
this correction here.

We compare the 1500\AA\ LFs from the \Bau\ and \Bow\ models with the
observed far-UV LFs in Figs.~\ref{fig:lf-comp-lowz} and
\ref{fig:lf-comp-hiz}. Fig.~\ref{fig:lf-comp-lowz} shows redshifts
$z=3,4,5,6$ and Fig.~\ref{fig:lf-comp-hiz} shows $z=7,8,10$. The \Bau\
model is shown in blue, and the \Bow\ model in magenta. In both cases,
we plot the dust-extincted LF as a solid line, and the unextincted LF
as a dashed line. The \Bau\ model fits the observational data well
over the whole redshift range $z=3-10$, when we include dust
extinction. This is remarkable when we consider that we have not
changed any model parameters from those published in \citet{Baugh05},
apart from $\zreion$ and $\Vcrit$ (as described in
\S\ref{sec:GALFORM}), whose effect is anyway quite small. We used the
observed far-UV LBG LF at $z=3$ to constrain the model parameters in
\citet{Baugh05}, but we did not use observations of LBGs at any other
redshift. In contrast, the \Bow\ model is in much worse agreement with
the observed LBG LFs at almost all of the redshifts plotted,
predicting far more very luminous galaxies than observed. The \Bow\
model in its original form can therefore be excluded based on this
observational data. 

By comparing the solid and dashed lines, we can assess the effect of
dust extinction on the predicted LFs. For the \Bau\ model, the effect
of dust extinction is very large over the luminosity range covered by
the observational data, causing the LF to be shifted faintwards by
$\sim 1.5-2.5$ mag at the bright end. In contrast, the effects of dust
extinction are generally smaller in the \Bow\ model, especially at the
highest luminosities. An important factor in this is that the gas
metallicity of the LBGs is typically much larger in the \Bau\ than
\Bow\ model, leading to higher dust-to-gas ratios. The higher
metallicities, in turn, result from the top-heavy IMF assumed for
starbursts in the \Bau\ model, which leads to a higher yield of heavy
elements from Type~II supernovae. {However, another important
factor is the much shorter timescale for bursts in the \Bow\ model
($\sim 0.001$~Gyr or less as against $\sim 0.01-0.1$~Gyr in the \Bau\
model). A consequence of this is that the bright end of the far-UV LF
in the \Bow\ model is dominated by bursts which have terminated within
the last $\sim 0.01-0.1$~Gyr. For these objects, most of the massive
stars formed during the burst are still shining, but the dust which
shrouded the burst while it was ongoing has gone. As a result, they
have large far-UV luminosities but very little dust extinction.}

%% {The dust extinction in LBGs
%% also has a different dependence on luminosity and redshift in the two
%% models. This reflects the different nature of the galaxies in the two
%% cases (typically starbursts in \Bau\ and quiescent disks in \Bow),
%% and the fact that the gas masses, metallicities and radii vary
%% differently.}

{Using a different semi-analytical model, \citet{LoFaro09}
  predicted a significant excess of faint LBGs compared to
  observational data at $z\sim 4-5$. We do not see such an excess in
  our own model, even though we compare to similar observational
  data. This is most likely due to the different treatments of
  supernova and photoionization feedback in the \citeauthor{LoFaro09}
  model. The effects of varying the treatment of supernova and
  photoionization feedback in our own model are discussed below in
  \S\ref{sec:phot-feedback} and \S\ref{sec:SN-feedback}.}

%%%%%%%%%%%%%%%%%%%%%%%%%%%%%%%%%%%%%%%%%%%%%%%%%%%%%%%%%%%%%%%%%%%%%%%%%%%%%%%%%%
% Fig.4
% LFs at z=3 comparing Bau05 variants: zcut & Vcut
\begin{figure*}
\begin{center}

\begin{minipage}{7cm}
\includegraphics[width=7cm]{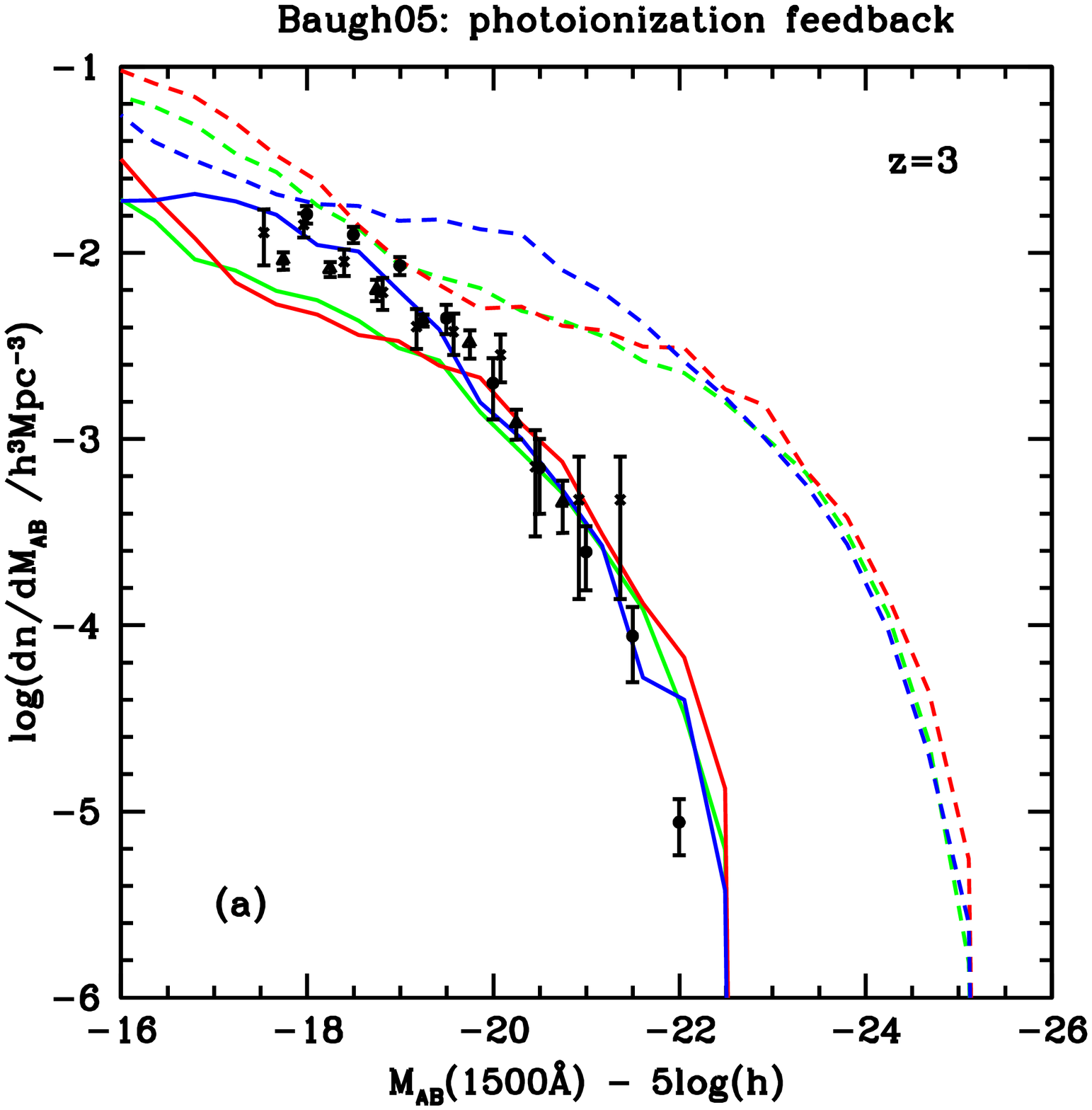}
\end{minipage}
%\hspace{1cm}
\begin{minipage}{7cm}
\includegraphics[width=7cm]{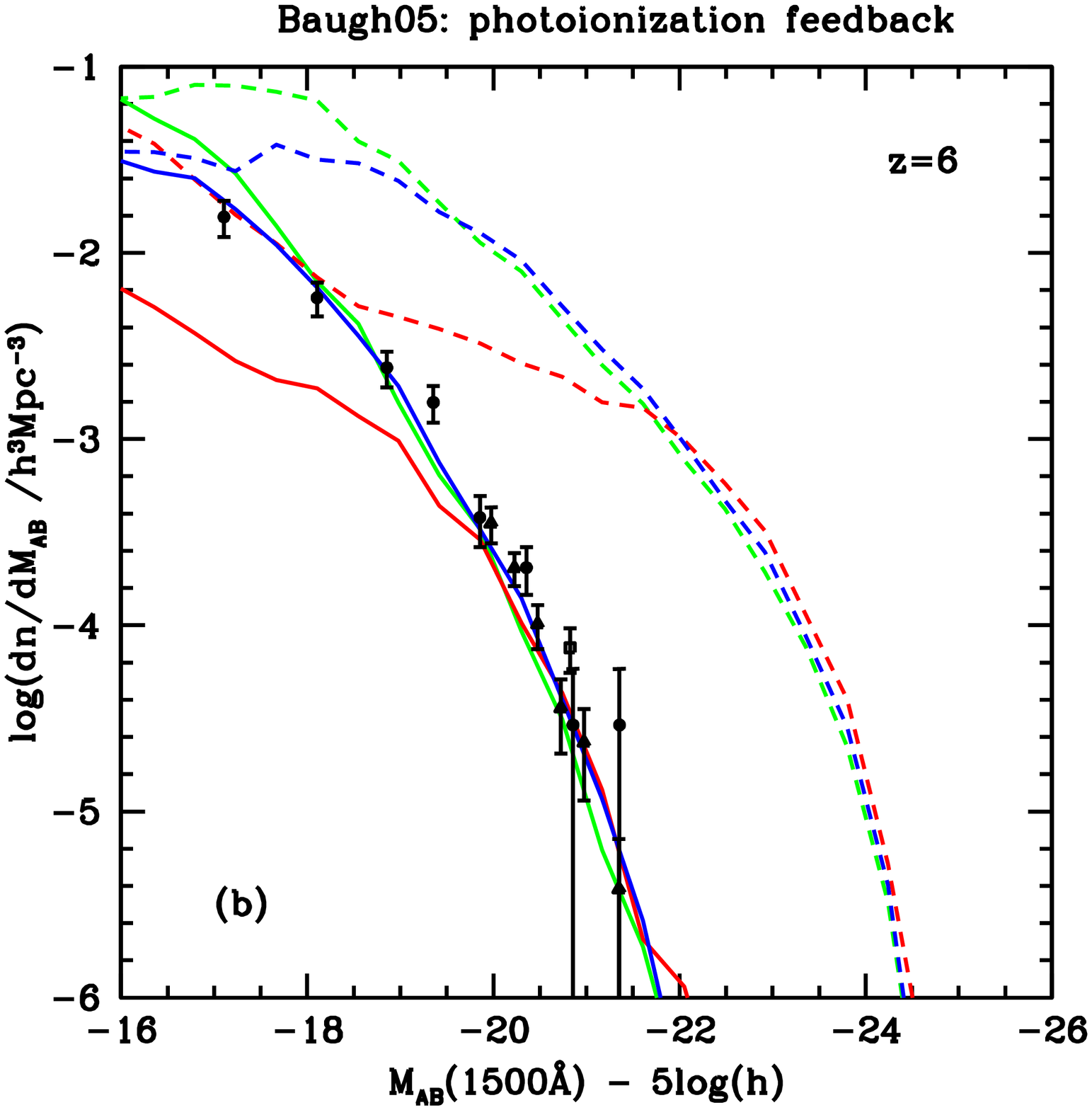}
\end{minipage}

\end{center}

\caption{Predicted rest-frame 1500\AA\ LFs for the \Bau\ model,
  showing the effect of varying the photoionization feedback
  parameters, at (a) $z=3$ and (b) $z=6$. The blue lines show the
  default model ($\zreion=10$, $\Vcrit=30\kms$), green lines show the
  case where $\zreion=6$, $\Vcrit=60\kms$, {and red lines show
  the case $\zreion=10$, $\Vcrit=60\kms$}. The solid and dashed lines
  show the LFs with and without dust extinction, as before, and the
  observational data are the same as plotted in
  Fig.~\ref{fig:lf-comp-lowz}.  }

\label{fig:lf-comp.reion}
\end{figure*}
%%%%%%%%%%%%%%%%%%%%%%%%%%%%%%%%%%%%%%%%%%%%%%%%%%%%%%%%%%%%%%%%%%%%%%%%%%%%%%%%%%

\subsection{Effects of varying model parameters}
\label{sec:vary_params}

We now consider the effects on the far-UV LFs of varying some of the
key parameters in \GALFORM. Since the default \Bau\ model has been
shown to agree much better with the observed LBG LFs than the default
\Bow\ model, we consider here only parameter variations around the
default \Bau\ model. Our purpose here is to understand the sensitivity
of the LBG predictions to different parameters. We emphasize that most
of the parameter variations we consider  do not lead to models
that we would consider acceptable overall, since we require our model
to fit a much wider range of observational data than just LBGs, as
discussed in \S\ref{sec:GALFORM}.

\subsubsection{Photoionization feedback}
\label{sec:phot-feedback}

Photoionization feedback is described by the parameters $\zreion$ and
$\Vcrit$. In Fig.~\ref{fig:lf-comp.reion}, we compare the far-UV LFs
for our default values of these parameters ($\zreion=10$,
$\Vcrit=30\kms$, {blue curves}) with the values used in the
original \citet{Baugh05} paper ($\zreion=6$, $\Vcrit=60\kms$,
{green curves}), {and with the case $\zreion=10$,
$\Vcrit=60\kms$ {(red curves)}.} We only show this comparison
at $z=3$ and $z=6$; for $z\geq 10$ the three models are identical. At
$z=6$, the model with $\zreion=6$ {and $\Vcrit=60\kms$} is
slightly above the default model at the faint end of the LF. This is
because photoionization feedback has only just turned on in the former
model, and so has not had time to have any effect on galaxy
luminosities. On the other hand, by $z=3$, the model with $\zreion=6$
and $\Vcrit=60\kms$ is somewhat below the default model at the faint
end. This is because the effect of photoionization feedback is much
stronger with $\Vcrit=60\kms$ than $\Vcrit=30\kms$ (the halo mass
afffected scales approximately as $\Vcrit^3/(1+z)^{3/2}$). {The
model with $\zreion=10$ and $\Vcrit=60\kms$ shows a large suppression
of the faint end of the LF at $z=6$ relative to the default model, but
by $z=3$ the predicted LF is very similar to that for $\zreion=6$ and
$\Vcrit=60\kms$. This shows that the predicted LF becomes insensitive
to the value of $\zreion$ at much later epochs.}  The reduction in
$\Vcrit$ for the present default model relative to \citet{Baugh05}
also causes a slight steepening in the present-day galaxy luminosity
function, but only for galaxies fainter than $M_B-5\log h \sim
-16$. For the \Bow\ model, the change in $\Vcrit$ and $\zreion$ from
the values used in \citet{Bower06} has negligible effects on either
the LBG LFs shown here or on the $z=0$ luminosity functions. This is
because the stronger SN feedback assumed in the \Bow\ model dominates
over the effect of photoionization feedback in low-mass galaxies.

%%%%%%%%%%%%%%%%%%%%%%%%%%%%%%%%%%%%%%%%%%%%%%%%%%%%%%%%%%%%%%%%%%%%%%%%%%%%%%%%%%
% Fig.5
% LFs at z=3,6,10 comparing Bau05 variants: bursts & IMF
\begin{figure}
\begin{center}
\includegraphics[width=7cm]{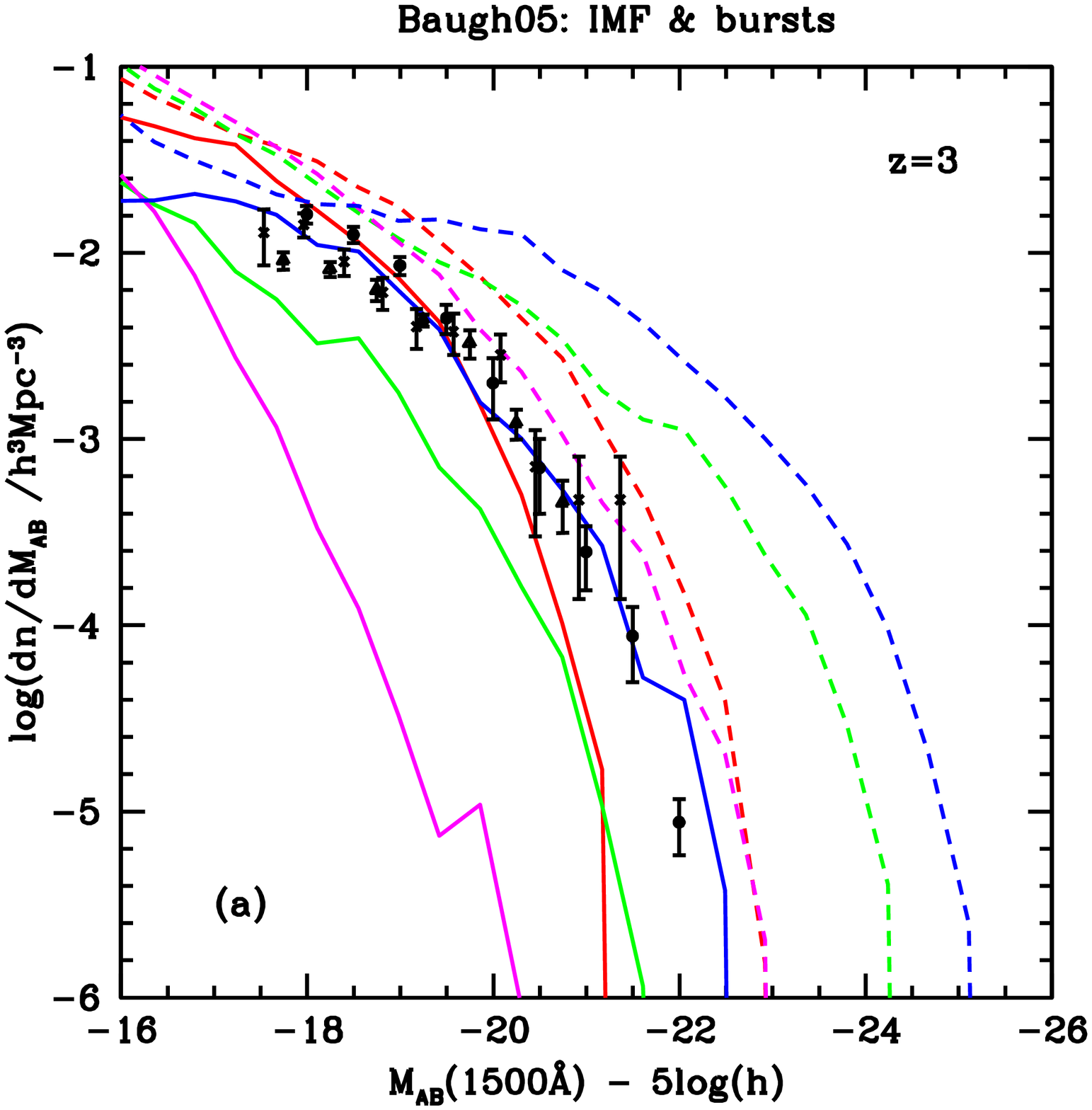}

\includegraphics[width=7cm]{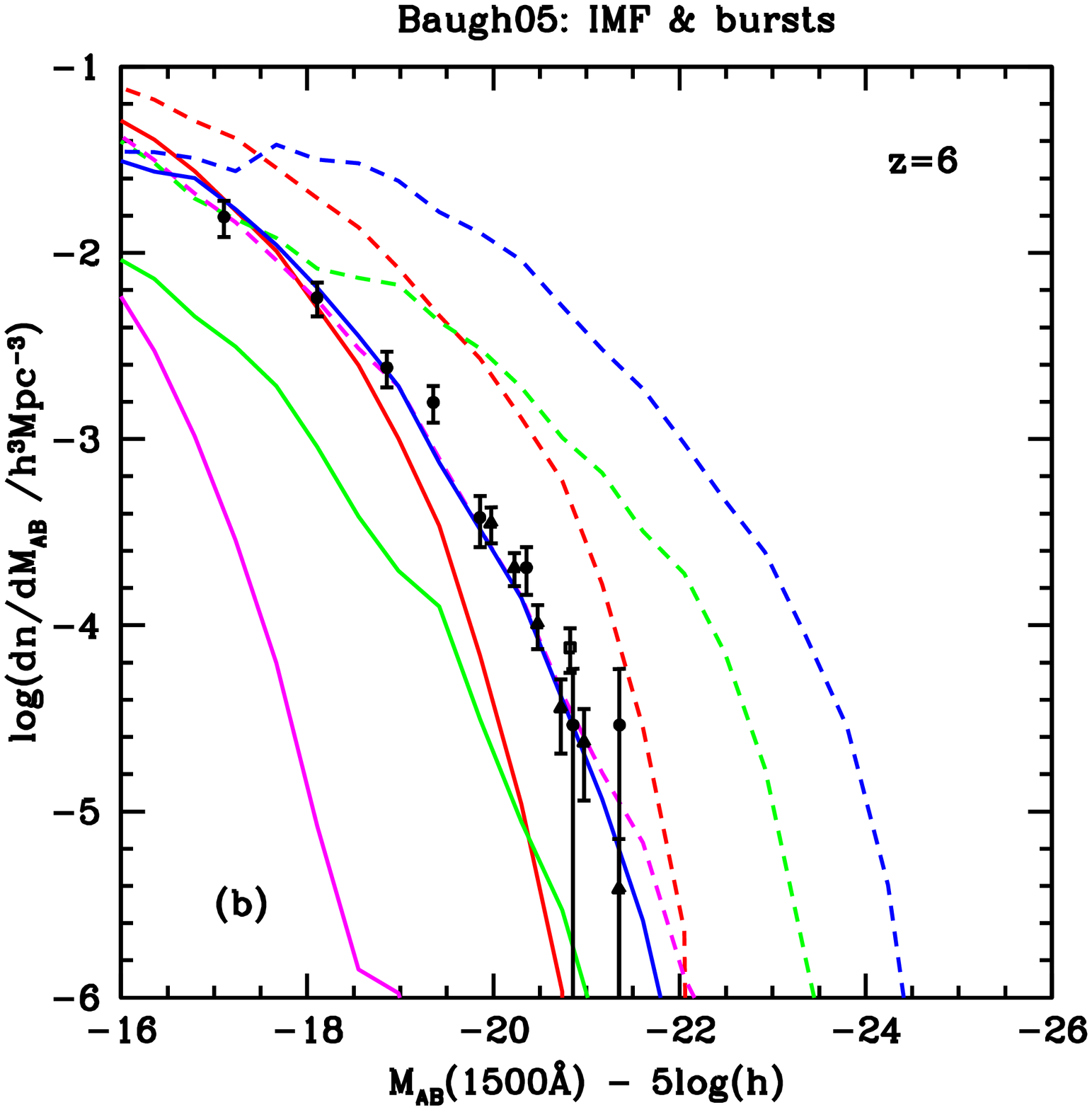}

\includegraphics[width=7cm]{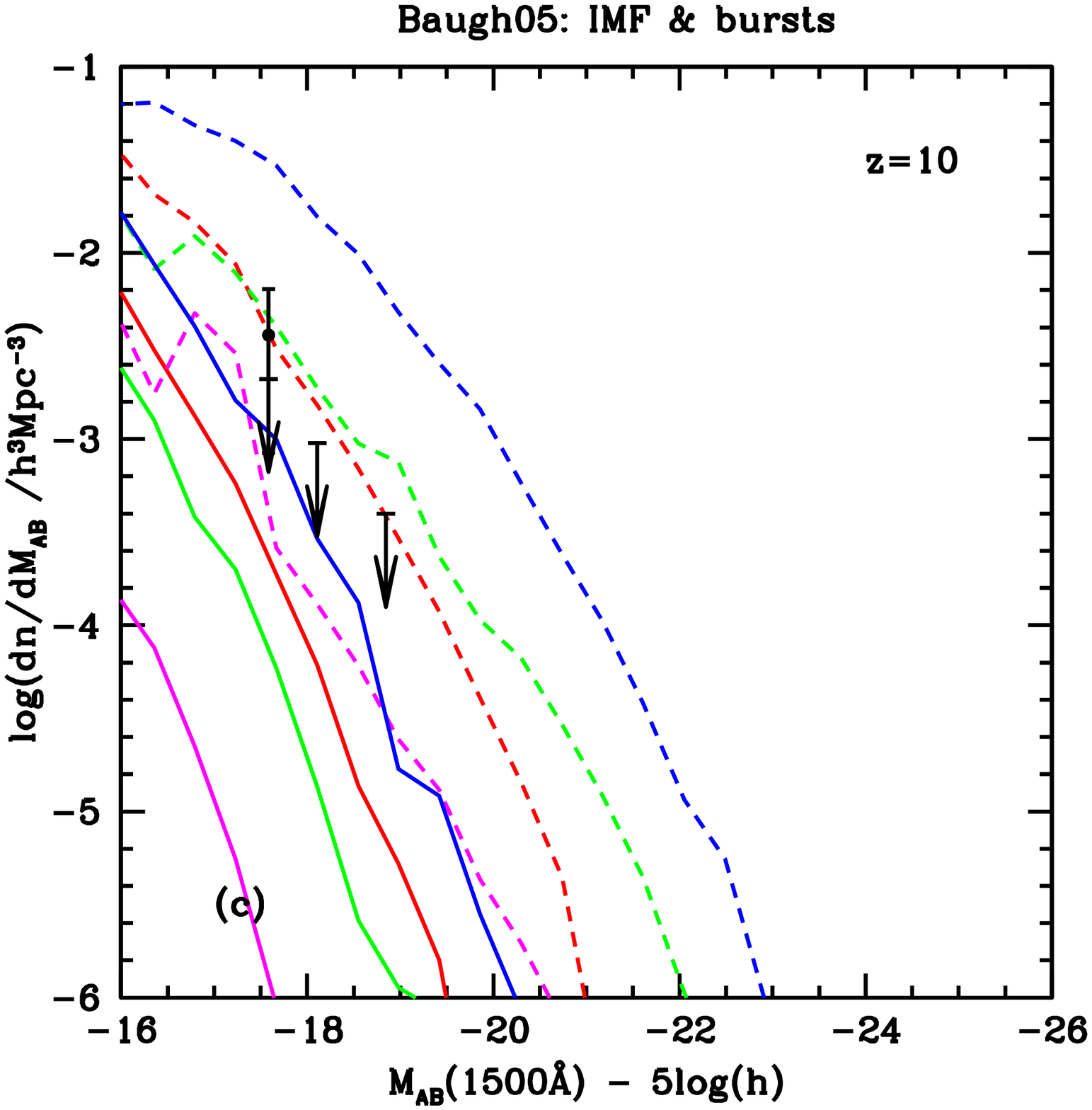}

\end{center}

\caption{Predicted rest-frame 1500\AA\ LFs for the \Bau\ model,
  showing the effects of varying the parameters for bursts.  (a)
  $z=3$. (b) $z=6$. (c) $z=10$. Blue lines -- default model; red --
  Kennicutt IMF in bursts; green -- bursts triggered by major mergers
  only; magenta -- no bursts. The observational data are  the same
  as plotted in Figs.~\ref{fig:lf-comp-lowz} and
  \ref{fig:lf-comp-hiz}.  }

\label{fig:lf-comp.burst}
\end{figure}
%%%%%%%%%%%%%%%%%%%%%%%%%%%%%%%%%%%%%%%%%%%%%%%%%%%%%%%%%%%%%%%%%%%%%%%%%%%%%%%%%%

\subsubsection{IMF and starbursts}
\label{sec:IMF_bursts}

Starbursts and a top-heavy IMF play crucial roles in the \Bau\ model,
so we next consider the effects of changes in these components in
Fig.~\ref{fig:lf-comp.burst}. We show comparisons of the far-UV LFs
for $z=3,6,10$ to span the range of current observational data on
LBGs. In these panels, the blue line shows the default model with
starbursts triggered by all major galaxy mergers and some minor
mergers, with a top-heavy IMF in all of these starbursts. We emphasize
that, acording to the results of \cite{Baugh05}, all of these
ingredients are required in order that the model reproduce the number
counts and redshift distributions of sub-mm galaxies. The magenta line
shows the effect of turning off starbursts completely. In this case,
the predicted LFs are far below the observations. This is because, in
the \Bau\ model, we deliberately chose a star formation timescale in
quiescent disks which is long compared to the Hubble time at high
redshift, so that disks at high redshift are gas-rich, providing more
fuel for star formation in starbursts. As a less extreme variation, we
consider allowing bursts to be triggered by major galaxy mergers only,
shown by the green lines. In this case, the changes in the LF are
smaller but still significant (factors of a few at all plotted
luminosities), implying that most LBGs in the model are bursts
triggered by minor mergers. Finally, we show by the red lines the
effects of assuming a Kennicutt IMF in bursts (the same as for
quiescent star formation), rather than a top-heavy IMF. In this case,
the effects on the LF are mainly at the bright end. Compared to the
default model, there is a large change in the unextincted LFs, but a
smaller change after dust extinction is included. This is because the
top-heavy IMF in bursts has two effects on the far-UV LF which partly
cancel each other: there are more massive stars, and so higher
intrinsic UV luminosities, but there is also more dust, due to
increased metal production, and so more dust extinction. The
relatively modest differences between the dust-extincted LBG LFs with
and without the top-heavy IMF show that fitting these data does not by
itself provide a strong argument for introducing variations in the
IMF, although the fit is clearly better with the top-heavy burst
IMF. The slope of the top-heavy IMF assumed in the \Bau\ model, $x=0$
(\S\ref{sec:GALFORM}), is quite extreme. However, the exact value of
the slope is not important; any IMF with a low-mass cutoff $\gsim
5\Msol$ would produce similar results.

%%%%%%%%%%%%%%%%%%%%%%%%%%%%%%%%%%%%%%%%%%%%%%%%%%%%%%%%%%%%%%%%%%%%%%%%%%%%%%%%%%
% new Fig.6
% LFs at z=3,6,10 comparing Bau05 variants with different dust extinction
\begin{figure}
\begin{center}
\includegraphics[width=7cm]{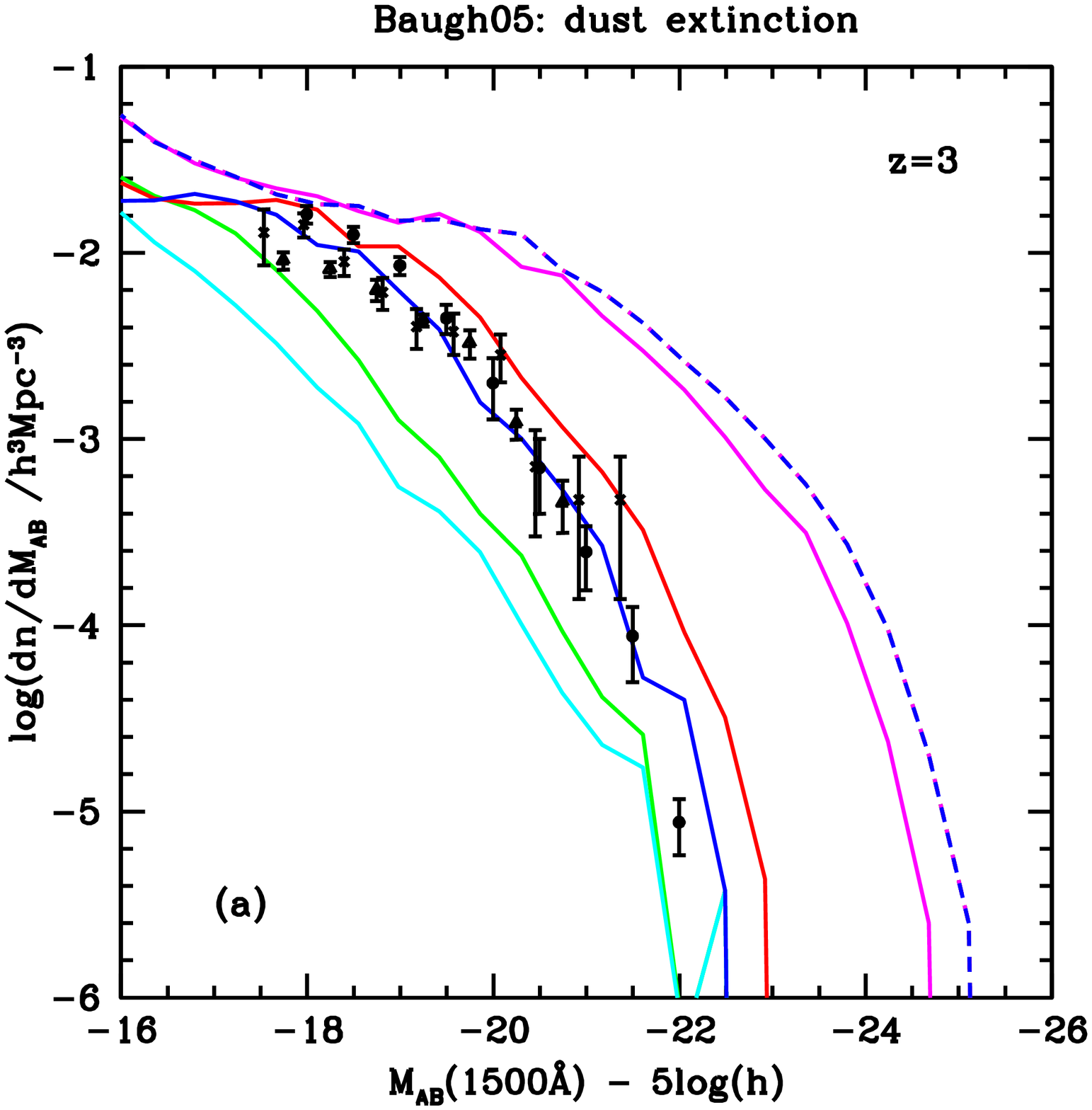}

\includegraphics[width=7cm]{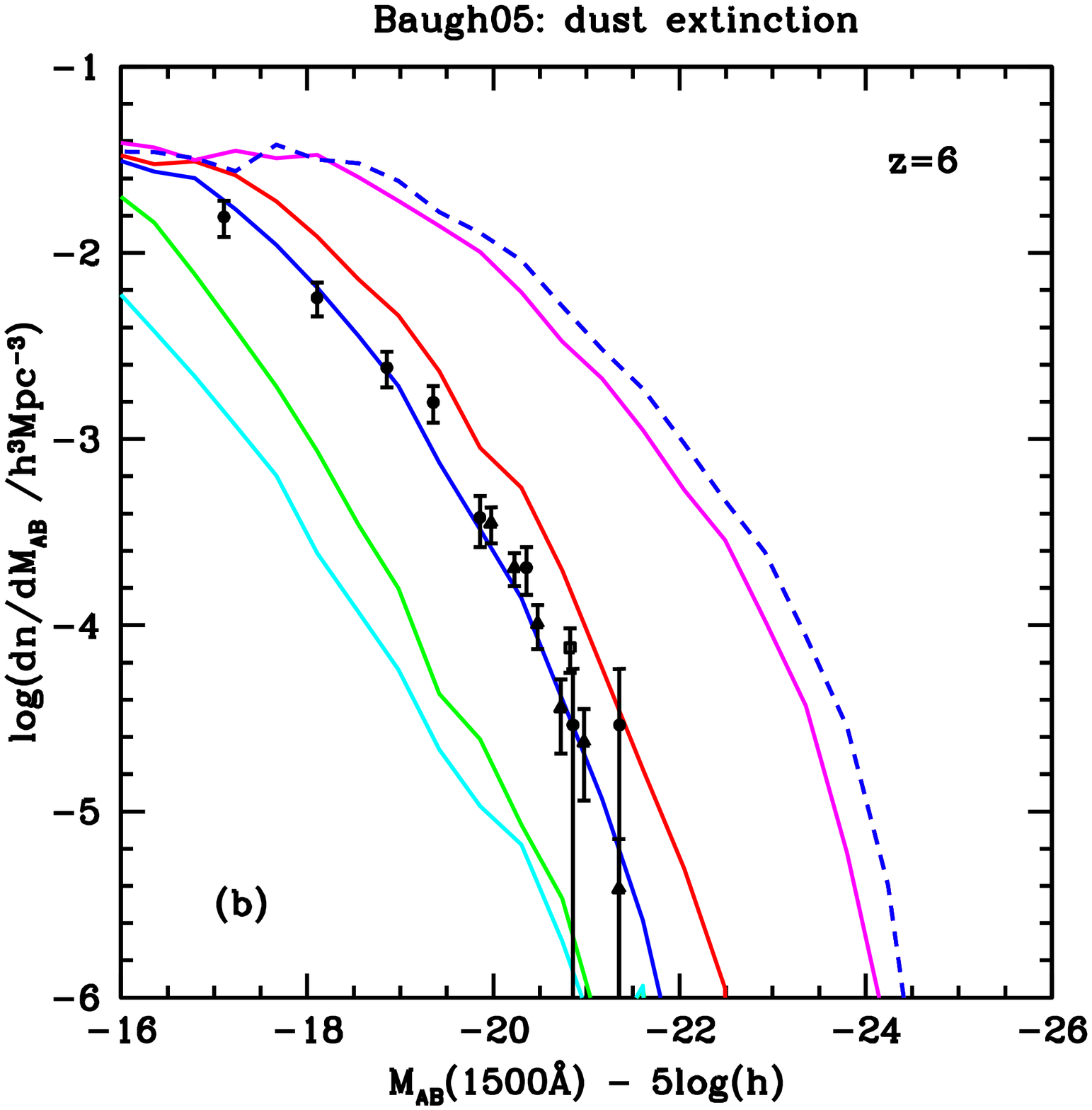}

\includegraphics[width=7cm]{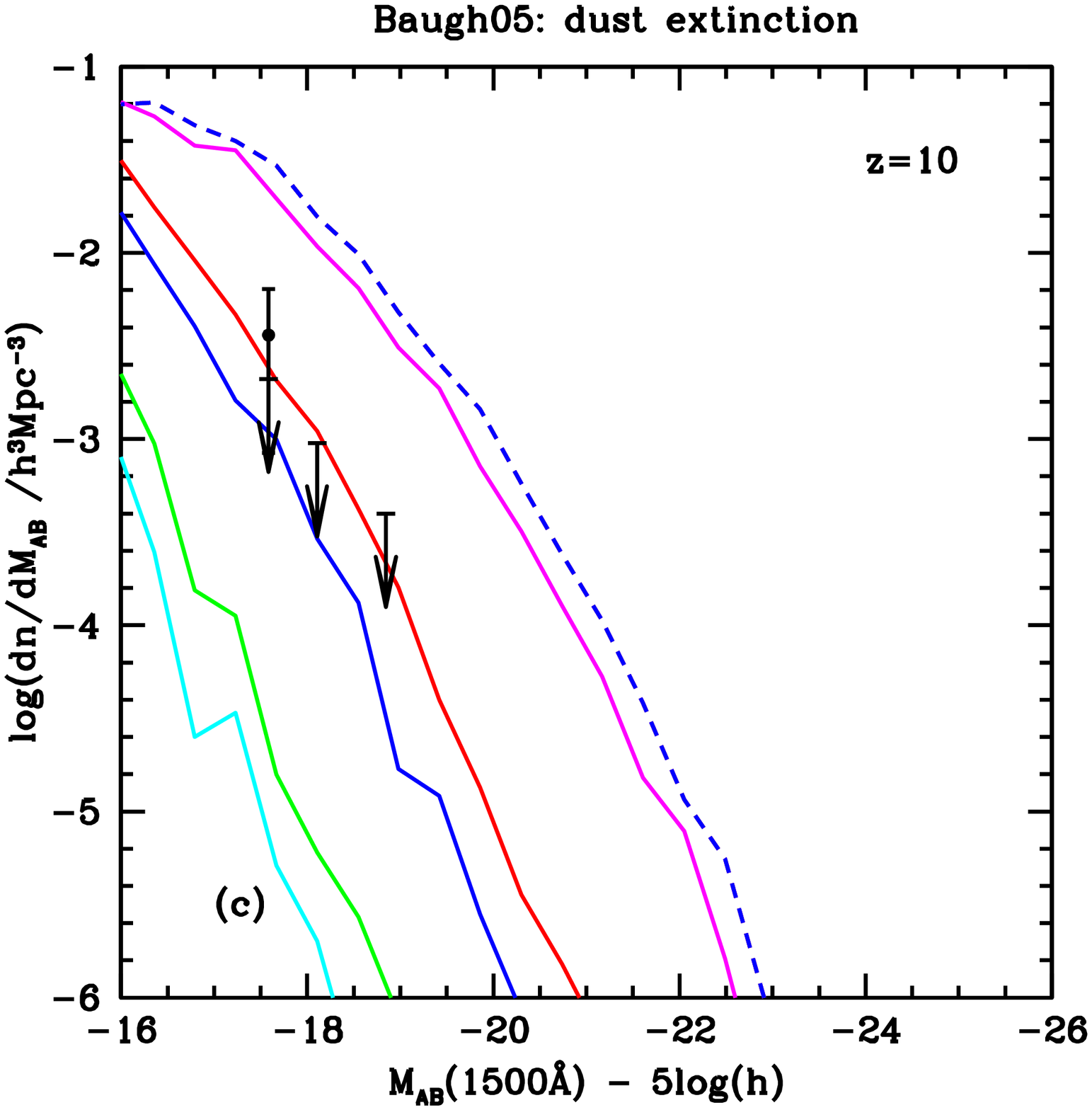}

\end{center}

\caption{{Predicted rest-frame 1500\AA\ LFs for the \Bau\
  model, showing the effects of varying parameters controlling dust
  extinction in the model. (a) $z=3$. (b) $z=6$. (c) $z=10$. The
  dashed blue line in each panel shows the LF without dust extinction,
  while the solid lines show LFs with dust extinction parameters as
  follows: blue -- default model ($\tesc=1\Myr$, $\fcloud=0.25$);
  green -- $\tesc=3\Myr$; cyan -- $\tesc=10\Myr$; red --
  $\fcloud=0.75$; magenta -- $\fcloud=1$.} }

\label{fig:lf-comp.dust}
\end{figure}
%%%%%%%%%%%%%%%%%%%%%%%%%%%%%%%%%%%%%%%%%%%%%%%%%%%%%%%%%%%%%%%%%%%%%%%%%%%%%%%%%%

\subsubsection{Dust extinction}
\label{sec:vary_dust}

{As already discussed, dust extinction has a large effect on
the predicted LBG LF. In Fig.~\ref{fig:lf-comp.dust}, we show the
effects of varying the two key adjustable parameters in our dust
model, $\fcloud$, the fraction of the ISM in molecular clouds (which
affects the attenuation by the diffuse dust), and $\tesc$, the
timescale for stars to escape from the clouds in which they are born
(which affects the attenuation by molecular clouds). The blue curves
show the LFs for the default values ($\fcloud=0.25$ and
$\tesc=1\Myr$). The green and cyan curves show the effect of
increasing $\tesc$ to 3~Myr and 10~Myr respectively, while keeping
$\fcloud$ fixed at 0.25, while the red and magenta curves show the
effect of increasing $\fcloud$ to 0.75 and 1 respectively, while
keeping $\tesc$ fixed at 1~Myr.  The dust-extincted LF is seen to be
quite sensitive to increasing the escape time from its default value,
reflecting the fact that a larger fraction of the UV light is absorbed
within the clouds when $\tesc$ is increased. This effect is especially
strong in the \Bau\ model due to the top-heavy IMF assumed for bursts,
which means that the UV emission is dominated by even higher mass
stars than for a normal IMF. For the $x=0$ IMF, 90\% of the 1500\AA\
light (integrated over time) is emitted by stars with lifetimes $\lsim
10\Myr$, as compared to lifetimes $\lsim 100\Myr$ for the Kennicutt
IMF. The LF is also seen to be insensitive to modest increases in
$\fcloud$ from its default value, but to become more sensitive as
$\fcloud$ approaches 1. This is because the optical depth of the
diffuse dust is proportional to $1-\fcloud$.}

{It can be seen that the effects of $\tesc$ and $\fcloud$ on
the far-UV LF are to some extent degenerate, with a decrease in
$\tesc$ having a similar effect to an increase in $\fcloud$. In
\citet{Baugh05}, we chose to try to match the far-UV LFs of LBGs by
reducing $\tesc$ in bursts from the value 10~Myr assumed in
\citet{Granato00} to 1~Myr, while keeping the same value of
$\fcloud=0.25$ as in the latter paper. We could probably obtain a
similarly good fit to the observed LBG LFs with a larger value of
$\tesc$ combined with a larger value of $\fcloud$, but we do not
pursue this here. In future, it might be possible to distinguish the
effects of $\tesc$ and $\fcloud$ by comparing with observed SEDs of
LBGs over a broader wavelength range, since the two components of the
dust extinction have different dependences on wavelength.}

{Individual galaxies in the \Bau\ model selected by their
  extincted 1500\AA\ luminosities are also predicted to have large
  far-UV extinctions, typically $\sim 1-2$~mag over the range of
  luminosity and redshift plotted in Figs.~\ref{fig:lf-comp-lowz} and
  \ref{fig:lf-comp-hiz}, though with a large scatter. There is a trend
  for the typical extinction to decrease somewhat with increasing
  redshift. Note that selecting galaxies by their extincted
  luminosities automatically biases against including the galaxies
  with the highest extinctions, which can cause the average extinction
  for such samples to differ from that of the underlying population as
  a whole. The values of dust extinction predicted by our model are
  similar to the observational estimates for LBGs by \citet{Bouwens09}
  at $z\sim 3$, but somewhat larger at $z\sim
  6$. \citeauthor{Bouwens09} estimate extinctions from observed UV
  continuum slopes, assuming the same relation between UV slope and
  extinction as found by \citet{Meurer99} for local UV-selected
  starbursts. However, the UV continuum slope is also sensitive to the
  age, metallicity and IMF of the stellar population, which could be
  different in high redshift galaxies compared to local
  samples. Furthermore, even many local galaxies show large deviations
  from the \citeauthor{Meurer99} relation \citep[e.g][]{Bell02,
  Buat10}. For these reasons, we defer a detailed comparison with
  observational estimates of dust extinction in LBGs to a future paper
  in which we directly compare predicted and observed far-UV colours
  in LBG samples.}

%%%%%%%%%%%%%%%%%%%%%%%%%%%%%%%%%%%%%%%%%%%%%%%%%%%%%%%%%%%%%%%%%%%%%%%%%%%%%%%%%%
% Fig.
% LFs at z=3,6,10 comparing Bau05 variants: burst timescale
%  tauburstmin & fdyn
\begin{figure}
\begin{center}
\includegraphics[width=7cm]{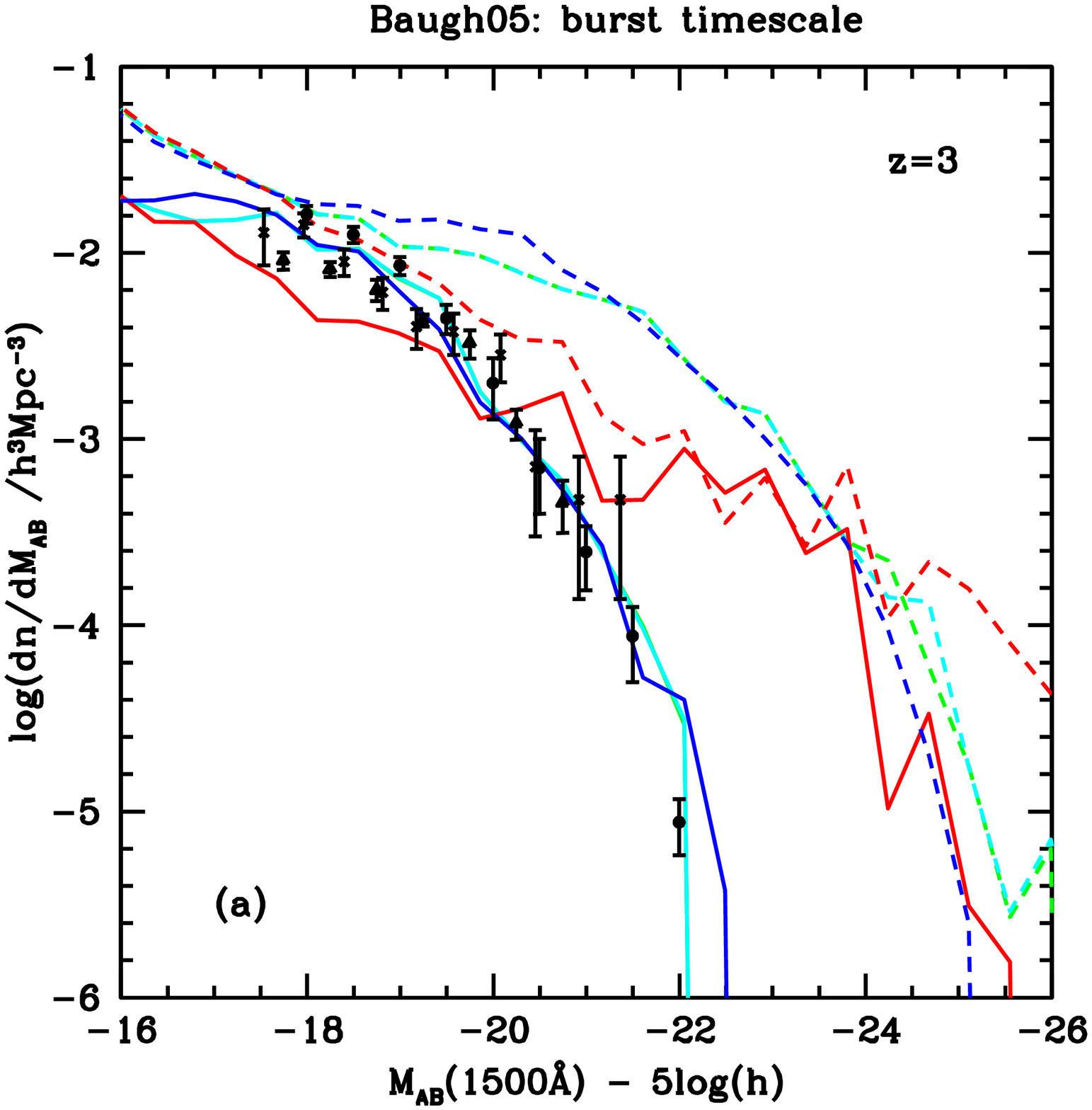}

\includegraphics[width=7cm]{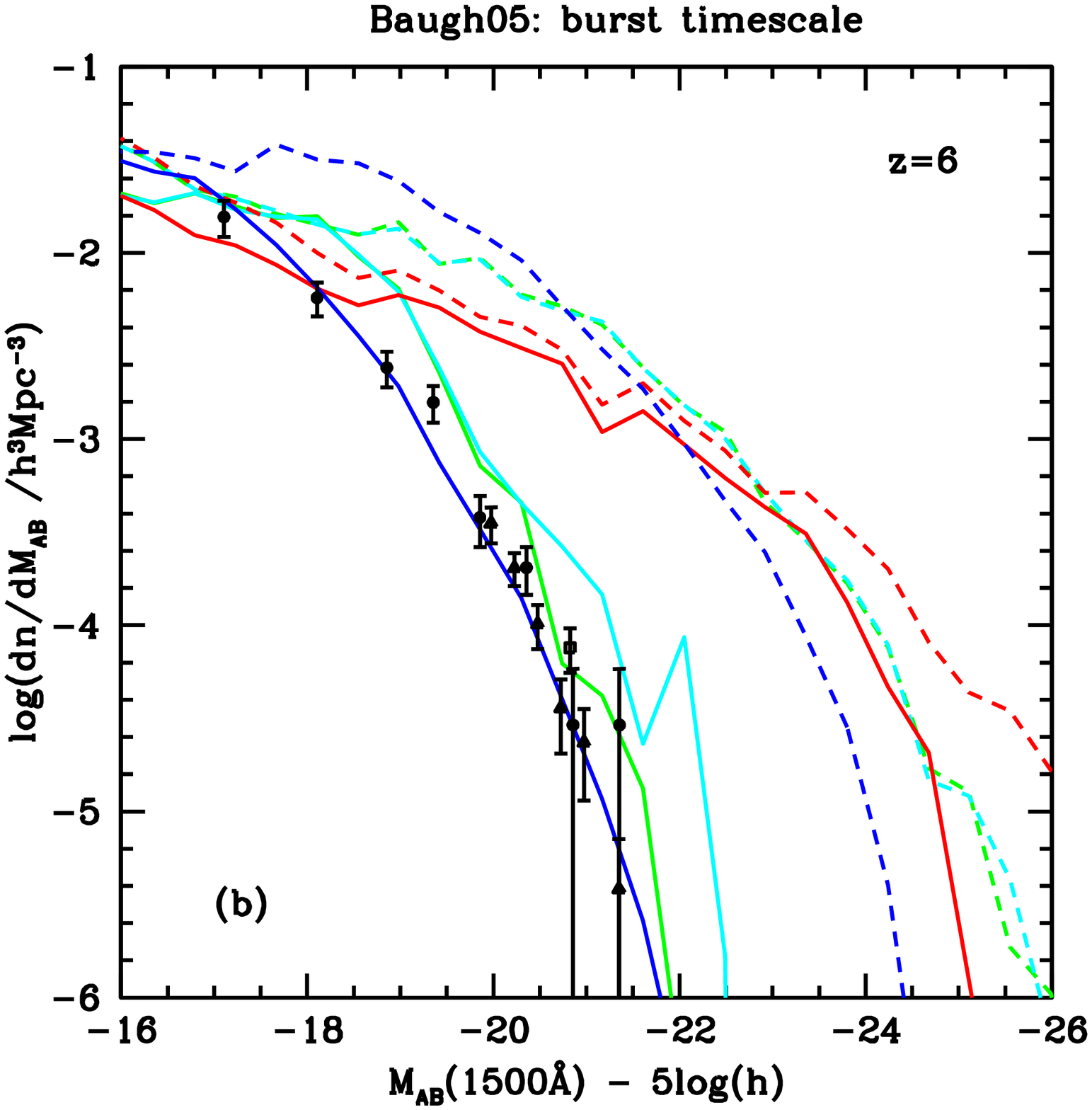}

\includegraphics[width=7cm]{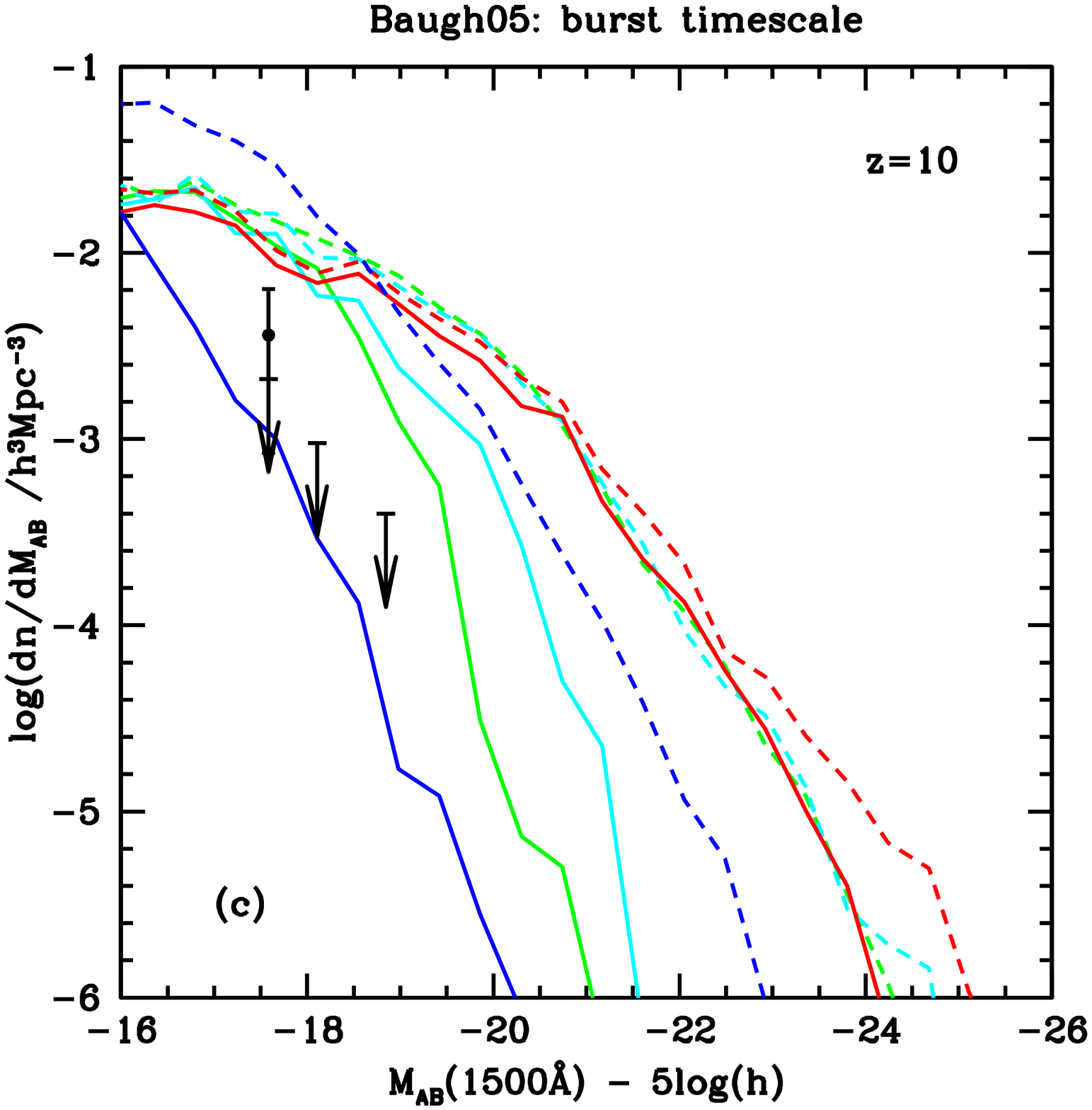}

\end{center}

\caption{Predicted rest-frame 1500\AA\ LFs for the \Bau\ model,
  showing the effects of varying the star formation timescale in
  bursts.  (a) $z=3$. (b) $z=6$. (c) $z=10$. Blue lines -- default
  model ($\tauburstmin=0.2 \Gyr$, $\fdyn=50$); green --
  $\tauburstmin=0.02 \Gyr$; cyan -- $\tauburstmin=0.005\ Gyr$; red --
  $\tauburstmin=0.005 \Gyr$, $\fdyn=2$. }

\label{fig:lf-comp.tauburst}
\end{figure}
%%%%%%%%%%%%%%%%%%%%%%%%%%%%%%%%%%%%%%%%%%%%%%%%%%%%%%%%%%%%%%%%%%%%%%%%%%%%%%%%%%

%%%%%%%%%%%%%%%%%%%%%%%%%%%%%%%%%%%%%%%%%%%%%%%%%%%%%%%%%%%%%%%%%%%%%%%%%%%%%%%%%%
% Fig.
% LFs at z=3 comparing Bau05 variants: efold
\begin{figure}
\begin{center}
\includegraphics[width=7cm]{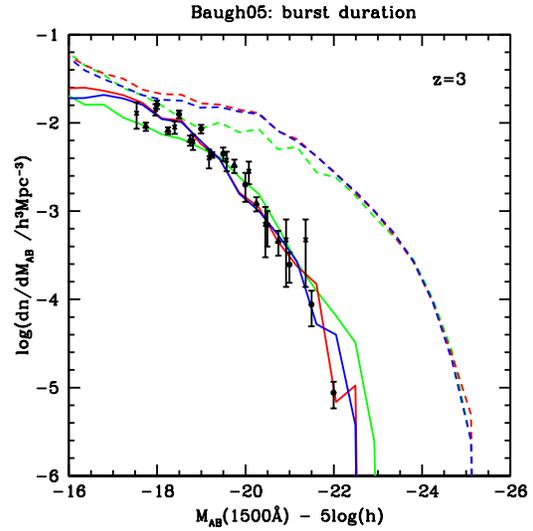}

\end{center}

\caption{Predicted rest-frame 1500\AA\ LF at $z=3$ for the \Bau\
  model, showing the effects of varying the burst duration as a
  multiple of its e-folding time. Blue lines -- default ($\ntau=5$);
  green -- $\ntau=1$; red -- $\ntau=5$.  }

\label{fig:lf-comp.efold}
\end{figure}
%%%%%%%%%%%%%%%%%%%%%%%%%%%%%%%%%%%%%%%%%%%%%%%%%%%%%%%%%%%%%%%%%%%%%%%%%%%%%%%%%%

\subsubsection{Burst timescale and duration}
Since starbursts play a very important role in the \Bau\ model, we
investigate the effects on the LBG LFs of changing the burst timescale
and duration. In Fig.~\ref{fig:lf-comp.tauburst} we show the effect of
varying the parameters $\fdyn$ and $\tauburstmin$ which control the
SFR timescale in bursts. The blue lines show the default \Bau\ model,
which assumes $\fdyn=50$ and $\tauburstmin=0.2$~Gyr. The green and
cyan lines show the effect of reducing $\tauburstmin$ to 0.02~Gyr and
0.005~Gyr respectively, while keeping $\fdyn=50$. This is seen to have
only a small effect on the dust-extincted LF at $z=3$ and $z=6$, but
to have a large effect at $z=10$, where reducing $\tauburstmin$
results in a larger number of bright LBGs. The larger sensitivity to
$\tauburstmin$ at higher redshifts is because the spheroid dynamical
time $\taudyn$ (which enters in the burst timescale as shown in
eq.(\ref{eq:tauburst})) is typically shorter at high-$z$. The red line
in Fig.~\ref{fig:lf-comp.tauburst} shows the effect of reducing both
$\fdyn$ and $\tauburstmin$ to the values $\fdyn=2$ and
$\tauburstmin=0.005$~Gyr assumed in the \Bow\ model. In this case, the
effect on the dust-extincted far-UV LF is dramatic at all redshifts -
there are many more very luminous LBGs and slightly fewer faint LBGs
compared to the default \Bau\ model. The effects of dust extinction on
the LF are also much smaller, which is the main reason for the larger
number of bright galaxies in the extincted LF. The reason for the
smaller dust extinction when the burst timescale is greatly reduced is
that the stars that emit the 1500\AA\ light have typical lifetimes
$\sim 0.01-0.1$~Gyr, longer than the burst e-folding time in this
case. Therefore the dust associated with the burst has mostly been
consumed or ejected while these stars are still shining. {This
is the same effect as discussed earlier in connection with the \Bow\
model.} The difference in burst timescales, rather than the top-heavy
IMF, seems to be the main reason why the \Bau\ model is much more
successful than the \Bow\ model in explaining the observed LBG LFs.

Fig.~\ref{fig:lf-comp.efold} shows the effect of changing the duration
of bursts while keeping the star formation timescales and e-folding
times fixed. The blue lines show our default model, in which the burst
duration is $\ntau=3$ e-folding times, while the green and red curves
show the effect of decreasing this to $\ntau=1$ or increasing it to
$\ntau=5$. The effects are seen to be quite small, showing that the
predictions for the far-UV LF are insensitive to details of the burst
duration, but are more sensitive to the star formation timescale.

%%%%%%%%%%%%%%%%%%%%%%%%%%%%%%%%%%%%%%%%%%%%%%%%%%%%%%%%%%%%%%%%%%%%%%%%%%%%%%%%%%
% Fig.
% LFs at z=3 comparing Bau05 variants: Vhot & alphahot
\begin{figure}
\begin{center}
\includegraphics[width=7cm]{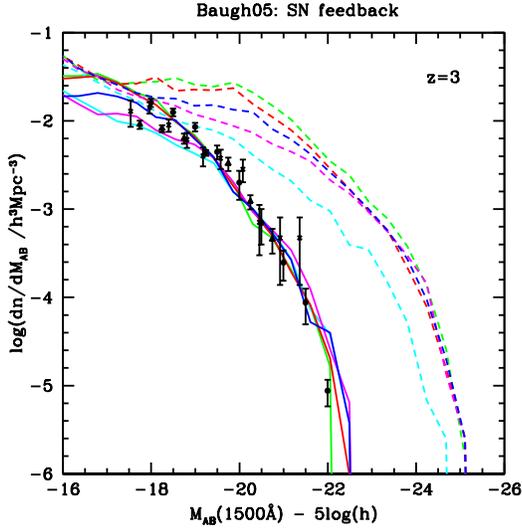}

\end{center}

\caption{Predicted  rest-frame 1500\AA\ LF at $z=3$ for the \Bau\ model,
  showing the effects of varying the parameters for supernova
  feedback. Blue lines --
default; green -- $\Vhot=150\kms$; cyan -- $\Vhot=600\kms$; red --
$\alphahot=1$; magenta -- $\alphahot=3$.   }

\label{fig:lf-comp.SN}
\end{figure}
%%%%%%%%%%%%%%%%%%%%%%%%%%%%%%%%%%%%%%%%%%%%%%%%%%%%%%%%%%%%%%%%%%%%%%%%%%%%%%%%%%

\subsubsection{Supernova feedback}
\label{sec:SN-feedback}

Fig.~\ref{fig:lf-comp.SN} shows the effect of varying the supernova
feedback parameters $\Vhot$ and $\alphahot$. We only show results for
$z=3$, since the changes for $z=6$ and 10 are similar or smaller for
the luminosity range plotted here. As before, the blue lines show the
default model with $\Vhot=300\kms$ and $\alphahot=2$. The green and
cyan lines respectively show the effect of reducing or increasing
$\Vhot$ by a factor 2, which causes the gas ejection rate for a given
star formation rate to decrease or increase by a factor 4. We see that
the LF increases or decreases at the faint end as $\Vhot$ decreases or
increases. The effect is modest for the dust-extincted LF, but larger
for the unextincted LF, showing that the effects of changing feedback
on the stellar luminosities are to some extent compensated by changes
in the dust extinction. The red and magenta lines show the effect of
decreasing $\alphahot$ to 1 or increasing it to 3. A larger
$\alphahot$ means that the feedback increases more strongly as the
galaxy circular velocity decreases. Again, the galaxy LF increases or
decreases at the faint end in the sense expected, though by a modest
amount. We conclude that the predicted LFs in the \Bau\ model are
relatively insensitive to the supernova feedback adopted, in the range
covered by current observational data.

%\clearpage

%%%%%%%%%%%%%%%%%%%%%%%%%%%%%%%%%%%%%%%%%%%%%%%%%%%%%%%%%%%%%%%%%%%%%%%%%%%%%%%%%%
% Fig.
% half-light radii vs M1500
\begin{figure}

\begin{center}

\includegraphics[width=6.5cm]{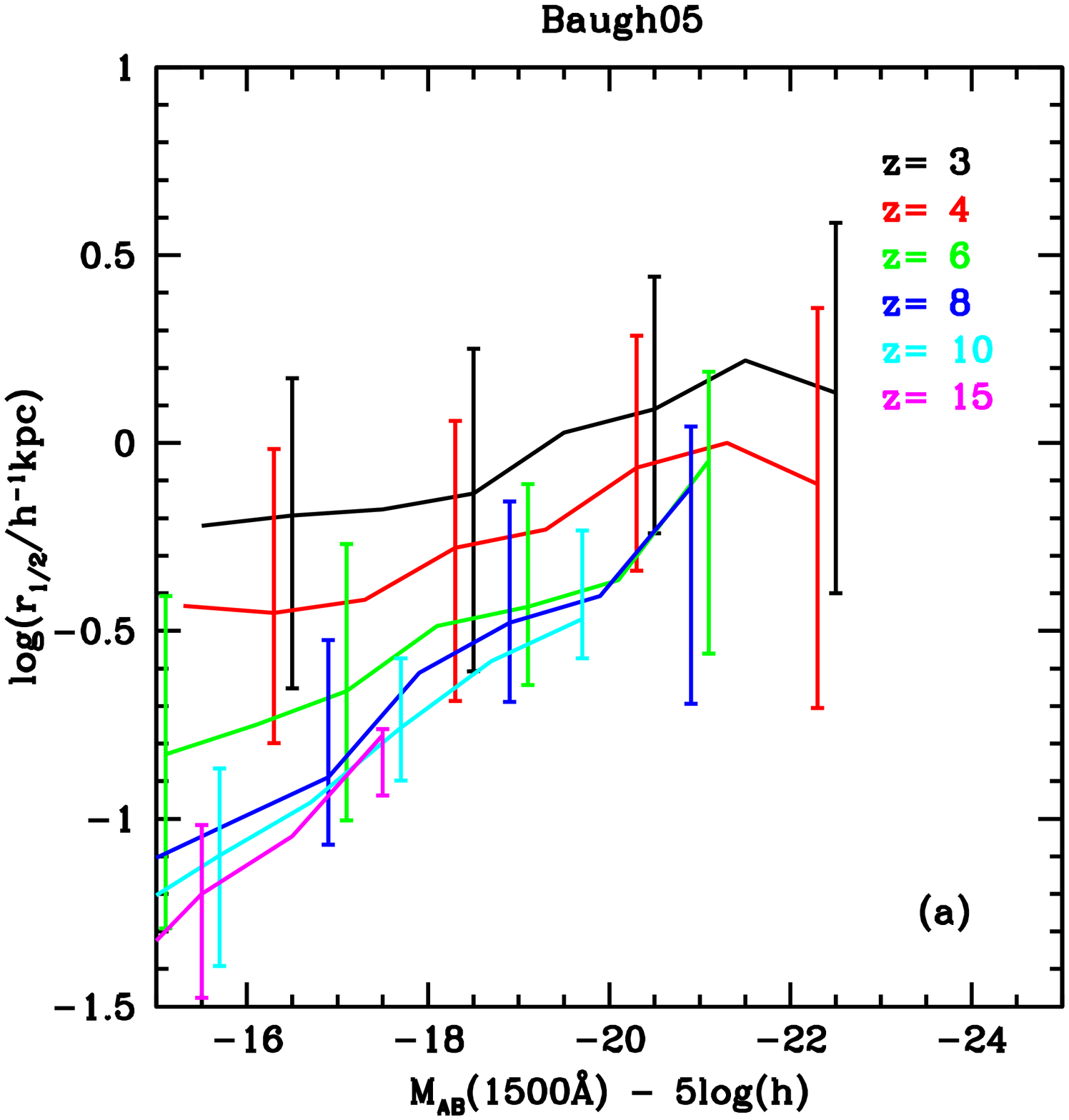}

\includegraphics[width=6.5cm]{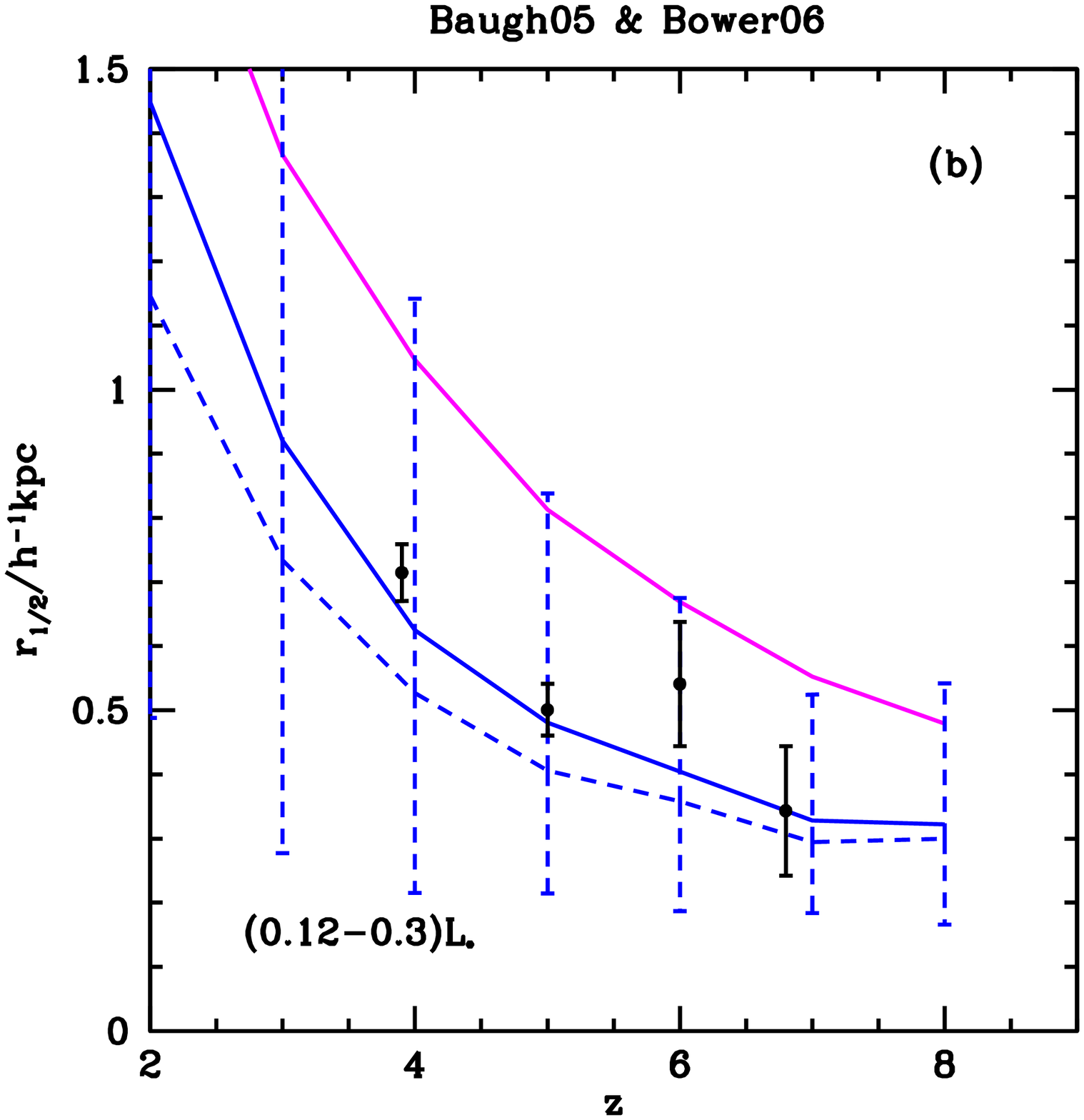}

\includegraphics[width=6.5cm]{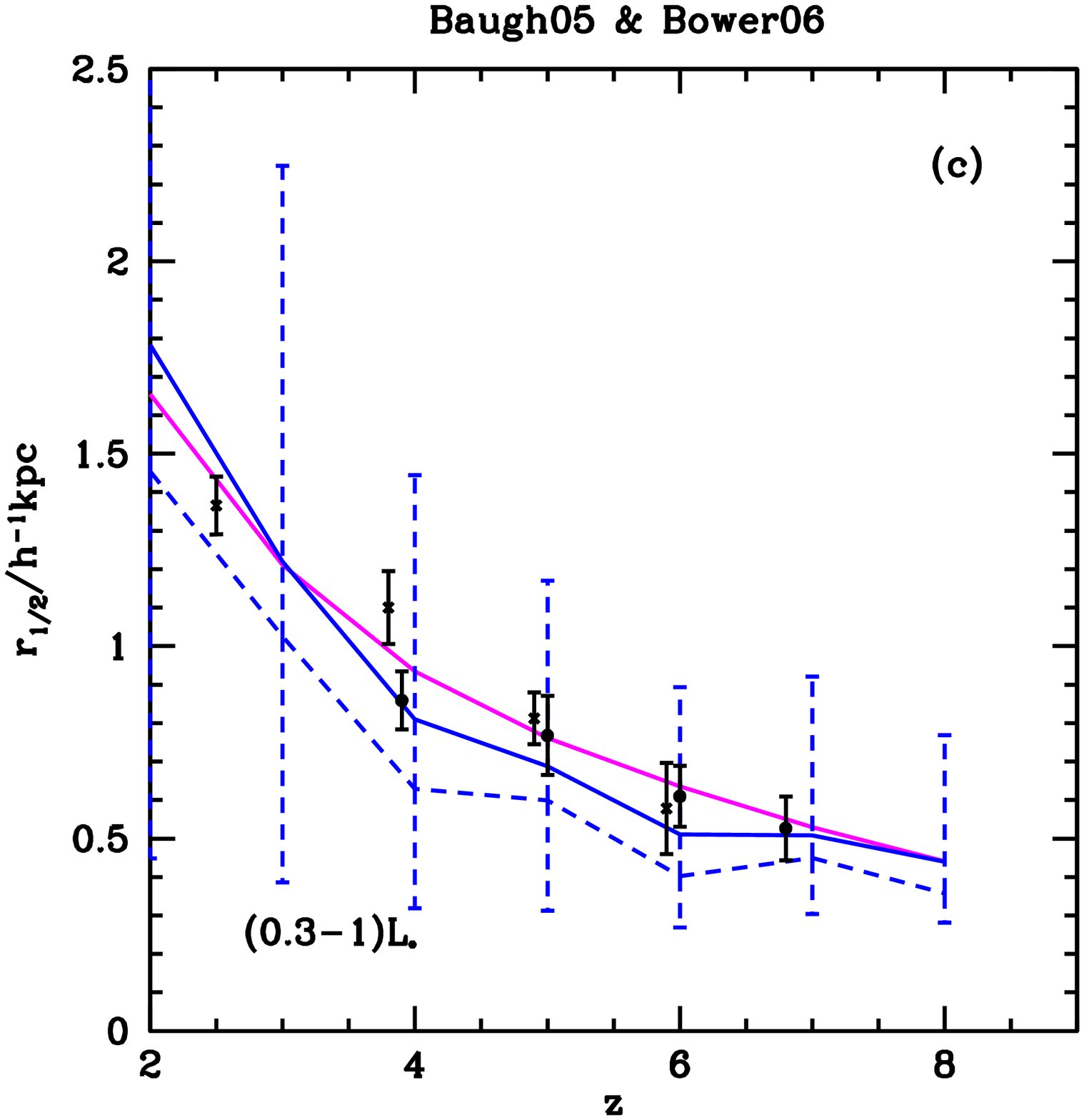}

\end{center}

\caption{(a) Predicted rest-frame UV (1500\AA) half-light radii of
  LBGs as a function of rest-frame 1500\AA\ absolute magnitude for the
  \Bau\ model. Both radii and absolute magnitudes include dust
  extinction. The lines show median values, and the error bars show
  the 10-90\% range. Different redshifts are shown in different
  colours, as indicated by the key, and have been slightly
  horizontally offset from each other for clarity. (b) and (c):
  Comparison of predicted rest-frame UV half-light radii for the \Bau\
  model (blue) and \Bow\ model (magenta) with observational from
  \citet{Oesch10b} (filled circles) and \citet{Bouwens04a} (crosses).
  Two luminosity ranges are shown, $0.12<L(UV)/L_* <0.3$ in (b) and
  $0.3<L(UV)/L_* <1$ in (c), equivalent to $-18.9< \MUV <-17.9$ and
  $-20.2< \MUV <-18.9$ respectively. In these two panels, the solid
  lines show the mean radius in that luminosity range, and the dashed
  lines show the medians and 10-90\% percentiles (\Bau\ model
  only). The observational data are the mean values, with the
  errorbars showing the error on the mean.}

\label{fig:rstar}
\end{figure}

%%%%%%%%%%%%%%%%%%%%%%%%%%%%%%%%%%%%%%%%%%%%%%%%%%%%%%%%%%%%%%%%%%%%%%%%%%%

\section{Sizes and other physical properties of LBGs}
\label{sec:props}

We now present predictions for other physical properties of LBGs as
functions of their rest-frame UV luminosities. For galaxy sizes, we
show some results for both the \Bau\ and \Bow\ models, but for the other
properties, we only show the \Bau\ model, since this model is in much
better agreement with observations of both LBG luminosity functions
and sizes. We make a detailed comparison here with observational data
on the sizes, but for reasons of space we make only brief comparisons
with observational constraints on the other properties, deferring more
detailed comparisons to future papers.

\subsection{Sizes}

Following their luminosity functions, the half-light radii of LBGs in
the rest-frame far-UV are their most directly observable physical
property, if HST imaging is available. {\GALFORM\ predicts
sizes for the disk and bulge components of each galaxy using the
method detailed in \citet{Cole00}. In summary, disk sizes are
calculated assuming conservation of angular momentum of the gas that
cools out of halos, while bulge sizes are calculated assuming energy
conservation when bulges form by mergers or by disk instabilities. The
method allows for the gravity of the disk, bulge and dark halo, and
includes the effects of halo contraction due to the gravity of the
disk and bulge. Half-light radii in different bands are then
calculated by combining the predicted half-mass radii and luminosities
for the disk and bulge components, assuming an exponential profile for
the disk and an $r^{1/4}$-law profile for the bulge.}

In the top panel of Fig.~\ref{fig:rstar}, we show the
predicted median half-light radii (at rest-frame 1500\AA) as a
function of rest-frame 1500\AA\ luminosity for galaxies at different
redshifts in the range $z=3-15$, for the \Bau\ model. The stellar
half-mass radii plotted in the same way are very similar to the
half-light radii, so we do not show them here. (In principle they can
differ if the disk and bulge components have different ages,
metallicities or IMFs.)  The galaxy sizes range from $\sim \kpc$ for
the brightest LBGs to $\sim 100\pc$ for the faintest and highest
redshift LBGs.  Both the \Bau\ and \Bow\ models predict that sizes
decrease with increasing redshift at a given far-UV luminosity,
although by factors that depend on the model and on the galaxy
luminosity. This size evolution reflects that for the host dark matter
halos, for which the size scales as $M_{\rm halo}^{1/3} (1+z)^{-3/2}$
at the redshifts shown here, but it differs in detail.
%% However, the size-luminosity relation for the model galaxies
%% evolves differently in detail from the size-mass relation for
%% halos. 
The ratio of galaxy to halo size is not fixed, but depends on the gas
cooling and merger history (see \citealt{Cole00} for more
details). The far-UV luminosity is also only indirectly related to the
halo mass, depending instead on the recent star formation rate, which
may correlate only weakly with halo mass, especially if bursts are
important, as in the \Bau\ model.

The dependence of size on luminosity is distinctly different in the
two models. In the \Bau\ model, it approaches $r \propto L^{1/3}$ at
the higher redshifts plotted, but is shallower at the lower
redshifts. On the other hand, the \Bow\ model shows a flat or even
declining dependence of size on luminosity. These dependences are
similar to what the respective models predict for the dependence of
disk size on luminosity in the $r$-band at $z=0$, where the \Bau\
model was found to agree much better with the disk size-luminosity
relation measured in the SDSS \citep{Gonzalez09}.

In the lower two panels of Fig.~\ref{fig:rstar}, we compare the size
predictions from both models with observational data on LBGs from
\citet{Bouwens04a} and \citet{Oesch10b}, based on HST imaging in the
rest-frame far-UV, and covering redshifts $2 \lsim z \lsim 7$. The
observational data are given as mean half-light radii at different
redshifts for two different far-UV luminosity ranges, specified as
multiples of the observed characteristic far-UV luminosity $L_\ast$ at
$z=3$ (taken by \citeauthor{Oesch10b} to be $M_{AB}(1600\AA)=-21.0$
for $h=0.7$). We present the model predictions in the same way. The
two panels show the two luminosity ranges; the solid blue and magenta
lines show the mean sizes predicted by the \Bau\ and \Bow\ models
respectively. The blue dashed lines show the median sizes in the \Bau\
model, with the error bars indicating the 10-90\% ranges. For the
higher luminosity range ($0.3<L/L_\ast<1$), both models are in
reasonable agreement with the observations, but in the lower
luminosity range ($0.12<L/L_\ast<0.3$), only the \Bau\ model matches
the observed sizes. This shortcoming of the \Bow\ model results from
the flat size-luminosity relation it predicts. The prediction by the
\Bau\ model of LBG sizes in agreement with observations is a
significant success of the model, since the only observed size
information originally used in fixing the model parameters was disk
sizes at $z=0$.

\begin{figure*}

\begin{center}

\begin{minipage}{7cm}
\includegraphics[width=7cm]{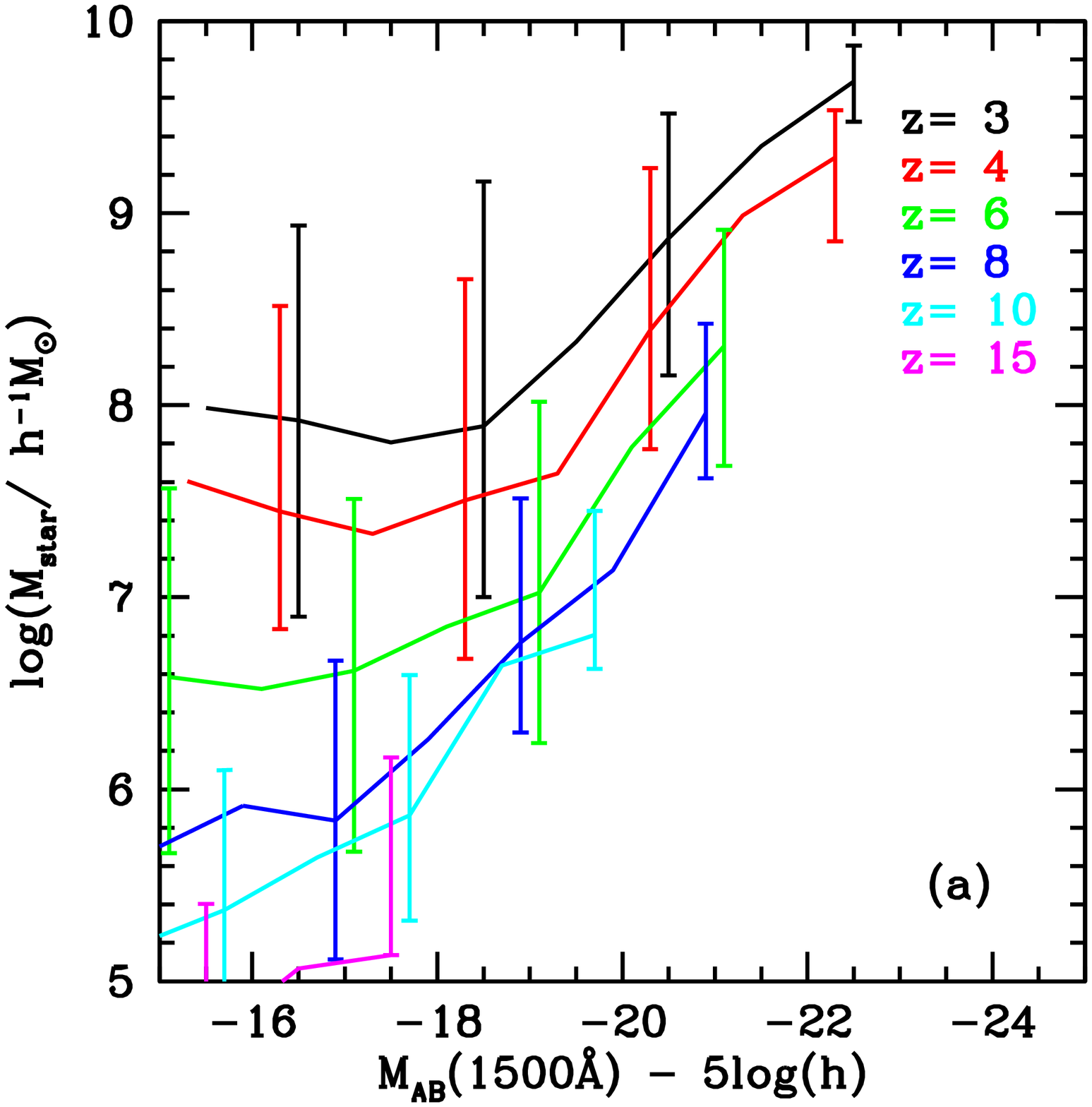}
\end{minipage}
\begin{minipage}{7cm}
\includegraphics[width=7cm]{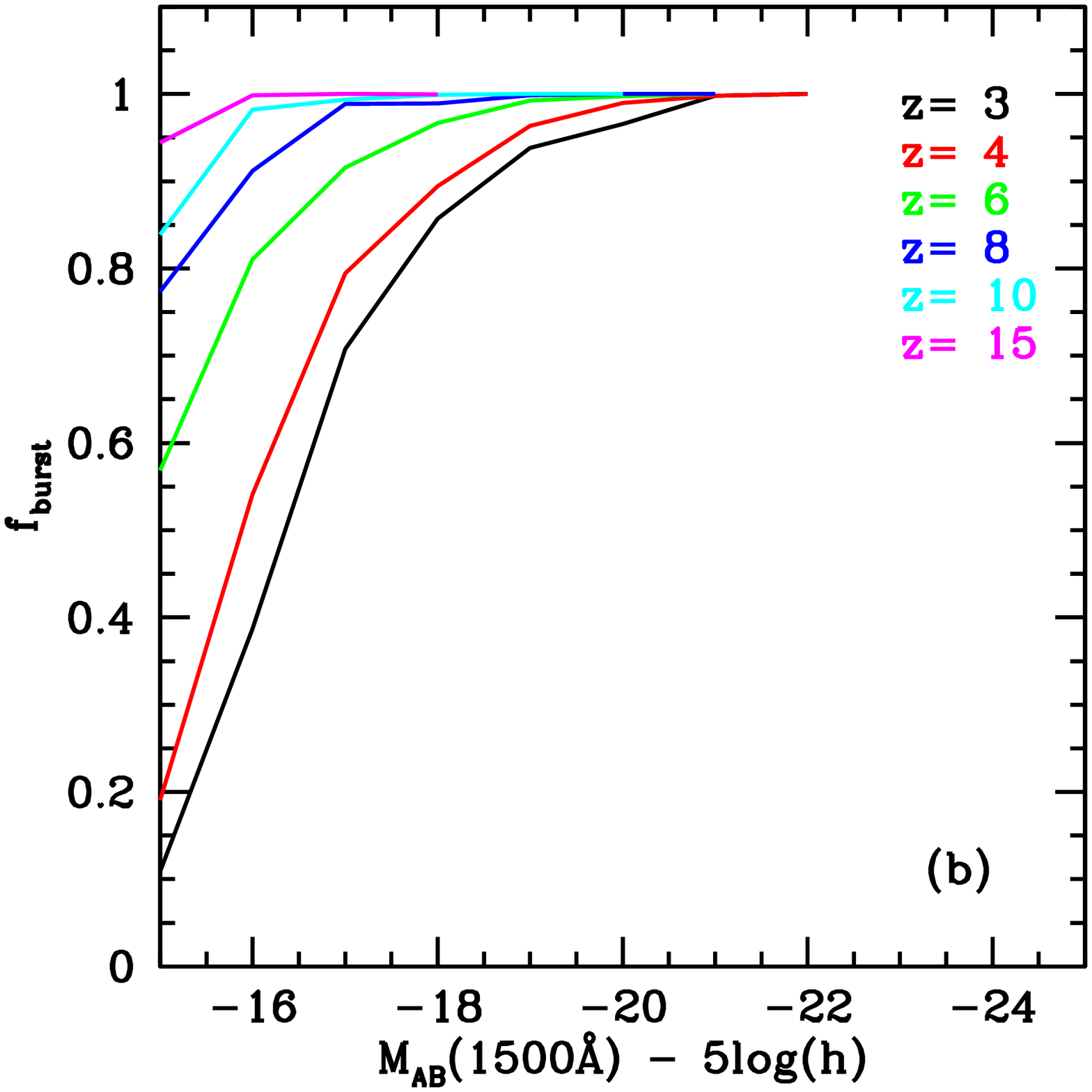}
\end{minipage}

\begin{minipage}{7cm}
\includegraphics[width=7cm]{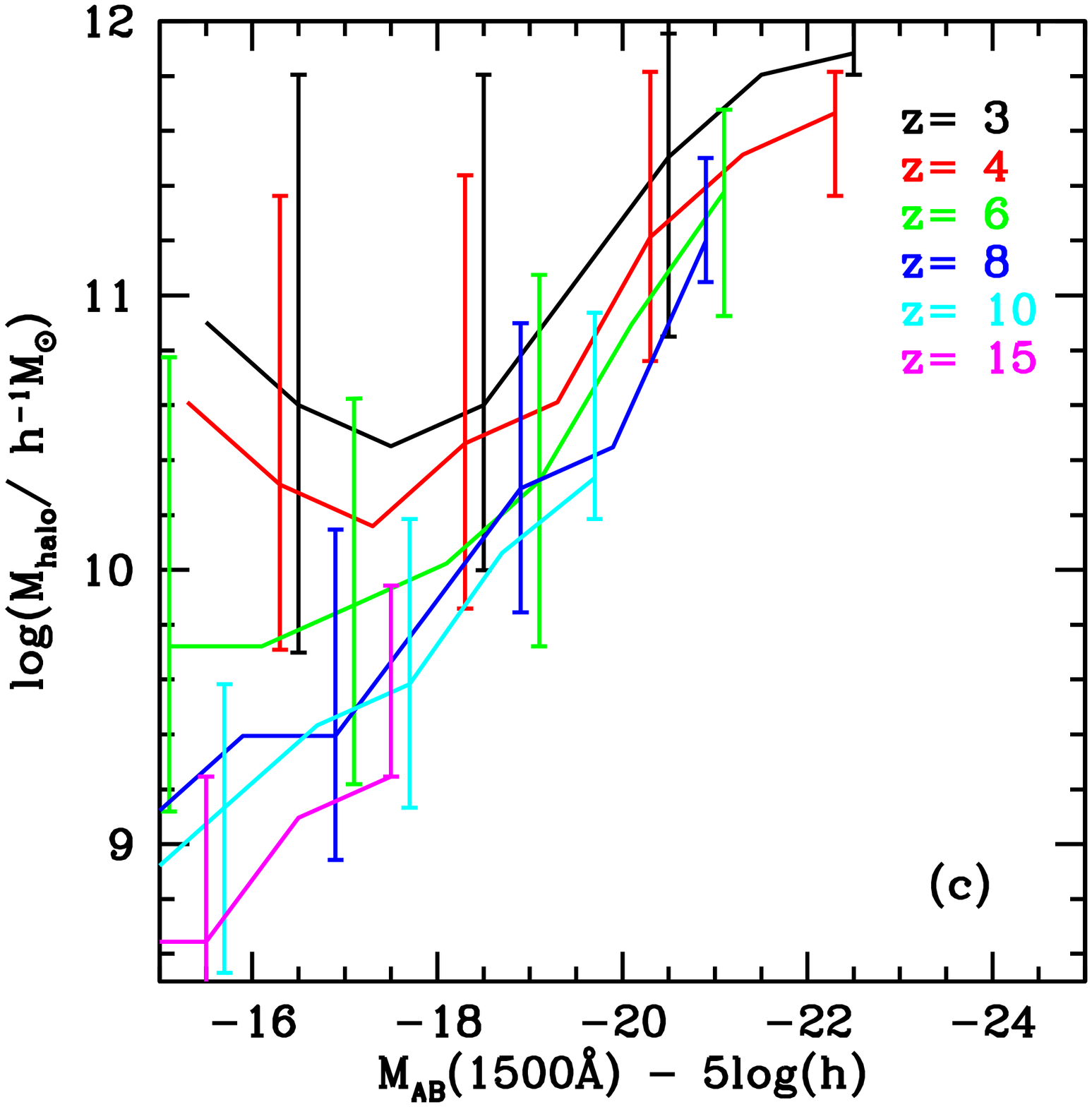}
\end{minipage}
\begin{minipage}{7cm}
\includegraphics[width=7cm]{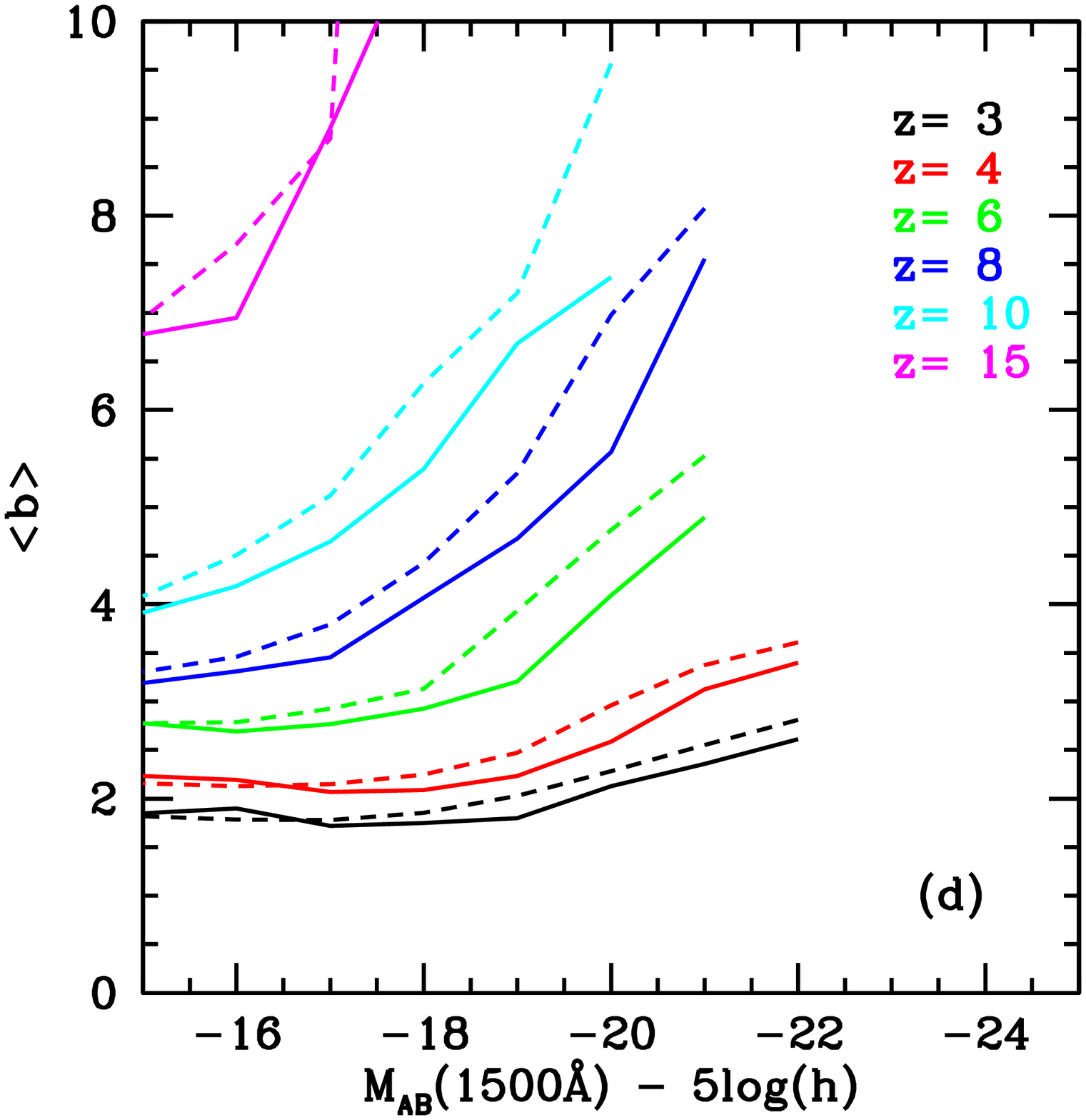}
\end{minipage}

\begin{minipage}{7cm}
\includegraphics[width=7cm]{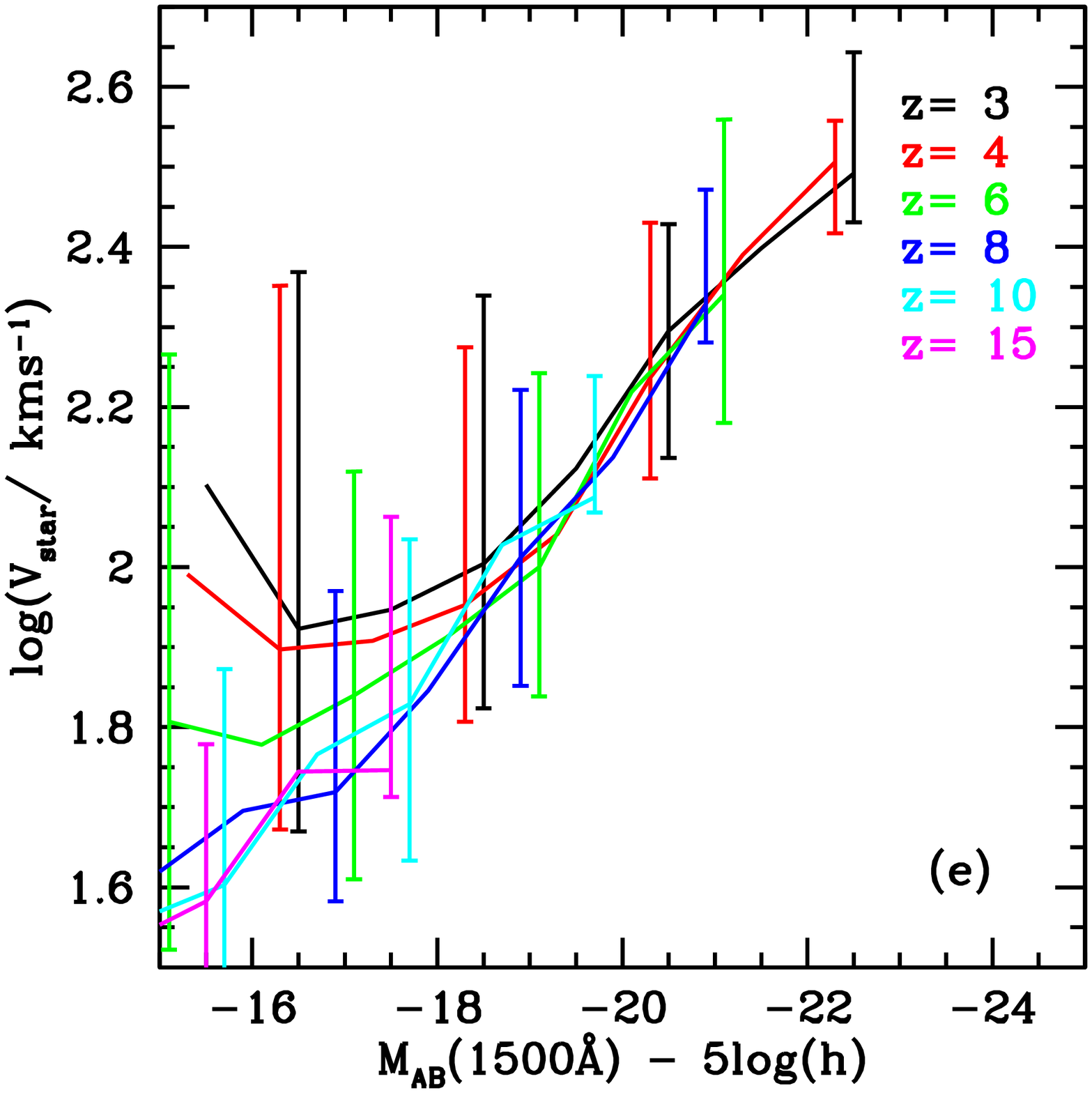}
\end{minipage}
\begin{minipage}{7cm}
\includegraphics[width=7cm]{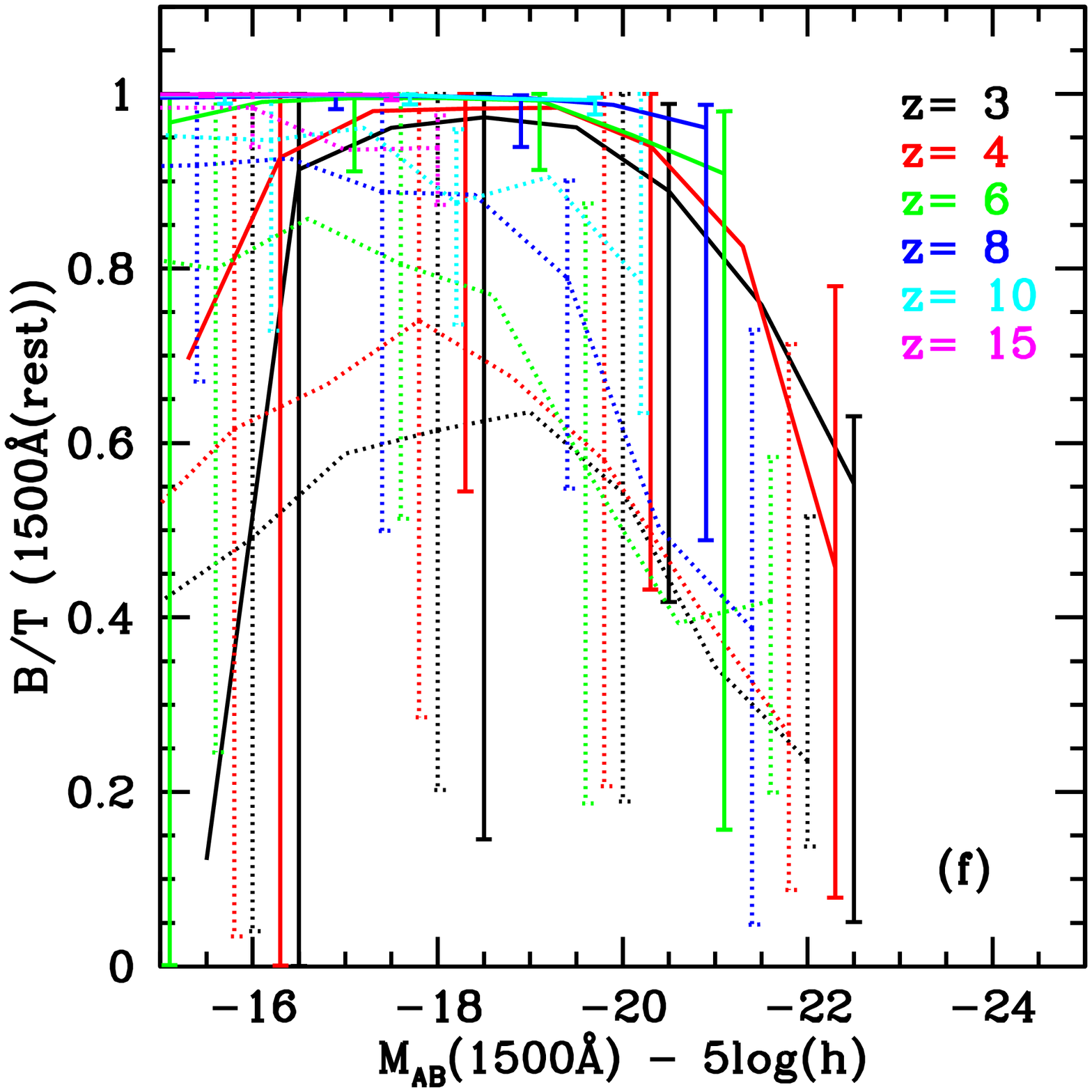}
\end{minipage}

\end{center}

\caption{Predicted physical properties of LBGs as functions of
  dust-extincted rest-frame 1500\AA\ absolute magnitude in the \Bau\
  model. Different redshifts are shown by different colours, as
  indicated by the key. Unless stated otherwise, all properties are
  plotted as medians, with error bars showing the 10-90\% range.  The
  different panels are as follows: (a) stellar mass; (b) fraction of
  LBGs currently undergoing a burst; (c) dark matter halo mass; (d)
  mean bias (solid and dashed lines respectively show mean bias in a
  luminosity bin and for galaxies brighter than a given luminosity);
  (e) circular velocity of stars; (f) dust-extincted bulge-to-total
  luminosity ratio at 1500\AA\ (with bulge-to-total stellar mass ratio
  shown by dotted lines). }

\label{fig:propsA}
\end{figure*}

We now consider a range of other physical properties for the \Bau\
model only, shown as functions of dust-extincted rest-frame far-UV
luminosity for redshifts $3<z<15$ in Figs.\ref{fig:propsA} and
\ref{fig:propsB}

%%%%%%%%%%%%%%%%%%%%%%%%%%%%%%%%%%%%%%%%%%%%%%%%%%%%%%%%%%%%%%%%%%%%%%%%%%%%%%%%%%
% Fig.
% physical properties of LBGs

%%%%%%%%%%%%%%%%%%%%%%%%%%%%%%%%%%%%%%%%%%%%%%%%%%%%%%%%%%%%%%%%%%%%%%%%%%%

%%%%%%%%%%%%%%%%%%%%%%%%%%%%%%%%%%%%%%%%%%%%%%%%%%%%%%%%%%%%%%%%%%%%%%%%%%%%%%%%%%
% Fig.
% mag(3.6mu) vs mag(1500A rest-frame)
\begin{figure}

\begin{center}
\includegraphics[width=7cm]{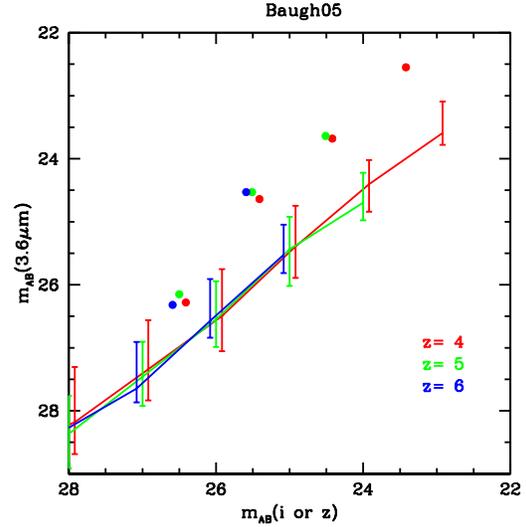}
\end{center}

\caption{Relation between \SPITZER\ 3.6$\mum$ and optical ($i$ or $z$)
  apparent magnitude, for LBGs at $z=4$ (red), $z=5$ (green) and $z=6$
  (blue). The lines show the predicted median values in the \Bau\
  model and the error bars show the 10-90\% range. The filled circles
  show the observed median values (binned in optical magnitude) from
  \citet{Stark09}, for the same redshifts as the model. For clarity,
  small horizontal offsets have been applied to lines for different
  redshifts, and the same offsets have been applied to the
  observational datapoints.}

\label{fig:mag3p6}
\end{figure}

%%%%%%%%%%%%%%%%%%%%%%%%%%%%%%%%%%%%%%%%%%%%%%%%%%%%%%%%%%%%%%%%%%%%%%%%%%%

\subsection{Stellar masses}
Median stellar masses are shown in the top left panel of
Fig.~\ref{fig:propsA}. They cover a very wide range, from $\sim 10^{10}
\hMsol$ at the highest luminosities and lowest redshifts, down to
$\sim 10^5 \hMsol$ or less at low luminosities and high redshifts. The
median stellar mass generally increases with increasing far-UV
luminosity, but the dependence is flatter at lower luminosities and
lower redshifts. At a given luminosity, the mass decreases with
increasing redshift, but this dependence is much stronger at lower
luminosities.

In general, the stars in a galaxy are a mixture of populations formed
quiescently and in bursts, with different IMFs.  As we discuss next, a
large fraction of LBGs are predicted to be starbursts. In this case,
the far-UV luminosity is generally dominated by the burst component,
but the stellar mass need not be. For model LBGs at $z=3$, $\sim
50-80\%$ of the stellar mass was formed quiescently in disks, and even
for those LBGs which are ongoing bursts, only 10-30\% of the stellar
mass has formed in the current burst. At higher redshifts and lower
luminosities, the fraction of the stellar mass formed in bursts
increases for galaxies selected by their far-UV luminosities. For
example, at $z=6$, the fraction of stellar mass formed quiescently
increases with luminosity from 20\% to 80\% over the range plotted
here, while for the subset which are ongoing bursts, the fraction
formed in the current burst decreases from $\sim 70\%$ to $\sim 10\%$
over the same range. At $z=10$, the stellar mass in LBGs is even more
dominated by burst populations, with only $\sim 10-30\%$ formed
quiescently, and $\sim 50-90\%$ formed in a current burst.

The stellar masses of observed LBGs have been estimated in a number of
studies, starting with $z=3$ \citep{Sawicki98,Papovich01,Shapley01},
and later extending to $z=4-6$
\citep{Verma07,Eyles07,Stark07,Yabe09,Stark09}. These various studies
selected LBGs in different ranges of luminosity, using different
colour selections, and obtained results which appear in conflict in
some cases. However, all of them estimated stellar masses {\em
photometrically}, by fitting models for galaxy SEDs (with an assumed
IMF) to flux measurements at different wavelengths ranging, in the
rest frame, from the far-UV to the optical. These SED models depend on
a significant number of parameters in addition to the stellar mass,
including dust extinction, age, star formation history and metallicity
(and possibly redshift). In principle, all of these parameters should
be determined simultaneously from the SED-fitting. In practice, this
leads to degeneracies between different parameters, and significant
uncertainties in the stellar masses, by factors up to 10
\citep[e.g.][]{Papovich01,Verma07,Yabe09}. These studies all used a
Salpeter or other similar Solar neighbourhood IMF. If one allows
larger variations in the IMF (as assumed in our model), then the
uncertainties in the estimated stellar masses become even
larger. Since the top-heavy IMF plays an important role in our model,
it is not meaningful to compare stellar masses from the model directly
with observational estimates which assume a Salpeter or similar
IMF. Therefore we do not make a detailed comparison with values of
stellar mass estimated from observational data, but simply note a few
numbers. \citet{Papovich01} and \citet{Shapley01} find typical stellar
masses $\sim 10^{10}\Msol$ for LBGs at $z=3$ with $\MUV \sim
-20$. \citet{Verma07} find masses $\sim 2\times 10^9 \Msol$ for
similar luminosities at $z=5$. For LBGs at $z=4-6$, \citet{Stark09}
find stellar masses $\sim 10^9-10^{10} \Msol$ over the luminosity
range $-19 \lsim \MUV \lsim -21$, with only a weak dependence on
redshift at a given luminosity. Our model predicts values of the
stellar mass which are $\sim 30$ times smaller at a given luminosity
and redshift.

However, the observational estimates of stellar mass are driven mainly
by the fluxes measured at rest-frame optical wavelengths, so given the
effects of the IMF, it is more meaningful to compare directly with
these flux measurements. As an example, in Fig.~\ref{fig:mag3p6} we
compare our model predictions with the \SPITZER\ 3.6$\mum$ fluxes
measured by \citet{Stark09} for LBGs at $z=4-6$ as a function of the
optical flux. We have chosen the \citeauthor{Stark09} sample because
it is large, homogeneously selected, and covers a wider range of
redshift and luminosity than other studies. At the redshifts of these
galaxies, the 3.6$\mum$ band corresponds to a rest-frame wavelength of
$5000-7000\AA$, while the $i$- or $z$-band optical magnitude
corresponds to $1200-1600\AA$ in the rest frame. In the Figure, the
lines show the medians predicted by the model, while the filled
circles show the median values of 3.6$\mum$ magnitude measured by
\citeauthor{Stark09}. The predicted relation is seen to lie
0.8-1.6~mag (i.e. a factor 2-4) fainter in the 3.6~$\mum$ magnitude
than the observations. The discrepancy is thus $\sim 10$ times smaller
than the apparent discrepancy in stellar masses, confirming that the
latter is mostly due to the difference between the IMF in the model
and the IMF assumed in the observational estimates of stellar mass.
Interestingly, the model predicts a very similar trend of rest-frame
optical vs. far-UV luminosity as seen in the observational data, with
almost no dependence on redshift in the range $z=4-6$.  The weak
dependence on redshift in the model presumably results from the LBGs
mostly being starbursts. We will make a more detailed comparison of
the predicted rest-frame far-UV to optical SEDs of LBGs with
observations in a future paper.

\subsection{Burst fraction}
The top right panel of Fig.~\ref{fig:propsA} shows the fraction of
galaxies which are currently undergoing a burst. The burst population
dominates at higher luminosities and higher redshifts, with a
transition to more quiescent systems at lower luminosities. This is
important for understanding many of the other properties. {In
the \Bau\ model, bursts are triggered only by galaxy mergers. As
discussed in \S\ref{sec:IMF_bursts}, the bursts responsible for LBGs
in the model are typically triggered by minor, rather than major, galaxy
mergers.}  The burst e-folding timescale varies over the range $\sim
0.01-0.2 \Gyr$ for model LBGs in the luminosity range shown here, for
$z=3-10$. It typically increases with luminosity but decreases with
increasing redshift. Note that the burst e-folding time can be
significantly shorter than the burst SFR timescale due to the effect
of supernova feedback. Bursts are assumed to last for 3 e-folding
times in our fiducial model. However, the brightest LBGs are on
average seen only a small fraction of an e-folding time after the
burst began - this is presumably a selection effect due to a burst
being brightest in its early stages when the SFR is highest.

\subsection{Halo masses and clustering bias}
Dark halo masses are shown in the middle left panel of
Fig.~\ref{fig:propsA}. The trends with luminosity and redshift are
similar to those already discussed for the stellar mass. They also
cover a wide range, from $\sim 10^{12} \hMsol$ at the highest
luminosities and lower redshifts, down to $\sim 10^9 \hMsol$ or less
at low luminosities and high redshifts.

The masses of the dark halos hosting LBGs are not directly observable,
but they can be constrained from clustering measurements. In the
middle right panel of Fig.~\ref{fig:propsA}, we show the predicted
mean large-scale clustering bias of LBGs. We calculated this using the
analytical halo bias formula of \citet{Sheth01}, averaging the bias
over the distribution of host halo masses in each bin or range of
luminosity (see \citealt{Baugh98} for details). This bias applies on
scales larger than roughly the sum of the virial radii of the two host
halos. The solid lines show the mean bias within a luminosity bin,
while the dashed lines show the mean bias for galaxies brighter than a
given luminosity. The bias increases with increasing redshift, and
also generally with increasing luminosity, although the latter
dependence is weak at lower redshifts. A similar behaviour for bias of
LAEs in the same model was earlier found by \citet{Orsi08}. We predict
a bias $b \sim 2$ at $z=3$, nearly independent of luminosity,
increasing to $b\sim 4-7$ at $z=10$ and to $b\sim 8$ at $z=15$. This
increase of the bias with redshift reflects the fact that selecting at
a fixed luminosity picks out host halos further and further out on the
tail of the halo mass function at larger and larger redshifts, which
are more and more strongly clustered relative to the dark matter as a
whole.

Observational measurements of clustering of LBGs include those of
\citet{Adelberger98} and \citet{Giavalisco98} at $z=3$ and
\citet{Ouchi04} at $z=4-5$. \citeauthor{Ouchi04} combine their own and
previous clustering measurements with theoretical predictions of the
dark matter clustering to estimate $b\approx 2.7\pm 0.4$ at $z\approx
3$, $b\approx 3.5\pm 0.7$ at $z\approx 4$ and $b\approx 4.6 \pm 1.1$
at $z\approx 5$ for LBGs with $\MUV \lsim -20$. Our model predictions
are $b \approx 2.2, 2.8, 3.5$ respectively for the same redshifts and
luminosities, which are consistent with these estimates within their
errors. We will investigate the clustering in more detail in a future
paper.

\subsection{Circular velocity}
The circular velocity at the stellar half-mass radius is plotted in
the lower left panel of Fig.~\ref{fig:propsA}. While the trends with
luminosity and redshift are qualitatively similar to those for stellar
and halo mass, the dependence on redshift is weaker and the total
range of values is much smaller, from $\sim 40\kms$ at the lowest
luminosities and highest redshifts plotted, up to $\sim 250\kms$ at
the highest luminosities and lowest redshifts. {At higher
luminosities ($\MUV \lsim -19$), there is almost no dependence of
circular velocity on redshift.}

The velocity widths of LBGs at $z\sim 3$ have been measured from their
rest-frame optical emission lines by
\citet{Pettini98,Pettini01}. These lines (unlike emission and
absorption lines in the rest-frame UV) reflect the kinematics of the
star-forming gas in the galaxies, which should be related to the
rotation velocity of the galaxy. \citet{Pettini01} measure a median
1-D velocity dispersion $\sigma \approx 70\kms$ in LBGs with $\MUV
\sim -21$. Allowing for inclination and other effects in galaxy disks,
this corresponds to a typical circular velocity $\Vc \approx 120\kms$
\citep{Rix97}. The line widths measured from CO in two
gravitationally-lensed LBGs imply similar values for $\sigma$ and
$\Vc$ \citep{Baker04,Coppin07}. Our model predicts a median circular
velocity $\sim 200\kms$ at the same luminosity and redshift, somewhat
larger than the observational estimates.

\subsection{B/T and morphology}
The bottom right panel of Fig.~\ref{fig:propsA} shows the
bulge-to-total ratio $B/T$, either measured in dust-extincted
rest-frame 1500\AA\ luminosity (solid lines) or in stellar mass
(dotted lines). We see that LBGs are very bulge-dominated in their
rest-frame UV light at higher luminosities, but with a transition to
disk domination at low luminosities. This reflects the dominance of
bursts at the higher luminosities, since our model assumes that star
formation in bursts all occurs in the bulge component. The dominance
of the bulge is generally less extreme in terms of stellar mass. At
the highest luminosities, LBGs are predicted to be disk-dominated in
stellar mass even though they are bulge-dominated in rest-frame UV
light. This reflects the triggering of many starbursts by minor galaxy
mergers, which leave the stellar disk of the larger galaxy intact, but
sweep all of the cold gas into the bulge, where it forms stars in a
burst.

HST imaging has revealed a wide range of morphologies for
LBGs. \citet{Ravindranath06} fit Sersic profiles to the rest-frame
far-UV images of a large sample of LBGs at $z=3-5$ with
$\MUV \lsim -19.5$. Based on this, they classified
about 30\% as bulge-dominated, 40\% as exponential, and 30\% as
multiple cores, suggesting galaxy mergers. \citet{Lotz06} analysed a
smaller sample of LBGs at $z\sim 4$ (also in the rest-frame far-UV)
using completely different techniques, but arrived at similar
conclusions, finding $\sim 30\%$ undisturbed bulgelike morphologies,
$\sim 10-25\%$ major mergers, and $\sim 50\%$ exponential disks or
minor mergers. Our model predicts that most LBGs at these
luminosities should be minor and major mergers, with the far-UV light
dominated by a bulgelike burst component, which seems qualitatively
consistent with the observational results.

%%%%%%%%%%%%%%%%%%%%%%%%%%%%%%%%%%%%%%%%%%%%%%%%%%%%%%%%%%%%%%%%%%%%%%%%%%%%%%%%%%
% Fig.
% more physical properties of LBGs
\begin{figure*}

\begin{center}

\begin{minipage}{7cm}
\includegraphics[width=7cm]{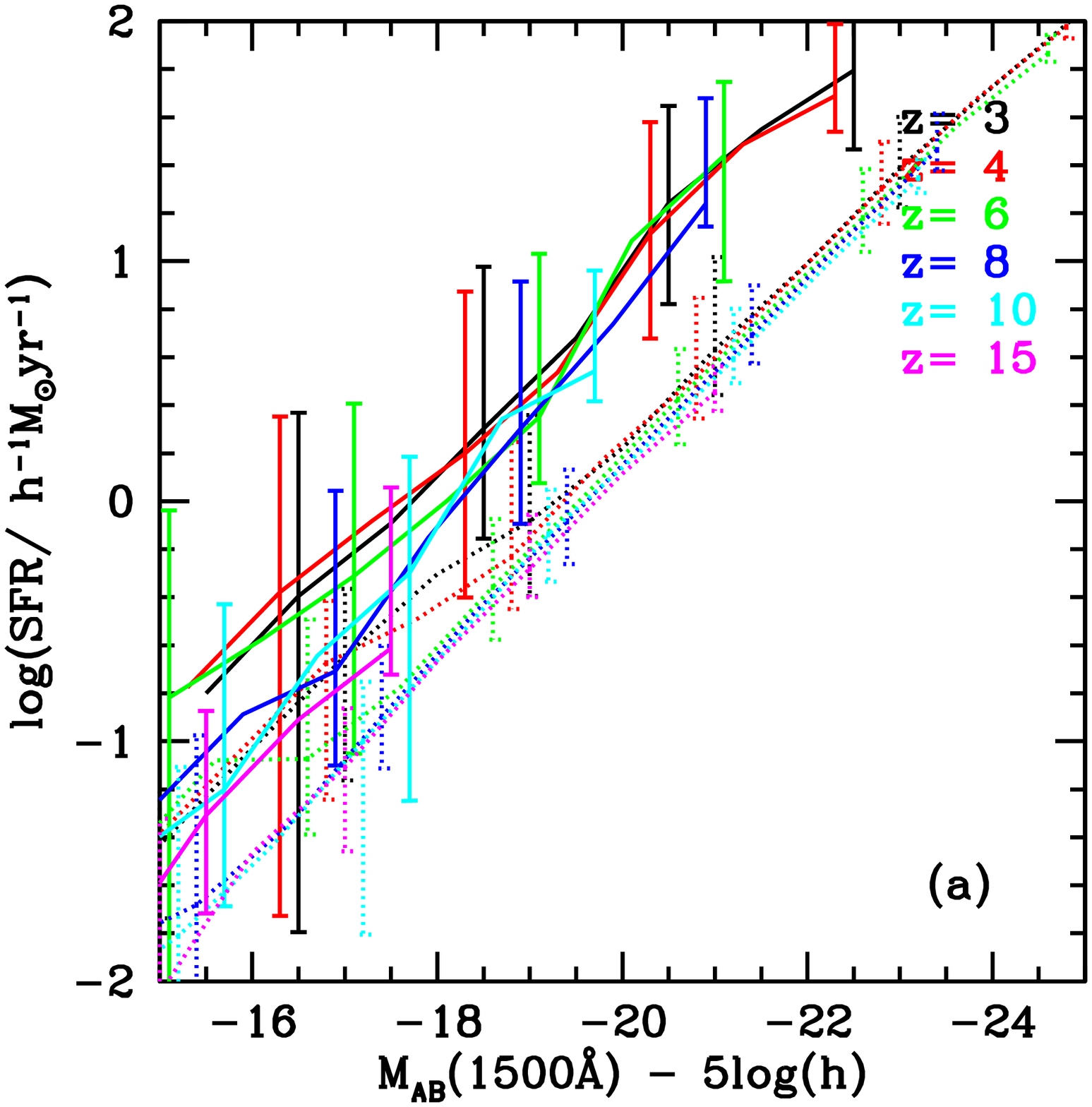}
\end{minipage}
\begin{minipage}{7cm}
\includegraphics[width=7cm]{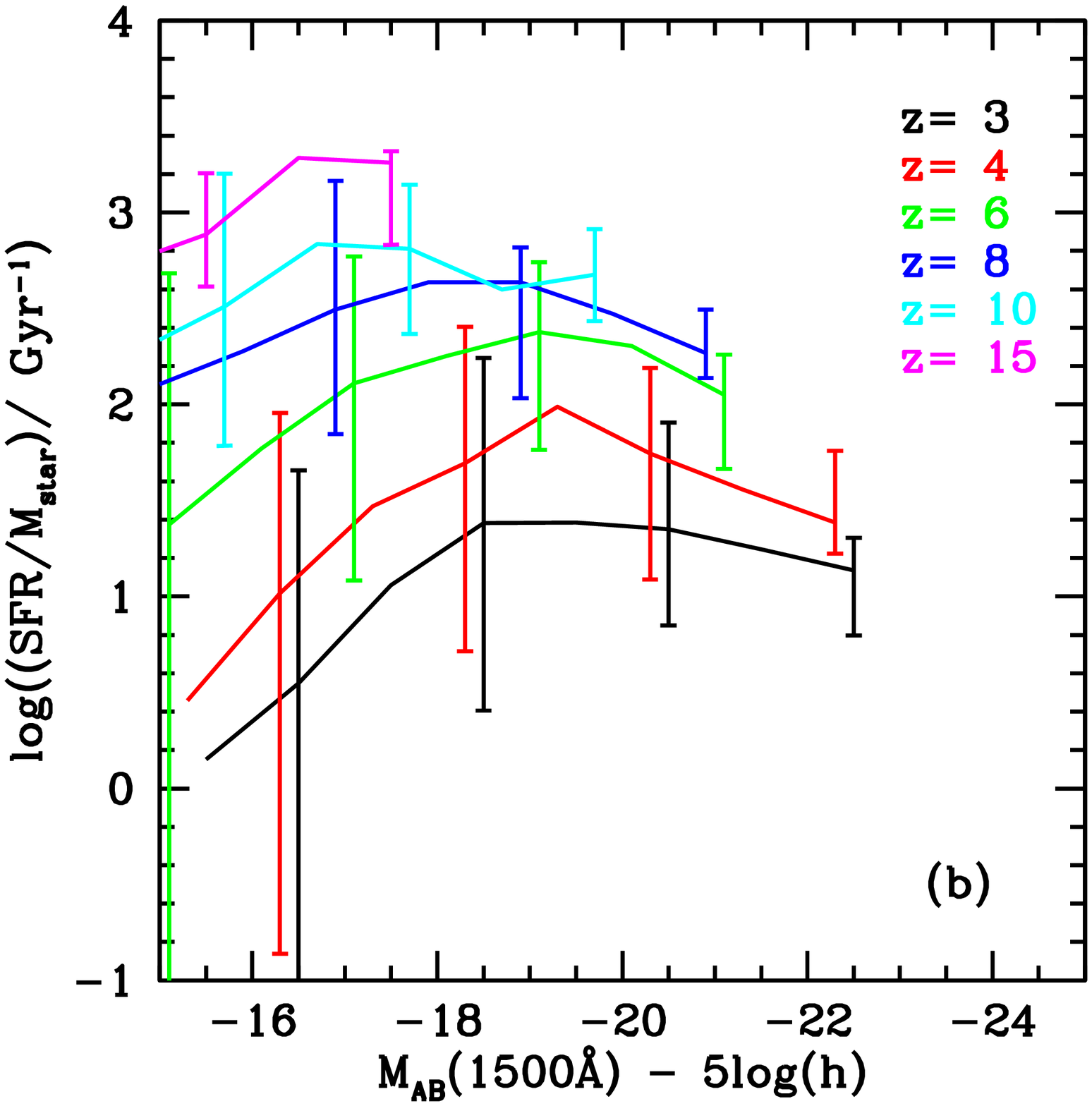}
\end{minipage}

\begin{minipage}{7cm}
\includegraphics[width=7cm]{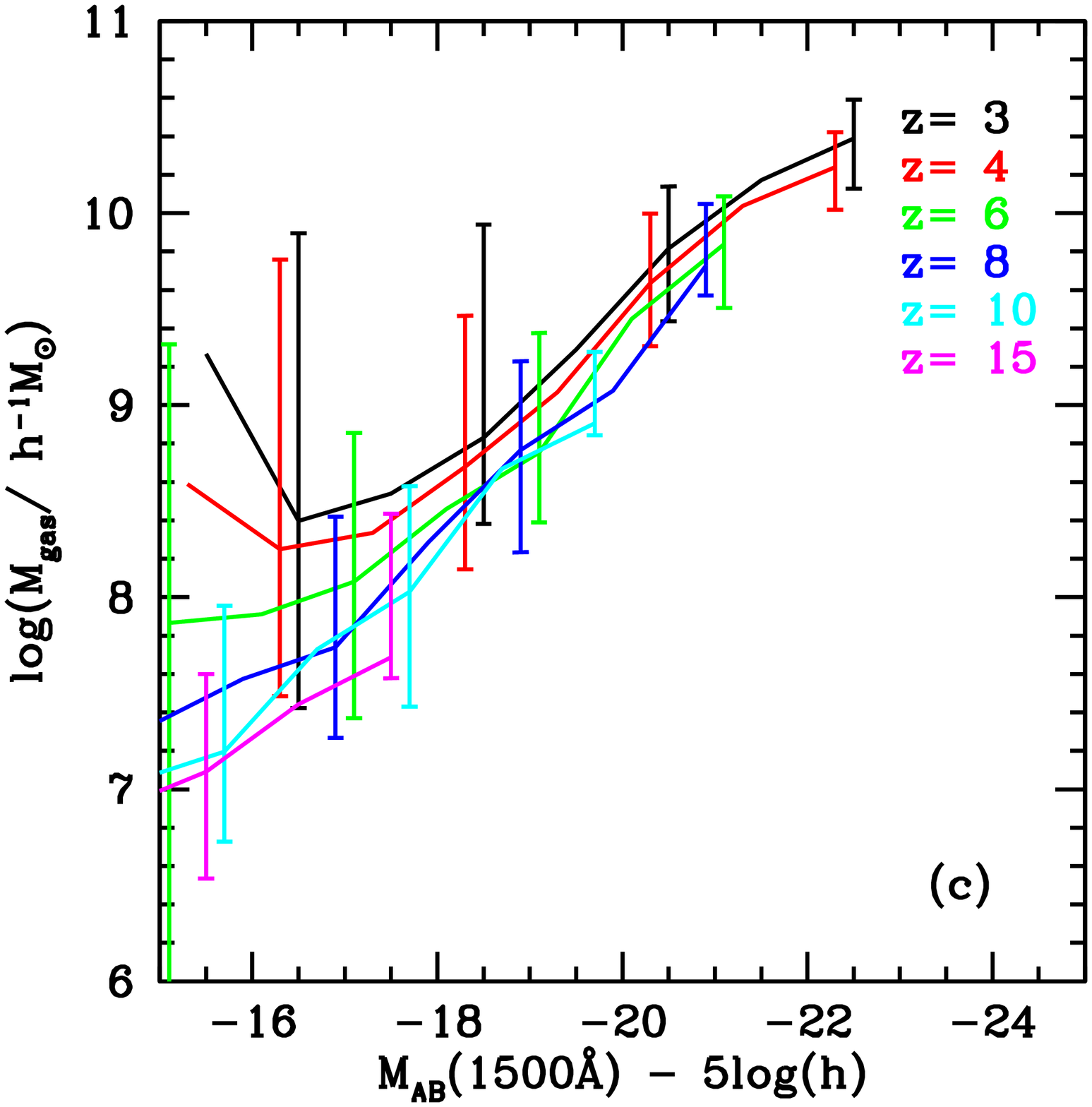}
\end{minipage}
\begin{minipage}{7cm}
\includegraphics[width=7cm]{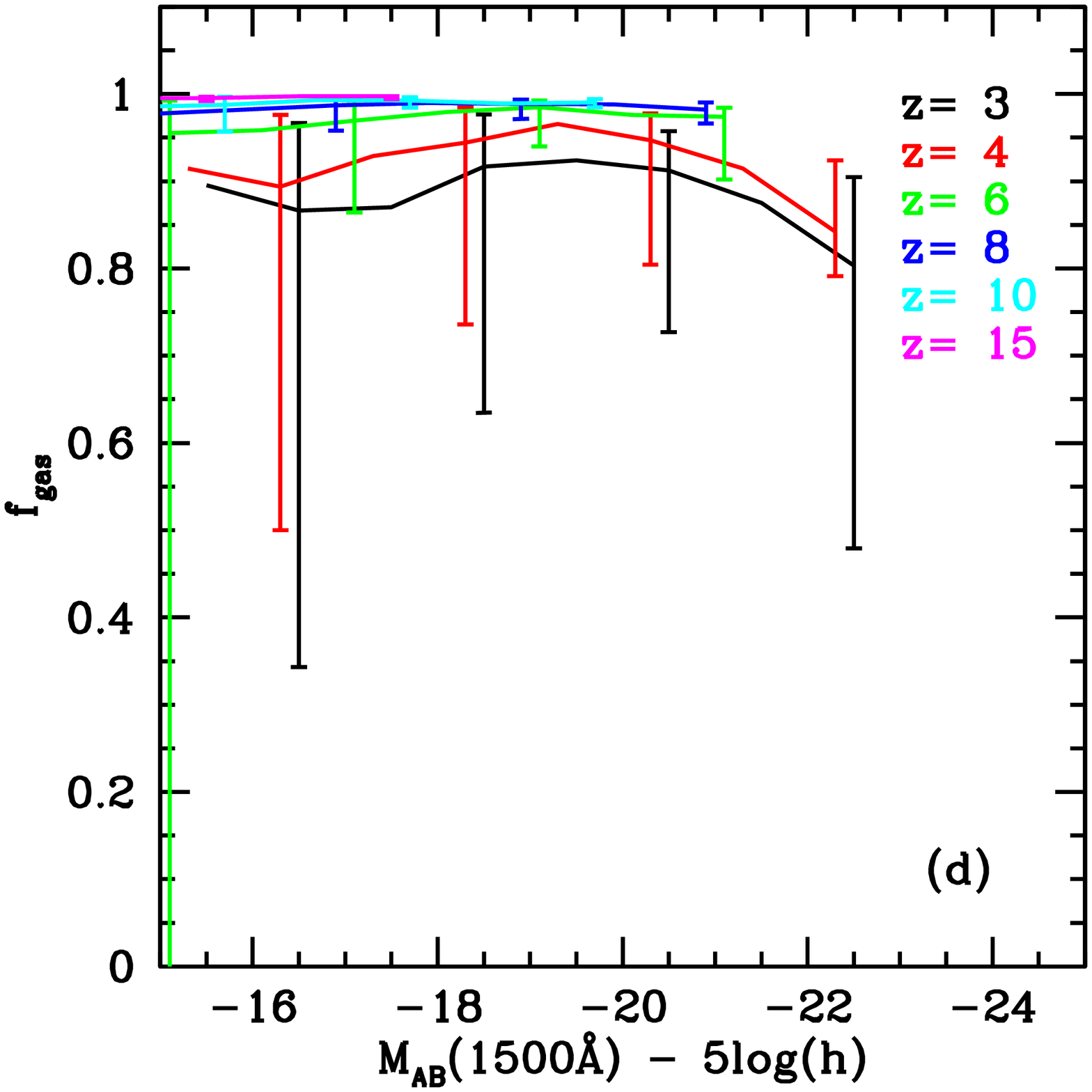}
\end{minipage}

\begin{minipage}{7cm}
\includegraphics[width=7cm]{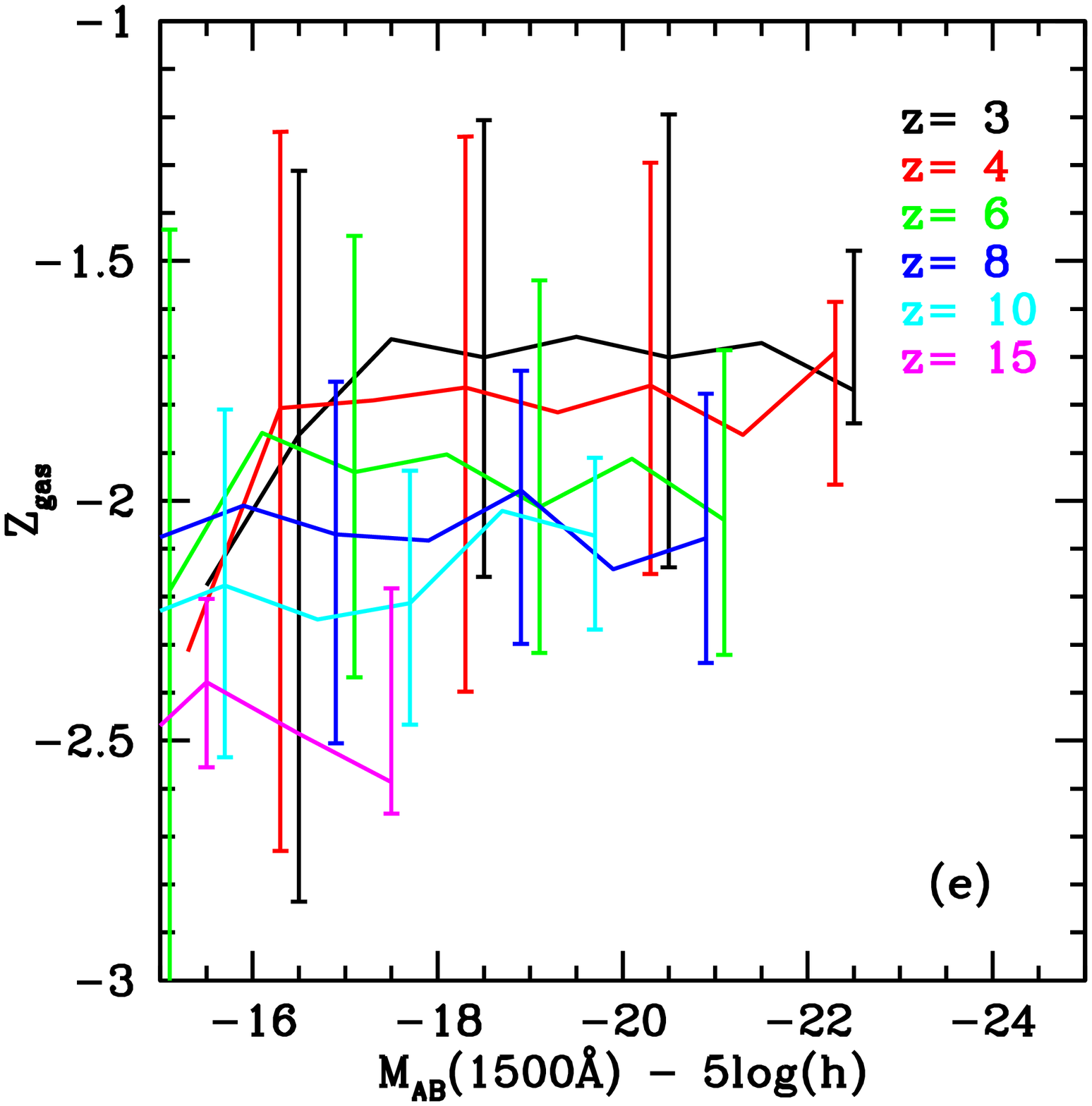}
\end{minipage}
\begin{minipage}{7cm}
\includegraphics[width=7cm]{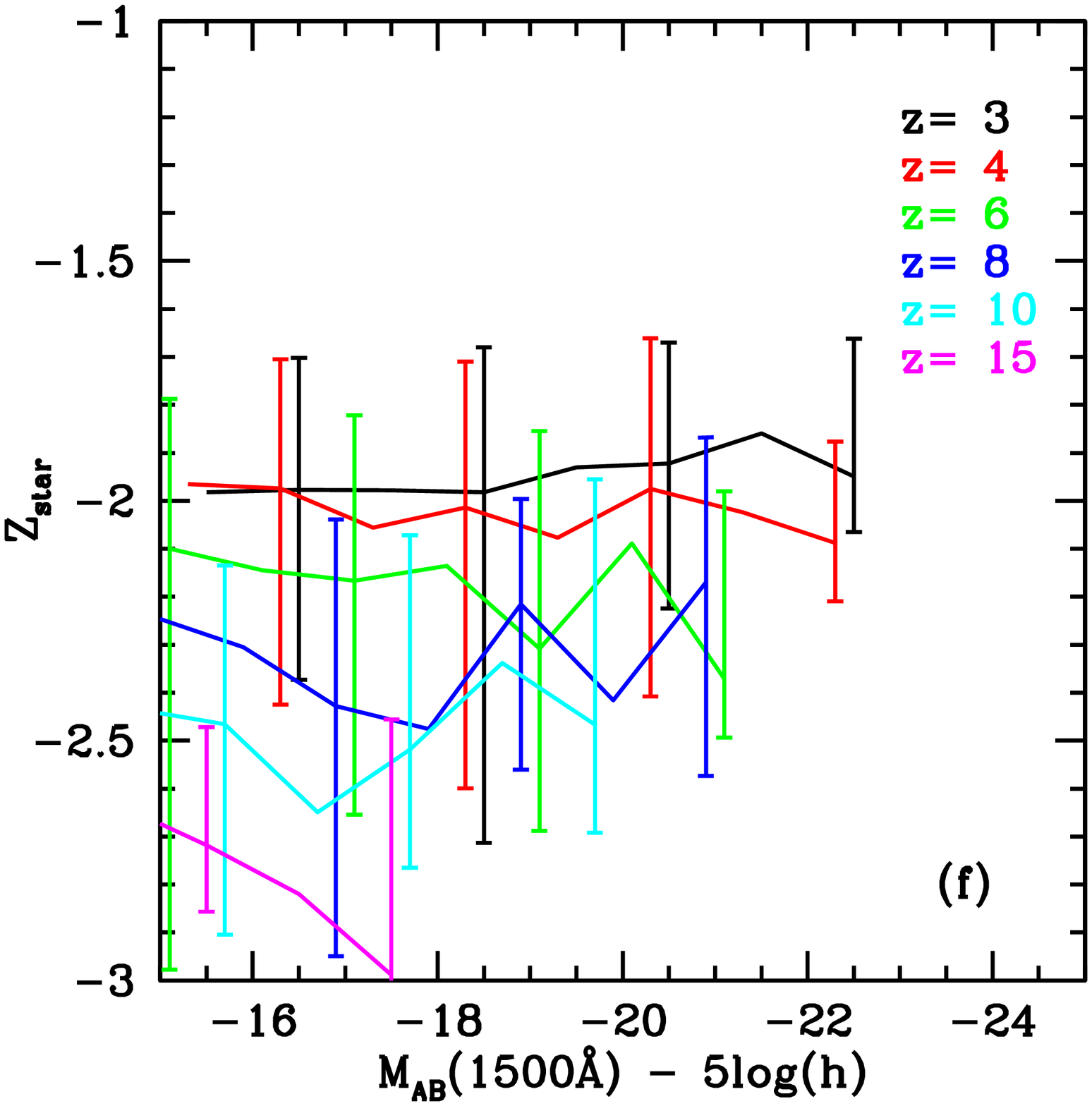}
\end{minipage}

\end{center}

\caption{Further predicted physical properties of LBGs as functions of
  dust-extincted rest-frame 1500\AA\ absolute magnitude in the \Bau\
  model. The lines show medians and the error bars indicate the
  10-90\% range. The different panels are as follows: (a) SFR -- the
  solid and dotted lines show the relations with and without dust
  extinction in the UV luminosity; (b) specific SFR; (c) cold gas
  mass; (d) cold gas fraction; (e) gas metallicity; (f) stellar
  metallicity.  }

\label{fig:propsB}
\end{figure*}

%%%%%%%%%%%%%%%%%%%%%%%%%%%%%%%%%%%%%%%%%%%%%%%%%%%%%%%%%%%%%%%%%%%%%%%%%%%

\subsection{Star formation rates and specific SFRs}
We show the SFRs of LBGs in the top left panel of
Fig.~\ref{fig:propsB}. The solid lines show SFR plotted against
dust-extincted far-UV luminosity, and the dotted lines against the
unextincted luminosity. There is an almost linear relation between the
SFR and the far-UV luminosity, with the lines for different redshifts
lying almost on top of each other. A constant linear relation between
rest-frame 1500\AA\ luminosity and SFR would be expected under the
following conditions: (a) a single IMF dominates; (b) dust extinction
is zero or constant; (c) SFR varies on timescales $\gsim 10^8 \yr$. In
the \Bau\ model, stars form with a top-heavy ($x=0$) IMF in bursts and
a Kennicutt IMF in quiescent disks. For a constant SFR and Solar
metallicity, the ratio $L_{\nu}(1500\AA)/SFR$ is 3.4 times larger for
the $x=0$ IMF than for the Kennicutt IMF (and 2.6 times larger for 0.2
times Solar metallicity). The LBG LF changes from being dominated by
bursts at high luminosities to being dominated by quiescent galaxies
at low luminosities, and this causes a corresponding shift in the
unextincted SFR vs $L_{\nu}(1500\AA)$ relation, which can be seen for
the lower redshifts in the plot. The effects of changes in metallicity
with redshift on the unextincted relation seem to be small. Comparing
the solid and dotted lines, dust extinction is seen to introduce more
scatter into the SFR vs $L_{\nu}(1500\AA)$ relation, but the average
effect of the extinction depends only modestly on luminosity and
redshift. Observationally, SFRs of LBGs are generally inferred
directly from their rest-frame far-UV luminosities, with or without a
correction for dust extinction, so such estimates do not provide an
independent test of the relation plotted here. Observational studies
typically assume a Salpeter IMF over the mass range $0.1-100 \Msol$
for converting luminosities to SFRs. This Salpeter IMF would require
an SFR 4.6 times larger than  our top-heavy burst IMF to produce
the same unextincted far-UV luminosity (for a constant SFR and Solar
metallicity).

The top right panel of Fig.~\ref{fig:propsB} shows the specific star
formation rate (SSFR), defined here as the ratio of the current SFR to
the stellar mass. There is a strong increase of the SSFR with
redshift, and a weaker trend with luminosity. LBGs at higher redshifts
are thus forming stars at a much larger fractional rate than lower
redshift LBGs, by factors up to $\sim 10^3$. The age, $t(z)$, of the
universe is, of course, much less at higher redshift (shrinking from
2.1~Gyr at $z=3$ to 0.26~Gyr at $z=15$), so it might be more
physically meaningful to plot $t(z) \times SFR/M_{star}$. This would
still increase by a factor $\sim 10^2$ from $z=3$ to $z=15$. Finally,
we should account for the fact that the rate of buildup of mass in
long-lived stars is actually $(1-R) \times SFR$, where $R$ is the
fraction of the initial stellar mass returned to the ISM by mass loss
from dying stars. $R$ has the value 0.41 for the Kennicutt IMF but
0.91 for the top-heavy IMF. So the current rate of buildup of stellar
mass compared to the past average rate is
$(1-R)t(z)SFR/M_{star}$. Allowing for the shift between burst and
quiescent domination with changing luminosity, this latter quantity
still increases by a factor $\sim 10$ over the range $z=3$ to $z=15$,
with the high-$z$ LBGs having $(1-R)t(z)SFR/M_{star} \sim 10$. Thus,
by any reasonable measure, the highest redshift LBGs are predicted to
be assembling stars extremely rapidly compared to their past average
rate.

\subsection{Cold gas masses and gas fractions}
The middle left panel of Fig.~\ref{fig:propsB} shows the cold gas
masses of LBGs, where by ``cold'' gas we mean all of the gas which has
condensed into the galaxy, as distinct from the ``hot'' gas which
remains distributed in the halo. Most of the cold gas will be in
either atomic or molecular form. The brightest LBGs are predicted to
have cold gas masses $\sim 10^{10} \hMsol$.  At higher luminosities
and higher redshifts, there is a very nearly linear relation between
the cold gas mass and the dust-extincted far-UV luminosity. This is a
consequence of two effects: the nearly linear relation between SFR and
dust-extincted luminosity already noted above, and the fact that the
more luminous LBGs are bursts, for which $SFR = M_{\rm
gas}/\tau_{\ast}$, with most of the bursts having very similar star
formation timescales, $\tau_{\ast}$, according to our model. The
linear relation between $M_{\rm gas}$ and $L_{\nu}(1500\AA)$ breaks
down at low luminosities where quiescent galaxies become
important. These have longer SFR timescales than bursts, and so must
have more gas to produce the same far-UV luminosity from young stars.

The middle right panel of Fig.~\ref{fig:propsB} shows the gas
fractions in LBGs, where we define this fraction as the ratio of cold
gas mass to total cold gas + stellar mass in a galaxy. Although the
gas fraction shows a very large scatter at lower luminosities, the
median value is predicted to be very high $\sim 90-99\%$. This results
from two effects in the model: the SFR timescale in disks is large
compared to the age of the universe at high-$z$, so that these disks
are gas-rich, and most of the bursts are triggered by minor mergers,
for which we require the gas fraction in the disk of the primary
galaxy to exceed 75\% for a burst to be triggered.

Molecular gas has been observed through its CO emission in two
gravitationally lensed LBGs at $z\sim 3$, both of which have
rest-frame luminosities $\MUV \sim -20$ \citep{Baker04,Coppin07}. The
two galaxies have CO luminosities which differ by a factor 7. The
conversion from CO luminosity to molecular gas mass is significantly
uncertain. \citeauthor{Coppin07} use a conversion factor estimated for
local starburst galaxies, and find molecular gas masses of 0.34 and
2.5$\times 10^9\Msol$ for the two galaxies. Using instead the
conversion factor estimated for the Milky Way would increase these
estimated gas masses by a factor $\sim 6$. Our model predicts a median
cold gas mass (including both molecular and atomic gas) $\sim 4\times
10^9 \Msol$ at the same luminosity and redshift. This seems compatible
with the current observations, given their uncertainties.

\subsection{Metallicities of gas and stars}
Finally, the bottom left and right panels of Fig.~\ref{fig:propsB}
show the metallicities of cold gas and stars respectively (the latter
being a mass-weighted mean value over stars of all ages). We see that,
at a given redshift and luminosity, the metallicity of the gas is
generally somewhat higher than that of the stars for LBGs that are
dominated by bursts. This reflects the buildup of gas metallicity by
self-enrichment in a current burst, while the stellar metallicity is
an average over past activity. For both gas and stars, the
metallicities typically depend only weakly on luminosity, and more
strongly on redshift, again for the LBGs which are dominated by
bursts. However, even at very high redshifts, the metallicities of
LBGs are predicted to be non-negligible (e.g. $Z_{\rm gas} \sim 0.003$
at $z=15$), due to self-enrichment by bursts which have large heavy
element yields due to the top-heavy IMF.

The gas metallicities of $z\sim 3$ LBGs have been estimated by
\citet{Pettini01} from rest-frame optical HII region emission lines,
for a sample of 4 LBGs with $\MUV \sim -21$. They find significant
uncertainties in the values for individual galaxies, but they are all
constrained to be in the range $\sim 0.1-1\Zsol$ (where the solar
metallicity $\Zsol=0.02$), with a preference for somewhat subsolar
values. Our model predicts roughly solar metallicities for the gas at
this luminosity and redshift, compatible with the upper end of the
observed range. The stellar metallicities of LBGs are only very weakly
constrained by SED-fitting to observed broadband fluxes, due to
degeneracies with other parameters \citep{Papovich01}, but are
compatible with our model predictions.

%\clearpage

%%%%%%%%%%%%%%%%%%%%%%%%%%%%%%%%%%%%%%%%%%%%%%%%%%%%%%%%%%%%%%%%%%%%%%%%%%%%%%%%%%
% Fig.
% LF evoln at high-$z$
\begin{figure}

\begin{center}
\includegraphics[width=7cm]{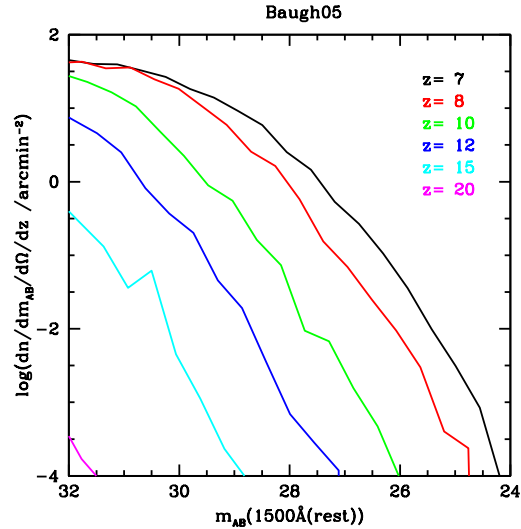}
\end{center}

\caption{Predicted evolution of the LBG luminosity function at
$z=7-20$ in the \Bau\ model, shown as a surface density of objects
(number per solid angle per unit redshift) {\em vs} dust-extincted
apparent magnitude at a fixed rest-frame wavelength of 1500\AA.}

\label{fig:ncts}
\end{figure}

%%%%%%%%%%%%%%%%%%%%%%%%%%%%%%%%%%%%%%%%%%%%%%%%%%%%%%%%%%%%%%%%%%%%%%%%%%%

%%%%%%%%%%%%%%%%%%%%%%%%%%%%%%%%%%%%%%%%%%%%%%%%%%%%%%%%%%%%%%%%%%%%%%%%%%%%%%%%%%
% Fig.
% selected properties vs apparent magnitude

\begin{figure}

\begin{center}

\begin{minipage}{7cm}
\includegraphics[width=7cm]{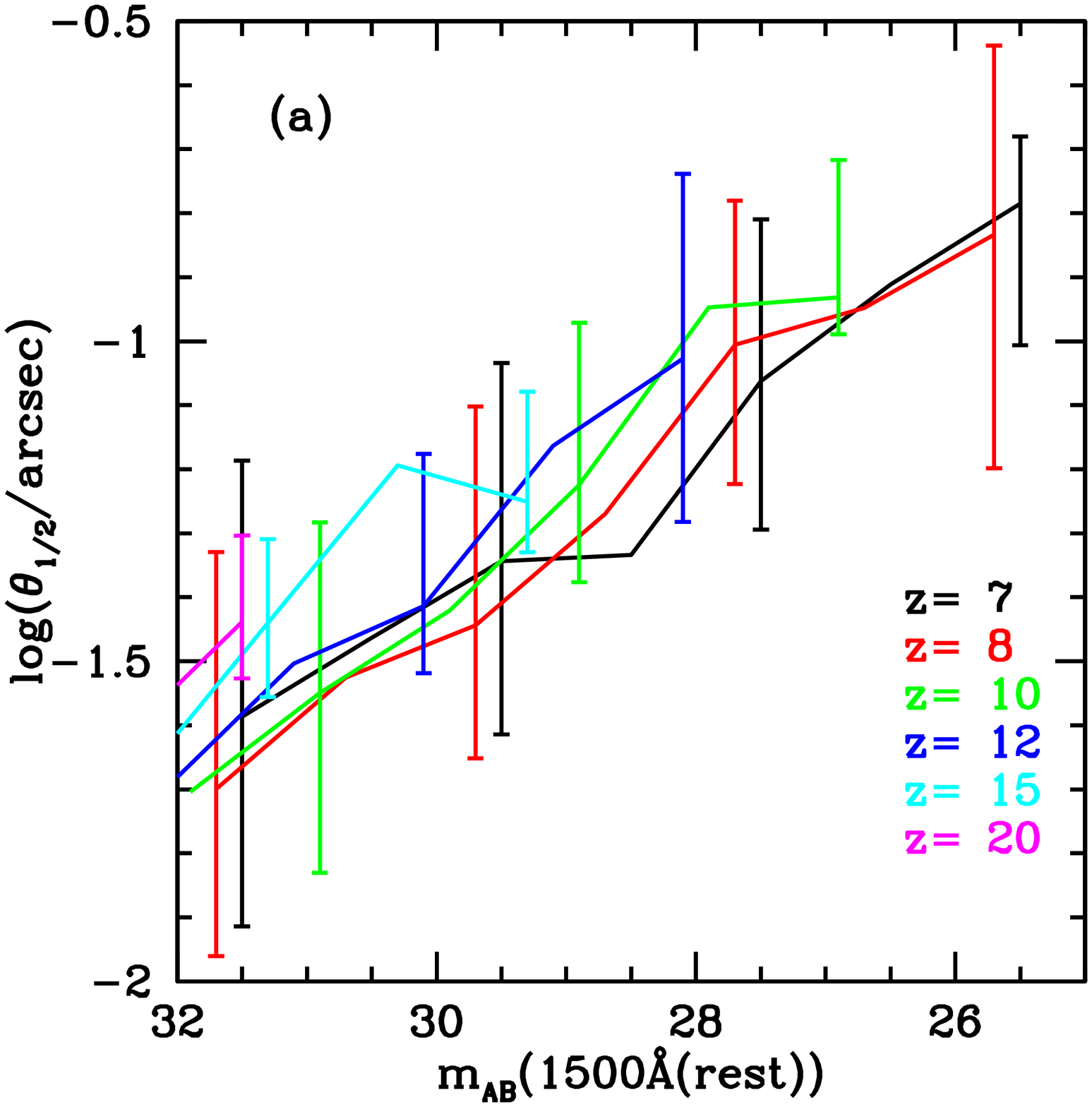}
\end{minipage}

\begin{minipage}{7cm}
\includegraphics[width=7cm]{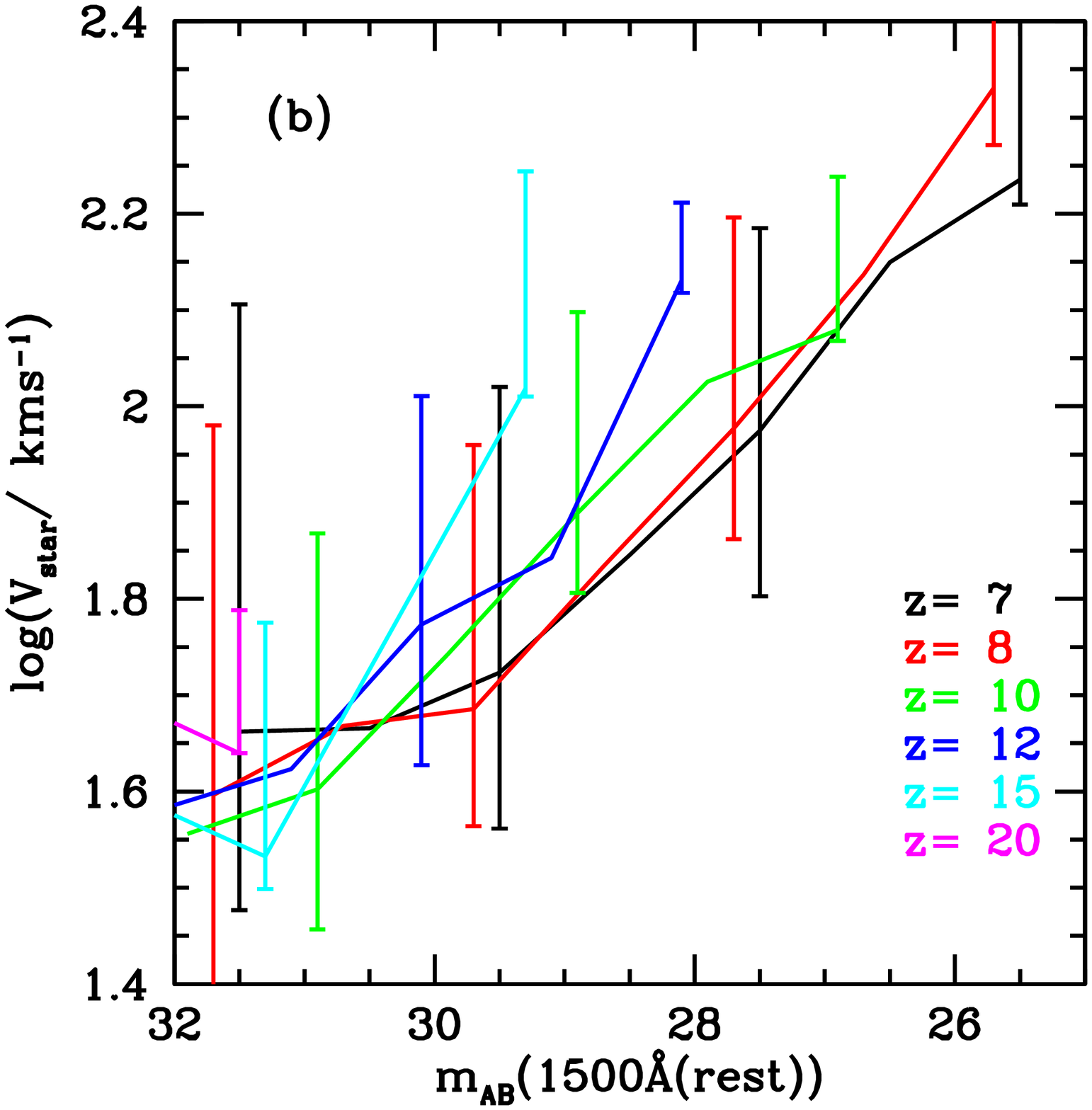}
\end{minipage}

\begin{minipage}{7cm}
\includegraphics[width=7cm]{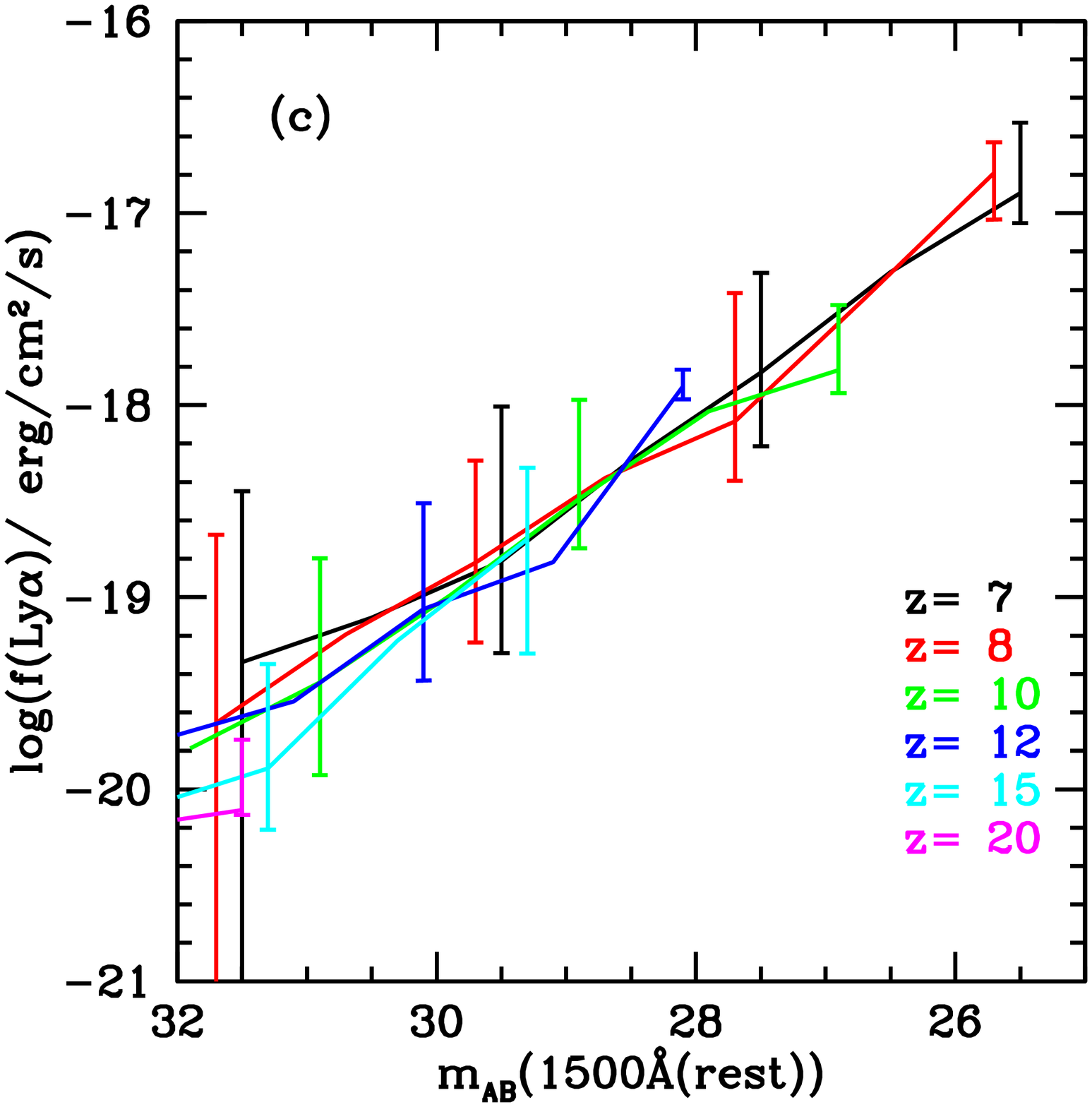}
\end{minipage}

\end{center}

\caption{Predicted observable properties of LBGs at $z\geq 7$ in the
  \Bau\ model as functions of dust-extincted apparent magnitude at a
  fixed rest-frame wavelength of 1500\AA. The lines show median values
  and the error bars show the 10-90\% ranges. (a) Angular half-light
  radius at rest-frame 1500\AA. (b) Circular velocity at stellar
  half-mass radius. (c) $Ly\alpha$ flux (assuming a $Ly\alpha$ escape
  fraction $f_{\rm esc}=0.02$). }

\label{fig:props_appmag}
\end{figure}

%%%%%%%%%%%%%%%%%%%%%%%%%%%%%%%%%%%%%%%%%%%%%%%%%%%%%%%%%%%%%%%%%%%%%%%%%%%

\section{Predictions for LBGs at very high redshifts}
\label{sec:high-z}

In this section, we present additional predictions for LBGs at very
high redshifts, $z > 7$, which are starting to be probed with HST, and
which in future will be probed to higher redshifts and fainter fluxes
by \JWST, as well as by future 20-40m ELTs on the ground, including
the {\em E-ELT}, {\em TMT} and {\em GMT}. NIRCam on \JWST\ will do
imaging over the wavelength range 0.6-5$\mum$, and so can in principle
detect LBGs over the whole redshift range $z=7-20$ and beyond, if they
are bright enough and numerous enough. Among the surveys planned with
NIRCam are the Deep-Wide Survey (DWS), to reach $m_{\rm AB}=30$ over
an area of 100~arcmin$^2$, and the Ultra-Deep Imaging Survey (UDS), to
reach $m_{\rm AB}=31$ over 10~arcmin$^2$ \citep{Gardner06}. 

We start by presenting in Fig.~\ref{fig:ncts} predictions from the
\Bau\ model for the number of $z>7$ LBGs, this time shown as the
surface density of objects (number per unit redshift per square
arcmin) {\em vs} apparent magnitude. {For simplicity, we
present results in this section in terms of the apparent magnitude for
a fixed rest-frame wavelength of 1500\AA, regardless of redshift. In
reality, future LBG surveys will use a variety of filters, depending
both on the telescope and on the redshift being targetted. However, in
order to detect the Lyman-break feature reliably, such surveys will
always include at least one filter probing wavelengths close to
1500\AA\ in the rest frame. Predictions for the number of LBGs in any
future survey can therefore be approximately read off from
Fig.~\ref{fig:ncts}, using the apparent magnitude limit for the filter
which is closest to rest-frame 1500\AA\ at the target redshift.}  We
see from Fig.~\ref{fig:ncts} that the planned DWS and UDS surveys on
\JWST\ should each detect a few LBGs at $z=15$, and many more at lower
redshifts. {For convenience in what follows, we note that for
the cosmology assumed in the \Bau\ model, the relationship between
apparent and absolute magnitudes at the same rest-frame wavelength is
$m_{\rm AB}-(M_{\rm AB}-5\log h) =
(46.16,46.37,46.71,46.97,47.29,47.69)$ at the redshifts
$z=(7,8,10,12,15,20)$ plotted in Figs.~\ref{fig:ncts} and
\ref{fig:props_appmag}. An apparent magnitude $m_{\rm AB}=31$ at
$z=15$ thus corresponds to an absolute magnitude $\MUV=-16.3$.}

We next present in Fig.~\ref{fig:props_appmag} a few directly
observable properties of high-$z$ LBGs as functions of rest-frame
1500\AA\ apparent magnitude. These particular properties are chosen
because they are directly relevant to the detectability of these LBGs
by different instruments on the \JWST\ and on future ELTs. The top
panel of Fig.~\ref{fig:props_appmag} shows the angular half-light
radius at a rest-frame wavelength of 1500\AA. This shows that at a
given apparent magnitude, LBGs have slightly larger angular sizes at
higher redshifts. This reversal of the trend seen for proper size vs
absolute magnitude shown in Fig.~\ref{fig:rstar} is due to the
combined effects of angular diameter and luminosity distances. LBGs at
$m_{AB}=30$ and $z=15$ are predicted to have angular radii $\sim
0.05$~arcsec, close to the diffraction limit of \JWST\ at the relevant
wavelength. The middle panel shows the predicted circular velocity at
the stellar half-mass radius. This affects the widths of emission and
absorption features in a galaxy spectrum, and so is relevant for
spectroscopic studies. The dependence of circular velocity on apparent
magnitude and redshift is again modest, varying from $\sim 40\kms$ to
$\sim 150\kms$ over the whole range plotted. At a fixed apparent
magnitude, LBGs have larger circular velocities at high redshift.

Finally, in the bottom panel we show the predicted $Ly\alpha$ line
flux. {This depends on the fraction of $Ly\alpha$ photons which
escape from galaxies, which is very difficult to calculate from first
principles due to the effects of resonant scattering of $Ly\alpha$ by
atomic hydrogen. We therefore follow the same approach as in our
previous work on $Ly\alpha$ emitters using \GALFORM, and assume a
constant escape fraction, tuned to match observations at lower
redshifts. \citet{LeDelliou06} and \citet{Orsi08} showed that the
observed numbers and properties of $Ly\alpha$ emitters at $z=3-6$
could be reproduced well in the \Bau\ model by assuming a constant
escape fraction of 2\%, so we use the same value here.} We see that
LBGs at different redshifts all fall on the same linear relation
between $Ly\alpha$ flux and rest-frame far-UV flux. LBGs with $m_{\rm
AB}=30$ are predicted to have $Ly\alpha$ fluxes $\sim 10^{-19}
\ergcms$. This is probably too faint to be detectable with \JWST\, but
should be within the reach of the {\em E-ELT}.

%%%%%%%%%%%%%%%%%%%%%%%%%%%%%%%%%%%%%%%%%%%%%%%%%%%%%%%%%%%%%%%%%%%%%%%%%%%%%%%%%%
% Fig.
% SFR density vs z
\begin{figure}

\includegraphics[width=7cm]{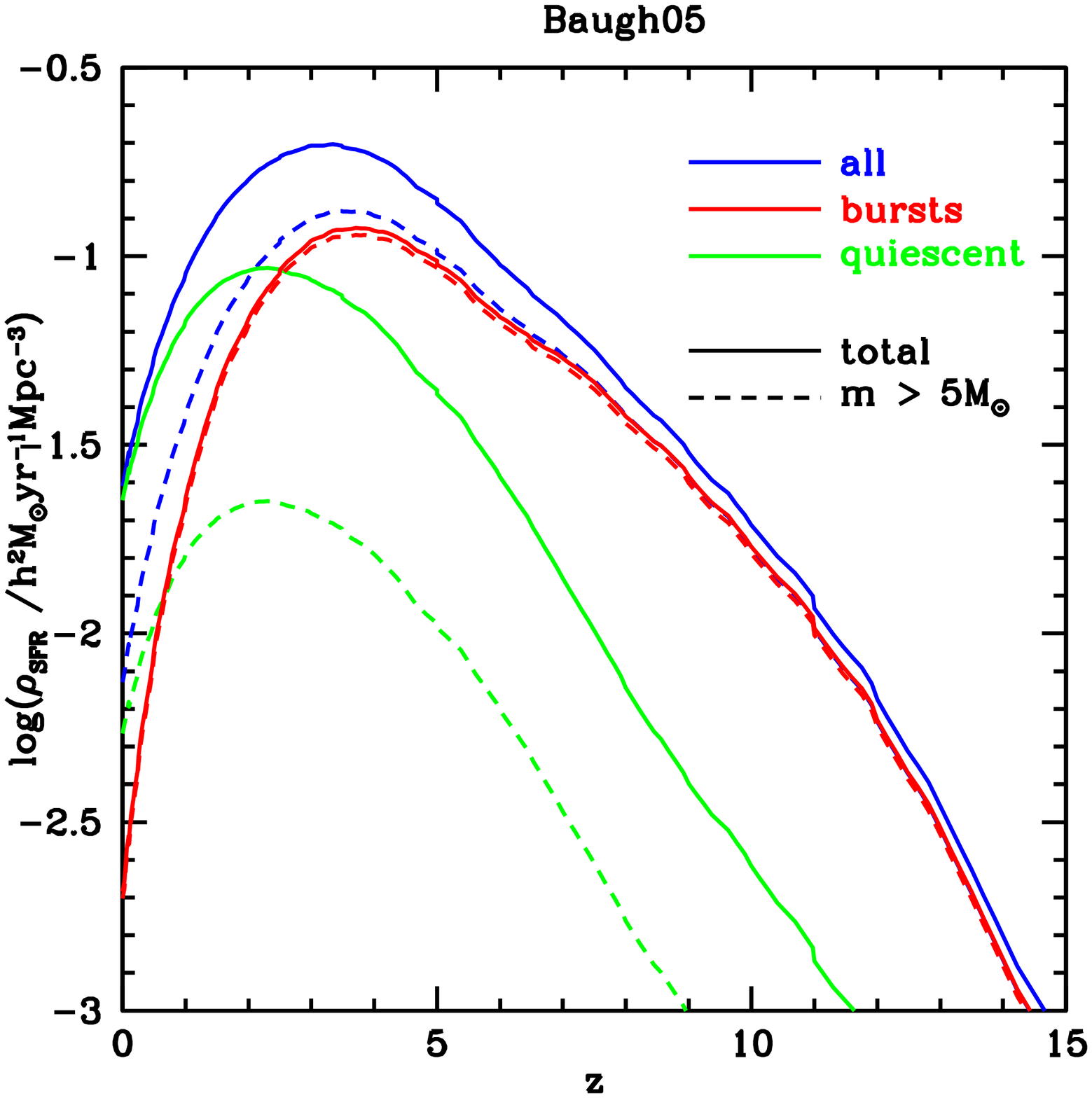}

\includegraphics[width=7cm]{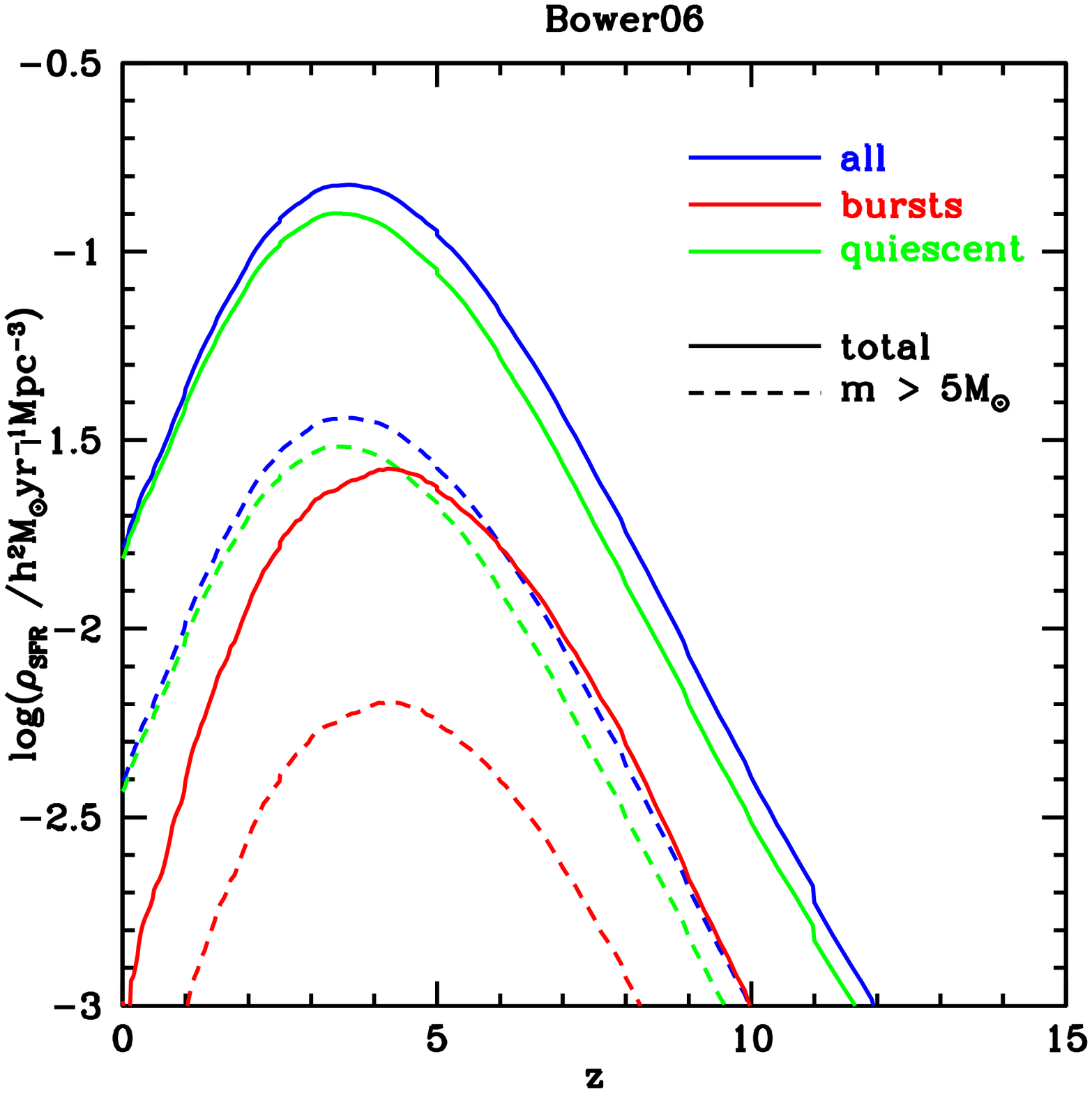}

\caption{Comoving SFR density {\em vs} redshift in the \Bau\
{and \Bow\ models (top and bottom panels). In each panel,} the
blue line shows the total, while the green and red lines show the
separate contributions of disks and bursts. The solid lines show total
SFRs integrated over all stellar masses, while the dashed lines show
the SFRs in high-mass ($m > 5\Msol$) stars only. }

\label{fig:rhoSFR}
\end{figure}

%%%%%%%%%%%%%%%%%%%%%%%%%%%%%%%%%%%%%%%%%%%%%%%%%%%%%%%%%%%%%%%%%%%%%%%%%%%

\section{Evolution of SFR and UV luminosity densities}
\label{sec:UVdens}

The final topic we consider is the cosmic star formation history and
how this is traced by the far-UV luminosity density.  We show in
Fig.~\ref{fig:rhoSFR} the comoving SFR density as a function of
redshift {for the \Bau\ and \Bow\ models (top and bottom
panels)}. The solid blue curves show the total SFR in all galaxies,
while the solid green and red curves show the separate contributions
from star formation in disks and in bursts.  

{In the \Bau\ model}, the total SFR density peaks at $z \approx
3$, with quiescent galaxies dominating at $z < 2.5$ and bursts
dominating above this. Similar results were shown in \citet{Baugh05}
and \citet{Lacey10}, but there are small differences of detail because
of the modified values for the photoionization feedback parameters,
$\zreion$ and $\Vcrit$, that we have used in this paper. As discussed
in \citet{Lacey10}, we also find it useful to show the SFR density in
massive stars only (dashed lines), which we define as stars with
masses $m>5\Msol$.  These stars, which have lifetimes $< 1\times
10^8\yr$, dominate the production of UV radiation.  The two IMFs in
our model (assumed to cover the stellar mass range $0.15<m<120\Msol$)
have very different fractions of their initial stellar mass in high
mass stars: $f(m>5\Msol)=0.24$ for the Kennicutt IMF assumed for
quiescent star formation, and $f(m>5\Msol)=0.96$ for the top-heavy IMF
assumed for bursts. The SFR density for massive stars evolves more
strongly than that for all stars, increasing by a factor $\approx 20$
from $z=0$ to its peak at $z \approx 3$, and then declining by a
factor $\approx 100$ to $z=15$.

{In the \Bow\ model, the total SFR density peaks at almost the
same redshift as in the \Bau\ model, though at a slightly lower
value. However, unlike in the \Bau\ model, the SFR density is
dominated by quiescent star formation at all redshifts, and it also
drops off more steeply with redshift beyond the peak.  The SFR density
for massive ($m>5\Msol$) stars in the \Bow\ model is roughly a factor
3 lower than in the \Bau\ model for $z<5$, but this difference
increases at higher redshifts, reaching a factor $\sim 20$ at $z=10$.}

%%%%%%%%%%%%%%%%%%%%%%%%%%%%%%%%%%%%%%%%%%%%%%%%%%%%%%%%%%%%%%%%%%%%%%%%%%%%%%%%%%
% Fig.
% UV luminosity density vs z
\begin{figure*}

\begin{minipage}{7cm}
\includegraphics[width=7cm]{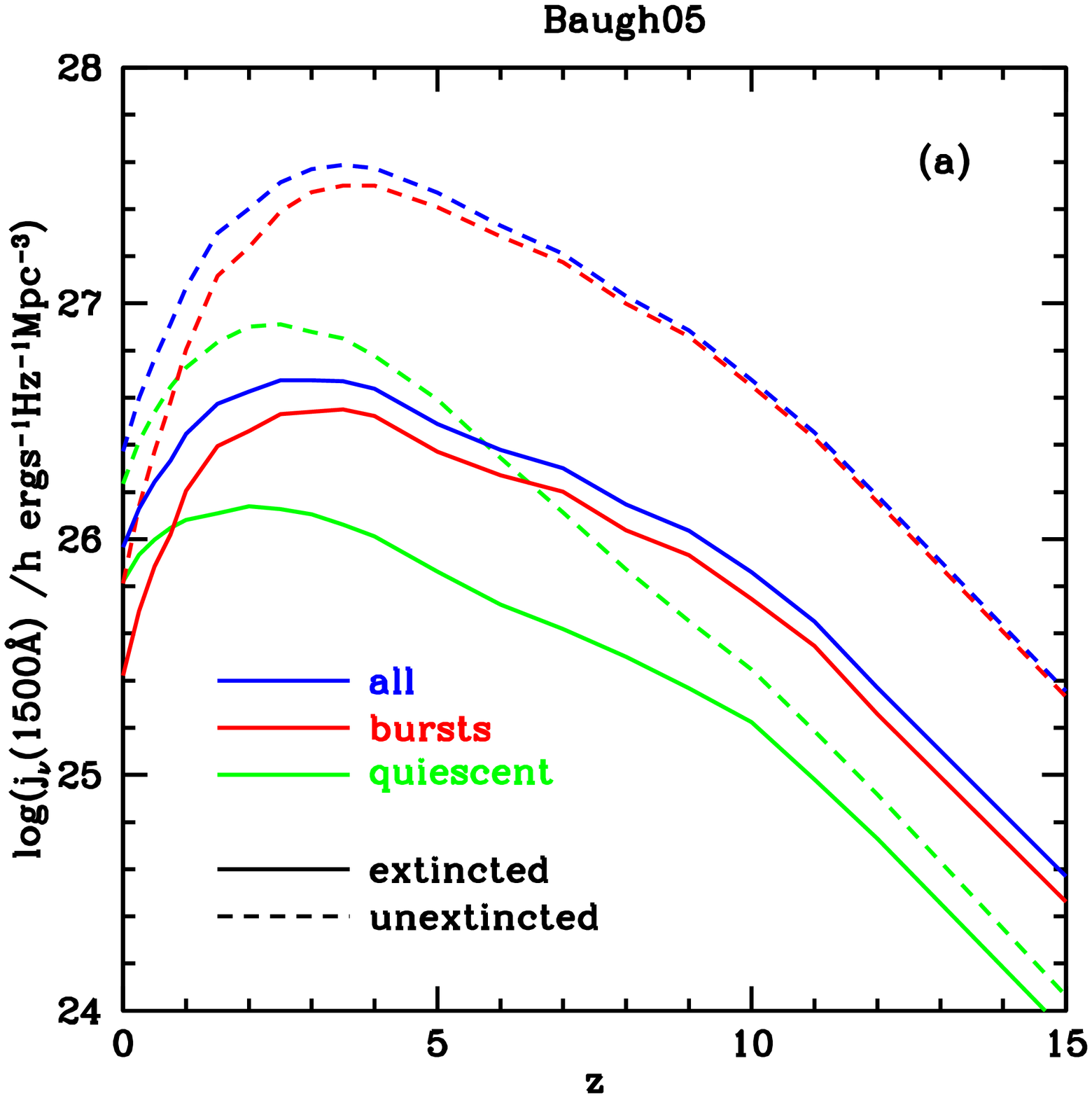}
\end{minipage}
\hspace{1cm}
\begin{minipage}{7cm}
\includegraphics[width=7cm]{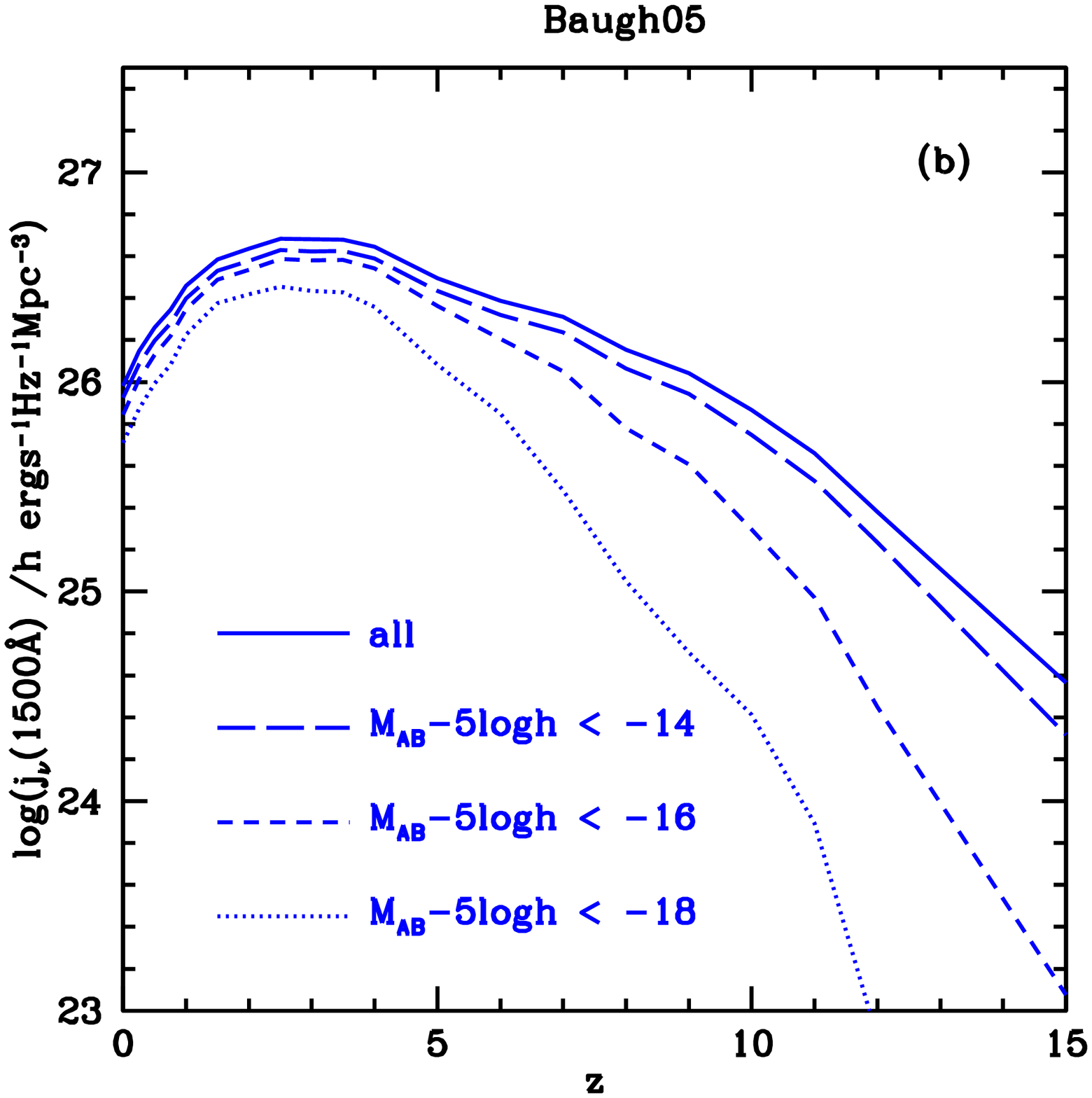}
\end{minipage}

\caption{Rest-frame 1500\AA\ luminosity density (in comoving units)
  {\em vs} redshift in the \Bau\ model. (a) The blue lines show the
  far-UV luminosity density due to all galaxies, while the green and
  red curves show the separate contributions of quiescent galaxies and
  ongoing bursts. Solid and dashed curves show luminosity densities
  with and without dust extinction. (b) The contribution to the
  dust-extincted far-UV luminosity density from galaxies with 1500\AA\
  luminosities brighter than different limits, as indicated in the
  key.  }

\label{fig:jnu1500}
\end{figure*}

%%%%%%%%%%%%%%%%%%%%%%%%%%%%%%%%%%%%%%%%%%%%%%%%%%%%%%%%%%%%%%%%%%%%%%%%%%%

In Fig.~\ref{fig:jnu1500}, we show the predicted evolution of the
rest-frame 1500\AA\ emissivity (i.e. luminosity density), {for
the \Bau\ model only.} In the left panel, we show the luminosity
densities with and without dust extinction (solid and dashed lines),
and also show the separate contributions of quiescent and bursting
galaxies to the total. Dust extinction has a large effect on the
far-UV emissivity, with the mean extinction (defined from the ratio of
emissivities with and without dust) increasing from $\approx 1$~mag at
$z=0$ to $\approx 2$~mag at $z=3-15$. The mean extinction in the
bursts is larger than that in quiescent galaxies at most redshifts,
but at $z \gsim 1$, the bursts dominate the far-UV emissivity with or
without dust extinction. As expected, the unextincted far-UV
emissivity approximately tracks the SFR density in high-mass stars
shown in Fig.~\ref{fig:rhoSFR} (although changes in the metallicity
and in the mix of the IMFs with redshift mean that the scaling is not
exact). {For comparison, in the \Bow\ model, the unextincted
far-UV emissivity is below that for the \Bau\ model at all
redshifts. However, the mean dust extinction is also lower, peaking at
$\approx 1$~mag at $z\sim 3$, and falling to $\approx 0.3$~mag at
$z=10$. As a result, the extincted emissivity is similar to the \Bau\
model at $z\lsim 7$, but falls below it at higher redshifts.}

In the right panel {of Fig.~\ref{fig:jnu1500},} we show the
contributions to the dust-extincted 1500\AA\ emissivity from galaxies
with rest-frame 1500\AA\ absolute magnitudes brighter than $\MUV =
-14, -16$ or $-18$, {again for the \Bau\ model.} An LBG survey
reaching down to $\MUV < -18$ (the typical limit for current surveys
at $z \gsim 7$) will detect the galaxies responsible for 50\% of the
far-UV emissivity at $z=5$, but only 5\% at $z=10$. Detecting 80\% of
the far-UV emissivity at $z=10$ requires detecting galaxies down to
$\MUV < -14$, {corresponding to an apparent magnitude $m_{\rm
AB}\approx 33$ at the same rest-frame wavelength.}

%%%%%%%%%%%%%%%%%%%%%%%%%%%%%%%%%%%%%%%%%%%%%%%%%%%%%%%%%%%%%%%%%%%%%%%%%%%%%%%%%%
% Fig.
% ionizing photon emissivity vs z
\begin{figure*}

\begin{minipage}{7cm}
\includegraphics[width=7cm]{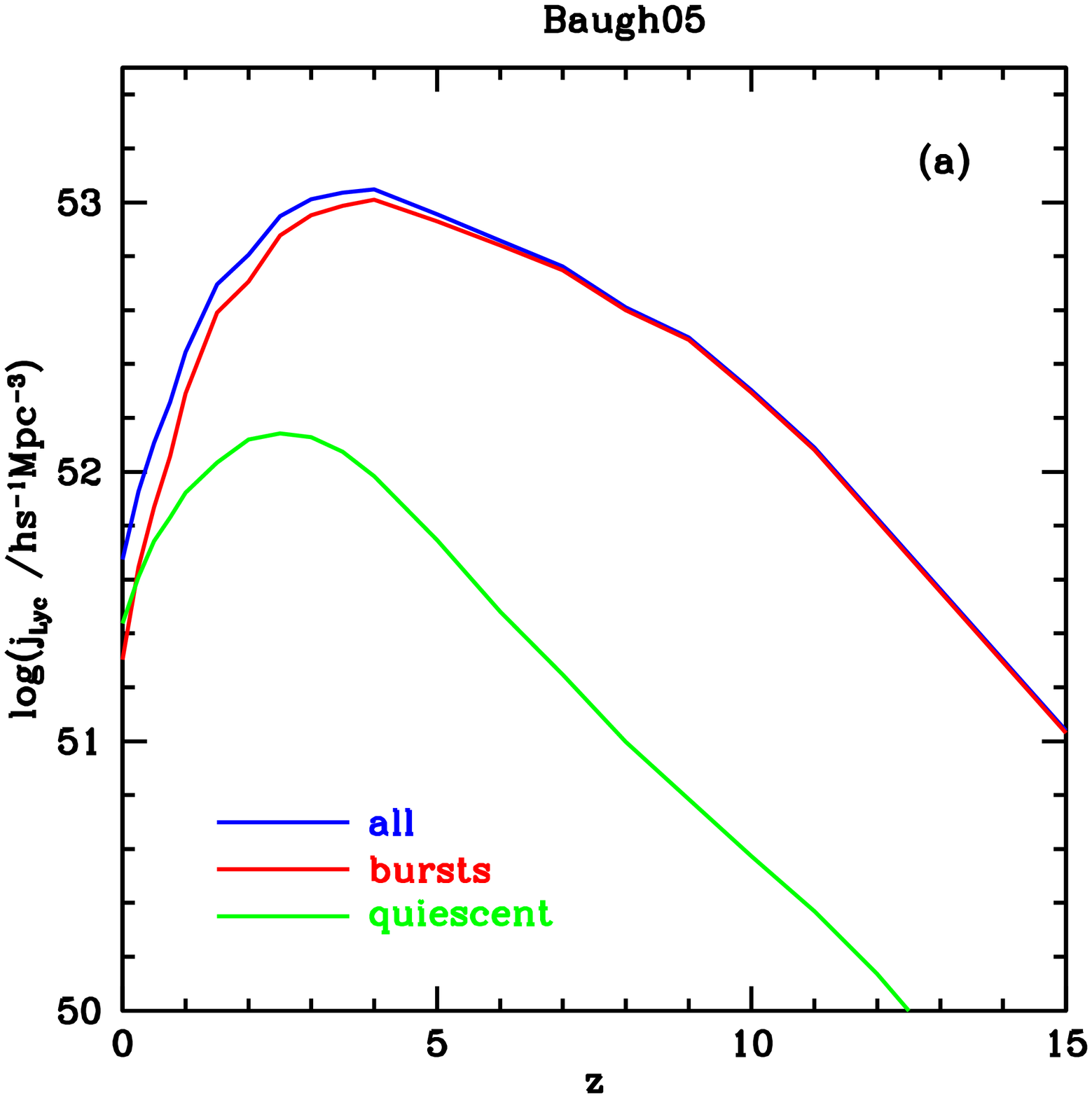}
\end{minipage}
\hspace{1cm}
\begin{minipage}{7cm}
\includegraphics[width=7cm]{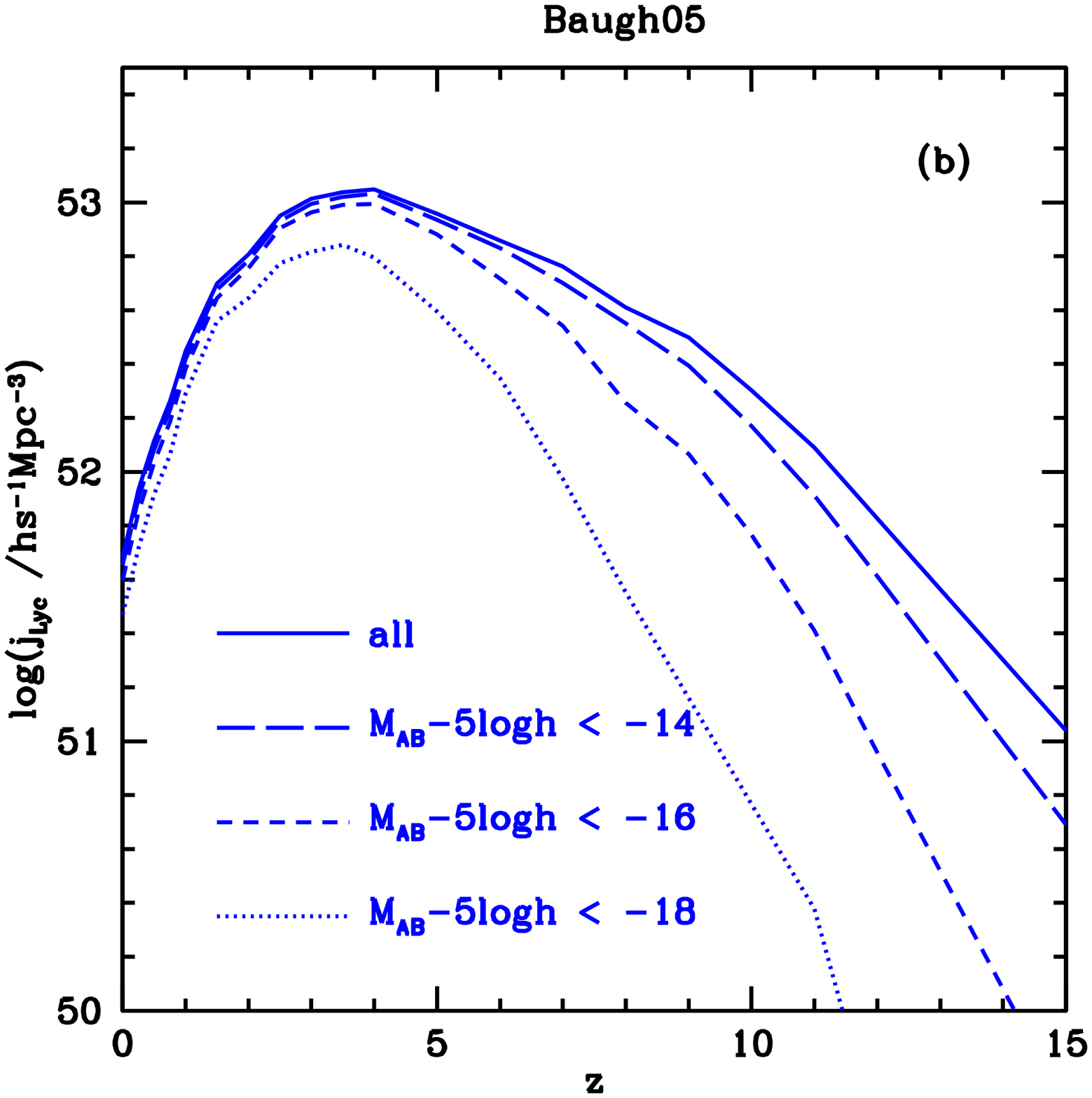}
\end{minipage}

\caption{Ionizing emissivity (number of Lyc photons per unit time per
  unit comoving volume) {\em vs} redshift in the \Bau\ model. (a) The
  blue curve shows the total emissivity (without dust extinction) and
  the green and red curves show the separate contributions of
  quiescent galaxies and ongoing bursts. (b) The contribution to the
  ionizing emissivity from galaxies with 1500\AA\ luminosities
  brighter than different limits, as indicated in the key.  }

\label{fig:jLyc}
\end{figure*}

%%%%%%%%%%%%%%%%%%%%%%%%%%%%%%%%%%%%%%%%%%%%%%%%%%%%%%%%%%%%%%%%%%%%%%%%%%%

Finally, we consider the production of hydrogen-ionizing Lyman
continuum (Lyc) photons, which is of critical importance for
reionizing the IGM. Fig.~\ref{fig:jLyc} shows the evolution of the Lyc
emissivity from galaxies predicted by the \Bau\ model. We plot the
emissivity without applying any correction for absorption of Lyc
photons by dust or gas in the galaxies, {({\em i.e.} assuming an
escape fraction of 100\%).} The actual escape fraction is probably
dominated by absorption on neutral hydrogen, but is currently very
uncertain both theoretically and observationally
\citep[e.g.][]{Loeb01,Benson06,Shapley06}. The left panel shows the
Lyc emissivities of quiescent and bursting galaxies as green and red
lines, with the total in blue. The Lyc emissivity is even more
dominated by bursts than is the 1500\AA\ emissivity (at least if
absorption by dust and gas are ignored). Bursts dominate at all
redshifts $z>0.2$, and the ionizing emissivity of bursts is 7--50
times larger than that of quiescent galaxies over the whole range
$z=3-15$. The reason for this is that stars formed with the top-heavy
$x=0$ IMF produce 11 times more Lyc photons per unit mass of stars
formed than the Kennicutt IMF (for Solar metallicity). For comparison,
the ratio is only 3.4 for production of 1500\AA\ photons. The right
panel shows the contributions to the Lyc emissivity from galaxies
brighter than various absolute magnitude limits in dust-extincted
1500\AA\ light. For example, galaxies brighter than $\MUV < -18$ are
predicted to emit 40\% of the hydrogen-ionizing photons at $z=5$, but
only 3\% at $z=10$. Resolving 80\% of the ionizing emissivity at
$z=10$ requires detecting galaxies down to $\MUV < -14$
{(corresponding to $m_{\rm AB}\approx 33$)}, as for the
1500\AA\ emissivity.

{In the \Bow\ model, ionizing emissivities are $\sim 7-40$
times lower than in the \Bau\ model over the redshift range $z=3-10$
(assuming a 100\% escape fraction in both cases). \citet{Raicevic10a}
have made a more detailed comparison of the emissivities in these two
models.}

\citet{Benson06} investigated reionization in the \Bau\ and other
\GALFORM\ models using an analytical Stromgren sphere model for the
growth of ionized regions in the IGM, and found that reionization was
predicted to occur at $z\approx 12$ for a Lyc escape fraction of
100\%.  Reionization in this model is studied in much greater detail
in \citet{Raicevic10a,Raicevic10b}, who use a radiative transfer
simulation to show that, for a Lyc escape fraction in the range
0.1--1, reionization should occur at $z\approx8-10$.

\section{Conclusions}
\label{sec:conc}

We have made a detailed investigation of the properties and evolution
of Lyman-break galaxies (LBGs) predicted by hierarchical models of
galaxy formation. We followed the galaxy formation process in the
framework of the $\Lambda$CDM cosmology using the \GALFORM\
semi-analytical model, which includes physical treatments of the
hierarchical assembly of dark matter halos, shock-heating and cooling
of gas, star formation, feedback from supernova explosions, AGN and
photoionization of the IGM, galaxy mergers and chemical
enrichment. The luminosities of galaxies are calculated from a stellar
population synthesis model, and dust extinction is then included using
a self-consistent theoretical model based on the results of radiative
transfer calculations. The dust mass is calculated from the predicted
mass and metallicity of the cold gas component, and this is combined
with the predicted galaxy radius to calculate the dust extinction
optical depth. The far-UV dust extinction is a critical component in
any model for LBGs.

We have presented predictions for two variants of the \GALFORM\
model. In the \citeauthor{Baugh05} (2005, \Bau) model, the formation
of very massive galaxies is inhibited by supernova-driven superwinds
which eject gas from halos, star formation at high redshifts is
dominated by starbursts triggered by galaxy mergers, and stars form in
these bursts with a top-heavy IMF. The top-heavy IMF was motivated by
the need to explain the number counts and redshift distributions of
the faint sub-mm galaxies. This model also matches a wide range of
other data on local galaxies (such as gas masses and disk sizes). In
the \citeauthor{Bower06} (2006, \Bow) model, the formation of very
massive galaxies is instead inhibited by AGN feedback which heats the
gas in halos, starbursts play a much smaller role in star formation,
and all stars form with a Solar neighbourhood IMF. The \Bow\ model
underpredicts the sub-mm number counts by more than an order of
magnitude, due to having too few very luminous and dusty star-forming
galaxies at high redshifts.  This shortcoming might be remedied by
introducing a top-heavy IMF in bursts, as we will explore in a future
paper. Both models match the present-day optical and near-IR
luminosity functions.

We first considered the evolution of the rest-frame far-UV luminosity
function (\S\ref{sec:lf-evoln}). Both models predict modest evolution
over the redshift range $3 \lsim z \lsim 8$, but a more rapid
evolution at higher redshifts, $z \gsim 8$, driven by the build-up of
dark matter halos. However, the models differ in that the \Bow\ model
predicts a more extended high-luminosity tail than the \Bau\ model,
once dust extinction is included. The effects of dust extinction on
the far-UV luminosity function are much larger in the \Bau\ model
($\sim 2$~mag) because the bright end of the luminosity function is
dominated by starbursts in which the dust content is enhanced by metal
production with the top-heavy IMF. We made a detailed comparison of
the predictions of both models with observed far-UV luminosity
functions of LBGs over the redshift range $z=3-10$. We found that the
\Bau\ model, without any modification of its parameters, predicts a
far-UV luminosity function in excellent agreement with current
observational data over this whole range. On the other hand, the \Bow\
model conflicts with the LBG observations at $z=3-7$ because it
predicts too many high-luminosity galaxies. We then investigated the
effect on the predicted luminosity functions of varying some of the
model parameters from their default values. Assuming a Solar
neighbourhood, rather than top-heavy, IMF in bursts has only a modest
impact on the far-UV luminosity function, because the effects of lower
intrinsic stellar luminosities are partly compensated by lower dust
extinctions. However, such a model predicts far too few sub-mm
galaxies. We find that the luminosity function over the range covered
by observational data is fairly sensitive to the assumed star
formation timescale in bursts, especially at higher redshifts, but is
less sensitive to the strength of supernova feedback, when these
parameters are varied over physically reasonable ranges. The inability
of the \Bow\ model to match the observed LBG luminosity function data
appears to be caused mainly by the short assumed star formation
timescale in bursts, rather than by the AGN feedback model or the
assumed IMF.

We next investigated a wide range of other physical properties of LBGs
predicted by the models (\S\ref{sec:props}). We first considered the
sizes of galaxies in the rest-frame far-UV, and compared to recent
observational measurements at $z \sim 2-7$. We found that both models
predicted sizes in reasonable agreement with observations at higher UV
luminosities, but only the \Bau\ model is consistent with observed
sizes at lower luminosities. We then presented predictions of the
\Bau\ model for stellar, halo and gas masses, clustering bias,
circular velocity, burst fractions, bulge-to-disk ratios, star
formation rates and gas and stellar metallicities, and made brief
comparisons with relevant observational data. The model predictions
appear to be broadly compatible with current observational constraints
(many of which are rather uncertain) in most cases. A particularly
interesting issue is the stellar masses - our predicted values are
well below observational estimates based on fitting stellar population
models to broad-band photometry. However, the observational estimates
all assume a Solar neighbourhood IMF, while in the \Bau\ model the LBG
population is dominated by starbursts forming stars with a top-heavy
IMF. The observationally inferred stellar masses therefore cannot be
directly compared with the values from the model. When instead we
compare the IR fluxes (which drive the observational stellar mass
estimates) directly, the model is much closer to the observations.  We
will investigate this important issue in more detail in a future
paper. We will also make a more detailed study of LBG clustering in
future work, since this provides constraints on the masses of the dark
matter halos hosting LBGs.

In \S\ref{sec:high-z}, we showed predictions for LBGs at very high
redshifts ($z=7-20$) {in the \Bau\ model,} including surface
densities of objects down to very faint apparent magnitudes
($m_{AB}=32$), relevant for observations with future telescopes such
as \JWST\ and ELTs. We find that
%% if the \Bau\ model remains accurate at $z>10$, then 
deep surveys planned with \JWST\ should be able to detect a few LBGs
at $z\sim 15$ and $m_{AB}\sim 30-31$, and many more at lower
redshifts. LBGs detected at $m_{AB}\sim 31$ are predicted to have
angular radii $\sim 0.02-0.05$~arcsec, depending only weakly on
redshift over this range, and to have circular velocities $\sim
30-100\kms$, again only weakly dependent on redshift.

In \S\ref{sec:UVdens}, we showed the predicted evolution of the far-UV
luminosity density, and its relation to the cosmic SFR history,
{again in the \Bau\ model}. The unextincted 1500\AA\ luminosity
density tracks the SFR density in high-mass stars ($m\gsim 5\Msol$)
more closely than the total SFR density, since the relative
contributions of quiescent and burst star formation (with Solar
neighbourhood and top-heavy IMFs respectively) change with
redshift. The effect of dust extinction on the far-UV luminosity
density is predicted to be large, $\approx 2$~mag at 1500\AA\ in the
range $3 \lsim z \lsim 15$, dropping to $\approx 1$~mag at
$z=0$. Finally, we considered the predicted contribution of galaxies
to the emissivity of ionizing photons which can reionize the IGM. For
a constant escape fraction of ionizing photons from galaxies, this
emissivity falls by a factor $\sim 100$ from its peak at $z\sim 5$ to
$z=15$. At high redshift, most of the ionizing photons are predicted
to come from very low luminosity galaxies, so that, for example, to
detect the galaxies responsible for $>50\%$ of the ionizing emissivity
at $z=10$ would require an LBG survey probing fainter than $\MUV \sim
-15$, {corresponding to an apparent magnitude $m_{AB}\sim 32$
at the same rest-frame wavelength.} The predictions of our model for
reionization of the IGM are discussed in much greater detail in
\citet{Raicevic10a,Raicevic10b}.

In conclusion, we find that the \citet{Baugh05} model, which was
originally constructed to match the far-UV luminosity function of LBGs
only at $z=3$, predicts results in remarkably good agreement with
subsequent observations of LBGs out to $z=10$. Further exploration of
whether this model provides a physically accurate description of LBGs
and other high-redshift galaxy populations will require more detailed
comparisons between the model predictions and observational data
(e.g. for stellar masses, clustering, and colours), but also new and
more sensitive observations.

\section*{Acknowledgements} 
This work was supported in part by the Science and Technology
Facilities Council rolling grant to the ICC. CSF acknowledges a Royal
Society Wolfson Research Merit Award. We thank Dan Stark,
{Milan Rai{\v c}evi{\'c} and Tom Theuns} for useful
discussions.

%----------------------------------------------
\bibliographystyle{mn2e}
\bibliography{paper}
%---------------------------------------------------------------------

\appendix

\section{Comparison of the Baugh et al. and Bower et al. models}
\label{sec:comp_models}

{The \Bau\ and \Bow\ models were originally constructed with
somewhat different aims, and their parameters for star formation,
feedback etc were chosen by comparison with different, though partly
overlapping, observational datasets. The parameters for both models
were constrained to reproduce the observed present-day B- and K-band
galaxy luminosity functions. The \Bau\ model was constructed to also
match a wide range of other properties of present-day galaxies,
including the far-IR (60$\mum$) luminosity function and
the dependence of disk size, gas content and metallicity on galaxy
luminosity. The \Bow\ model did not use any of these present-day
properties as constraints in choosing its parameters, but instead
aimed to reproduce the general form of the colour-magnitude
distribution of galaxies at $z=0$. In addition, the \Bau\ model was
designed to explain the observed number counts and redshift
distributions of the faint sub-mm (850$\mum$) galaxies, and the
rest-frame far-UV luminosity function of Lyman-break galaxies at
$z\approx 3$, while the \Bow\ model instead focused on matching the
observed evolution of the K-band luminosity function of galaxies at
$z\lsim 2$.} 

{Given these differences, it is not surprising that the two
models, although fundamentally quite similar, have different successes
and drawbacks, as discussed in subsequent papers. \citet{Gonzalez09}
compared both models side-by-side with observational data from the
SDSS on luminosity functions, colours, morphologies and sizes of
present-day galaxies. They found that both models are reasonably
consistent with the fraction of early-type galaxies as a function of
luminosity, but that the \Bow\ model is in better agreement with the
distribution of galaxy colours. Further, the \Bau\ model agrees much
better with the size-luminosity relation of disk galaxies, but both
models have problems explaining the sizes of spheroidal galaxies (as
also found by \citet{Almeida07}). \citet{Almeida08} found the \Bow\
model in better agreement with the luminosity function and correlation
function of Luminous Red Galaxies (LRGs) in the SDSS. \citet{Power10}
investigated the cold gas mass functions in both models, and found the
\Bau\ model in better agreement with data from the HIPASS 21cm survey
of the local universe.}

{At higher redshifts, the \Bau\ model seems in generally better
agreement with observations of actively star-forming galaxies at
$z\gsim 2$, while the  \Bow\ model is in better agreement with
observations of more passive galaxies at $z\lsim 2$. \citet{Bower06}
compared the evolution of the K-band luminosity function in both
models with observations at $z\leq 1.5$, and found that the \Bau\
model predicts somewhat too few high-luminosity galaxies at higher
redshifts. The \Bau\ model also predicts much lower number counts for
Extremely Red Objects (EROs) than \Bow, with the latter being in much
better agreement with observational data
\citep{Gonzalez-Perez09}. However, the \Bow\ model predicts number
counts of faint sub-mm galaxies which are too low by more than an
order of magnitude. The comparison of both models with observations of
LBGs is the topic of this paper.}

\end{document}